\documentclass[a4paper,fleqn,usenatbib]{mnras}

\usepackage[T1]{fontenc}

\usepackage{ae,aecompl}

\usepackage{graphicx}	
\usepackage{amsmath}	
\usepackage{amssymb}	
\usepackage{subfig}

\usepackage{hyperref}

\usepackage{etoolbox}
\makeatletter
\newcount\c@additionalboxlevel
\newcount\c@maxboxlevel
\patchcmd\@combinedblfloats{\box\@outputbox}{%
   \stepcounter{additionalboxlevel}%
   \box\@outputbox
}{}{\errmessage{\noexpand\@combinedblfloats could not be patched}}

\AtBeginShipout{%
   \ifnum\value{additionalboxlevel}>\value{maxboxlevel}%
     \typeout{Warning: maxboxlevel might be too small, increase to %
       \the\value{additionalboxlevel}%
     }%
   \fi
   \@whilenum\value{additionalboxlevel}<\value{maxboxlevel}\do{%
     \typeout{* Additional boxing of page `\thepage'}%
     \setbox\AtBeginShipoutBox=\hbox{\copy\AtBeginShipoutBox}%
     \stepcounter{additionalboxlevel}%
   }%
   \setcounter{additionalboxlevel}{0}%
}
\makeatother

\title[AGN specfic accretion-ratio distribution]{Observational constraints on
  the specific accretion-rate distribution of X-ray selected AGN} 

\author[Georgakakis et al.]{
A. Georgakakis,$^{1,2}$\thanks{E-mail: age@mpe.mpg.de}, J. Aird$^{3}$,
A. Schulze$^{4,5}$, T. Dwelly$^{1}$, M. Salvato$^{1}$, K. Nandra$^{1}$,
\newauthor A. Merloni$^{1}$, Donald P. Schneider$^{6,7}$
\\
$^1$Max Planck Institut f\"{u}r Extraterrestrische  Physik, Giessenbachstra\ss e, 85748 Garching, Germany\\ 
$^2$National Observatory of Athens, V.  Paulou  \& I.  Metaxa, 11532,  Greece\\ 
$^3$Institute of Astronomy, University of Cambridge, Madingley Road, Cambridge, CB3 0HA\\
$^{4}$National Astronomical Observatory of Japan, Mitaka, Tokyo 181-8588, Japan\\
$^{5}$Kavli IPMU (WPI), UTIAS, The University of Tokyo, Kashiwa, Chiba 277-8583, Japan\\
$^{6}$ Department of Astronomy and Astrophysics, The Pennsylvania
State University, University Park, PA 16802\\
$^{7}$Institute for Gravitation and the Cosmos, The Pennsylvania State University,University Park, PA 16802
}

\date{Accepted XXX. Received YYY; in original form ZZZ}

\pubyear{2016}

\begin{document}
\label{firstpage}
\pagerange{\pageref{firstpage}--\pageref{lastpage}}
\maketitle

\begin{abstract}This  paper  estimates   the  specific  accretion-rate
distribution of  AGN using a  sample of  4821 X-ray sources  from both
deep and shallow surveys.  The specific accretion-rate distribution is
defined as the  probability of a galaxy with a  given stellar mass and
redshift hosting an  active nucleus with a  certain specific accretion
rate.  We  find that  the  probability  of  a  galaxy hosting  an  AGN
increases with  decreasing specific accretion rate.  There is evidence
that   this  trend   reverses   at  low   specific  accretion   rates,
$\lambda\la10^{-4}-10^{-3}$  (in Eddington  units).  There  is also  a
break close to the Eddington limit,  above which the probability of an
accretion  event   decreases  steeply.  The   specific  accretion-rate
distribution  evolves such  that the  fraction of  AGN among  galaxies
drops toward lower  redshifts. This decrease in the AGN  duty cycle is
responsible for the  strong evolution of the accretion  density of the
Universe  from  redshift  $z\approx1-1.5$  to  the  present  day.  Our
analysis  also  suggests  that  this evolution  is  accompanied  by  a
decoupling of accretion events onto  black holes from the formation of
stars in  galaxies. There is also  evidence that at earlier  times the
relative  probability of  high vs  low specific  accretion-rate events
among  galaxies increases.  We argue  that this  differential redshift
evolution of the AGN duty cycle with respect to $\lambda$ produces the
AGN downsizing trend, whereby luminous  sources peak at earlier epochs
compared to less  luminous ones. Finally, we also  find a stellar-mass
dependence  of the  specific  accretion-rate  distribution, with  more
massive galaxies avoiding high specific accretion-rate events.
\end{abstract}




\section{Introduction}

X-ray surveys in  the last 15 years have provided  an excellent census
of  the Active  Galactic  Nuclei (AGN)  population  and describe  with
increasing accuracy how the space  density of these sources depends on
redshift, accretion luminosity and amount of line-of-sight obscuration
\citep[e.g.,][]{Brandt_Alexander2015}.   The  diverse studies  of  the
accretion  history  of  the  Universe have  established  a  number  of
observational facts and  trends.  We now know for  example, that about
80\% of the  black-hole growth in the Universe is  taking place behind
obscuring gas  and dust clouds and  that the obscured AGN  fraction is
higher at earlier  cosmic times \cite[e.g.,][]{Ueda2014, Buchner2015}.
The space density of AGN increases  rapidly from the local Universe to
redshift $z\approx1.5-2$  \citep[][]{Ueda2003, Hasinger2005, Aird2010,
Aird2015}, and then shows a strong  decline, at least for the luminous
part    of    the     population,    beyond    redshift    $z\approx3$
\citep[e.g.,][]{Aird2008,          Vito2014,          Georgakakis2015,
Marchesi_z3_2016}.   These evolutionary  patterns  are  not unique  to
X-ray AGN  but are also  found in  studies of optically  selected QSOs
\citep[][]{Osmer1982, Schmidt1995, Richards2006, Croom2009, Ikeda2011,
Ross2013, McGreer2013}.  X-ray selected AGN also appear to evolve in a
anti-hierarchical manner,  which is  often termed  downsizing, whereby
more luminous sources have space densities that peak at earlier epochs
compared  to  less  luminous ones  \citep[][]{Ueda2003,  Hasinger2005,
Barger2005,  Ebrero2009}.   Striking  similarities are  also  observed
between the  global evolution  of the  accretion-rate density  and the
star-formation  rate  density  of  the  Universe  \citep[][]{Ueda2003,
Aird2010, Zheng2009, Aird2015}.

Work is also ongoing to  interpret the evolutionary patterns described
above in a  physical context.  One approach is to  study the accretion
events  onto  supermassive  black  holes in  relation  to  their  host
galaxies to explore which galaxy  types and environments are conducive
to black-hole growth.   The diversity of AGN  host galaxy morphologies
for example, provides constraints on  the role of major-galaxy mergers
or secular processes as triggering  mechanisms of the observed nuclear
activity  \citep[e.g.,][]{Georgakakis2009,  Gabor2009,  Cisternas2011,
Kocevski2012, Cheung2015, Cisternas2015}.  The level of star-formation
in AGN  hosts is used  as a proxy of  gas availability to  explore the
necessary     conditions    for     black-hole    accretion     events
\citep[e.g.,][]{Santini2012, Rovilos2012,  Rosario2013, Mullaney2015}.
The position of AGN on the  cosmic web and their clustering properties
relative  to galaxies  may hold  important clues  on the  diversity of
black-hole fuelling  modes and  the role  of Mpc-scale  environment on
accretion      events     \citep[][]{Allevato2011,      Fanidakis2013,
Allevato2014}.

The studies  above explore AGN  host galaxy properties in  relation to
accretion  luminosity  measured  by either  X-ray  or  multiwavelength
observations.   An alternative  important parameter  of the  accretion
process that provides more physical insights into the fuelling mode of
AGN is the Eddington ratio, which measures the rate a black hole grows
relative to  its maximum  capacity.  Different  AGN fuelling  modes or
triggering  mechanisms  may  produce  events  with  similar  accretion
luminosities     but     distinct    Eddington-ratio     distributions
\citep[e.g.,][]{Fanidakis2012, Hirschmann2014}.  One approach to place
constraints on the  Eddington-ratio distribution of AGN  as a function
of  redshift is  via the  continuity equation  of the  black-hole mass
function  using  the AGN  luminosity  function  as boundary  condition
\citep[][]{Merloni_Heinz2008,  Shankar2013,   Aversa2015}.   A  direct
measurement of the Eddington ratio  distribution of AGN populations is
limited    to    broad     optical    emission-line    sources    only
\citep[e.g.,][]{Kelly2010, Schulze2015},   which  are   needed   to  estimate   the
black-hole mass of individual systems. A relatively recent development
that has opened new opportunities  in studies of black-hole growth has
been the determination of the specific accretion rate of large samples
of  AGN.   This  quantity  measures the  rate  a  black-hole  accretes
material relative  to the stellar  mass of  its host galaxy  and under
certain  assumptions it  can be  viewed as  a proxy  of the  Eddington
ratio.   The advantage  is that  the  specific accretion  rate can  be
measured  for AGN  over a  wide  range of  accretion luminosities  and
levels  of   line-of-sight  obscuration,  thereby  providing   a  more
representative  view  of  the  AGN  population.   Early  observational
studies suggested a  nearly universal shape of  the specific accretion
rate distribution of  X-ray selected AGN, approximated  by a power-law
with a  slope that does not  depend strongly on either  redshift or
host galaxy  stellar mass \citep[][]{Aird2012},  and a break  close to
the Eddington  limit \citep[][]{Bongiorno2012}.  However,  more recent
investigations suggest a more complex  picture, where the shape of the
specific accretion rate  distribution may depend on  the redshift, the
stellar    mass    \citep[][]{Bongiorno2016},   and    possibly    the
star-formation  rate  of AGN  hosts  \citep[][]{Kauffmann_Heckman2009,
Georgakakis2014, Azadi2015}.   Therefore a  detailed study of  how the
specific  accretion rate  distribution of  AGN changes  with redshift,
and/or  diagnostics of  the physical  conditions on  larger scales  is
important to get insights into the  processes at play. A limitation of
current observational studies outside the  local Universe is the small
sample size  that affects the  statistical reliability of  the results
and the ability to explore trends with redshift.

In this paper  we combine the largest sample to  date of X-ray sources
with stellar mass  estimates for their host galaxies  to constrain the
specific accretion-rate distribution of AGN as a function of redshift.
The sample is  compiled from X-ray survey fields  with different sizes
and depths, from deep and pencil-beam,  such as the 4\,Ms Chandra Deep
Field South with an area of about $\rm 0.08 \, deg^2$ \citep{Xue2011},
to  wide  and  shallow, such  as  a  subset  of  the XMM  Slew  Survey
\citep{Saxton2008}  with  approximately  6\,s exposure  over  about  a
thousand  square degrees.   A Bayesian  methodology is  developed that
uses the stellar mass function of  the galaxy population as a boundary
condition to estimate the specific accretion-rate distribution of AGN.
Our  analysis is  non-parametric, i.e.,  no  shape is  imposed on  the
specific accretion-rate  distribution of AGN.  This  approach also has
the advantage that  uncertainties in the properties  of individual AGN
that are used  in the analysis, such as stellar  masses, redshifts and
X-ray luminosities  are correctly  propagated into the  analysis.  The
combination of samples with diverse X-ray depths and sky areas provide
a satisfactory  coverage of  the luminosity-redshift plane  and yields
robust constraints on the specific  accretion rate distribution of AGN
over a range of redshifts and specific accretion rates.

Parallel to  the work presented  in this paper,  Aird et  al. (in
prep.) have  developed an independent  approach to infer  the specific
accretion-rate  distribution  of  black holes  within  galaxy  samples
binned in redshift, stellar mass  and star-formation rate. Although we
share some of the  X-ray data with the Aird et  al. (in prep.)  paper,
the two works differ in  the supporting multiwavelength photometry and
the overall  methodology adopted to infer  the specific accretion-rate
distribution, i.e., the probability of a  galaxy hosting an AGN with a
certain specific accretion rate. In our work we use ony X-ray selected
AGN and  require that the  convolution of the  specific accretion-rate
distribution and  the galaxy  stellar mass  function yields  the X-ray
luminosity function of  AGN. Aird et al. (in prep.)   start with large
galaxy samples, binned by stellar mass and redshift, and use the X-ray
data to infer the probability of a  galaxy hosting a black hole with a
certain  specific accretion  rate  \citep[see also][]{Aird2017}.   The
results  from these  two independent  approaches are  compared in  the
Appendix  of our  paper.  In  the  calculations that  follow we  adopt
cosmological parameters  $\rm H_0 = 70  \, km \, s^{-1}  \, Mpc^{-1}$,
$\Omega_M = 0.3$, $\Omega_\Lambda = 0.7$. This choice is to facilitate
comparisons with the results from previous studies.

\section{Data Products}

\subsection{Chandra and XMM-Newton survey fields}

The X-ray survey  fields used in this paper  include the 4\,Ms Chandra
Deep Field South  \citep[CDFS;][]{Xue2011, Rangel2013}, the deep ($\rm
\approx800\,ks$) Chandra  survey of the  All-wavelength Extended Groth
Strip \citep[AEGIS-XD,][]{Nandra2015},  the recently completed Chandra
COSMOS-Legacy survey  \citep[][]{Civano2016} and the  subregion of the
XMM-XXL  \citep{Pierre2016} equatorial  field that  overlaps  with the
VISTA  (Visible  and Infrared  Survey  Telescope  for Astronomy)  Deep
Extragalactic Observations survey \citep[VIDEO;][]{Jarvis2013}.

The Chandra observations of the  survey fields above are analysed in a
homogeneous way by  applying the data processing steps  and the source
detection described in  \cite{Laird2009} and \cite{Nandra2015} as well
as   the    sensitivity   map   generation    methods   described   in
\cite{Georgakakis2008_sense}.   In  this paper  we  use X-ray  sources
selected  in   the  0.5-7\,keV  (full)  band  with   a  Poisson  false
probability threshold  of $<4\times10^{-6}$.  The  source catalogue of
the XMM-XXL  equatorial field is presented by  \cite{Liu2016} based on
the    data   reduction   and    analysis   pipeline    described   in
\cite{Georgakakis_Nandra2011}. We  use sources in  that field selected
in  the 2-8\,keV  (hard) band  to the  same Poisson  false probability
threshold as above.

The  identification of  X-ray  sources with  optical and  mid-infrared
counterparts     follows      the     likelihood      ratio     method
\citep{Sutherland_and_Saunders1992, Ciliegi2003,  Brusa2007}. Specific
details on the identification of  sources in the Chandra survey fields
used in this work can be found in \cite{Aird2015}.  The association of
XMM-XXL  sources   with  multiwavelength  photometric   catalogues  is
described by \cite{Georgakakis2017xxl} and \cite{Menzel2016}. In
this paper  we use a  subregion of  the XMM-XXL equatorial  field with
total  area   of  $\rm 3.6\,deg^2$,   within  which  there   is  available
near-infrared photometry from the 4th  Data Release (DR4) of the VIDEO
survey produced by the Vista Science Archive \citep[VSA;][]{Irwin2004,
Hambly2008, Cross2012}.  We adopt  this approach because near-infrared
data  are needed  for the  estimation of  stellar masses  and specific
accretion rates for AGN host galaxies (see next section).

The compilation  of spectroscopic redshifts  for X-ray sources  in the
Chandra  fields  is presented  in  \cite{Georgakakis2015}  and in  the
XMM-XXL in \cite{Georgakakis2017xxl} and \cite{Menzel2016}.  These
papers also describe the estimation of photometric redshifts for X-ray
sources based on methods developed by \cite{Salvato2009, Salvato2011}.
For the new Chandra  COSMOS-Legacy survey, spectroscopic redshifts are
from  the   public  releases  of  the   VIMOS/zCOSMOS  bright  project
\citep{Lilly2009}, the  compilation of spectroscopy  for X-ray sources
presented  by \cite{Civano2012}  and \cite{Marchesi2016},  the MOSFIRE
Deep   Evolution   Field   \citep[MOSDEF,][]{Kriek2015},   the   PRIsm
MUlti-object Survey  \citep[PRIMUS,][]{Coil2011}, the VIMOS Ultra-Deep
Survey    \citep[VUDS,][]{LeFevre2015},     the    3D-HST    programme
\citep{Skelton2014,  Momcheva2015} and  the Sloan  Digital  Sky Survey
\citep[SDSS;][]{Alam2015}.   Photometric  redshifts  for  the  Chandra
COSMOS-Legacy  X-ray sources  are from  \cite{Marchesi2016}.   For the
small  number  of  X-ray  sources  in our  reduction  of  the  Chandra
COSMOS-Legacy  data  that do  not  appear  in the  \cite{Marchesi2016}
catalogue  photometric redshifts are  estimated following  the methods
described in \cite{Aird2015}.

\begin{figure}
\begin{center}
\includegraphics[height=0.85\columnwidth]{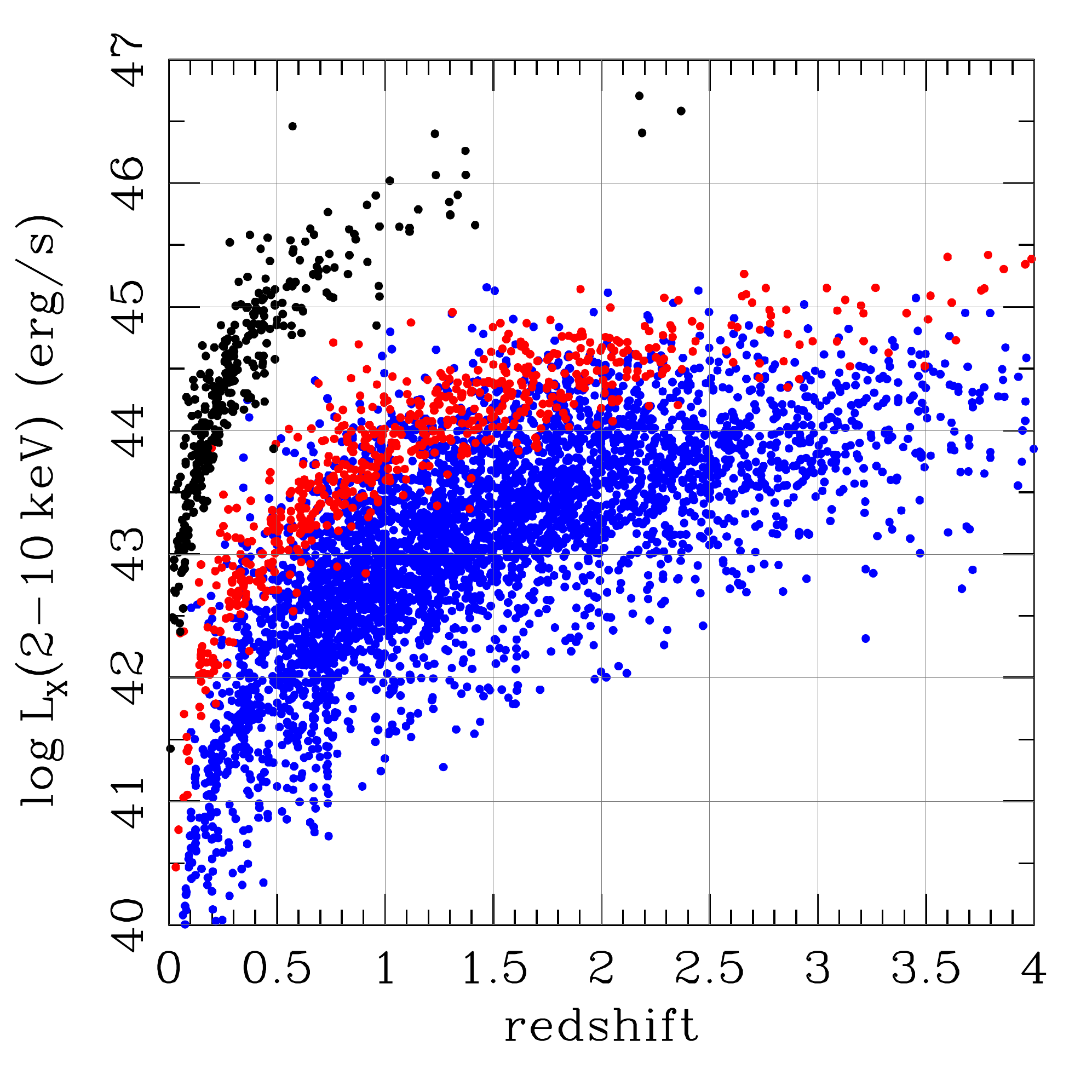}
\end{center}
\caption{X-ray luminosity  as a function redshift for  the sample 4821
X-ray selected AGN  used in this work. The blue  data points are X-ray
sources  detected  in  the   Chandra  surveys,  the  red  data  points
correspond to  the XMM-XXL sample and  the black ones to  the XMM Slew
survey.   Each data  point is  randomly drawn  from  the 2-dimensional
X-ray  luminosity and  redshift probability  distribution  function of
individual sources.  X-ray luminosities  in the 2-10\,keV energy range
are  estimated from  the observed  count-rates in  the  spectral bands
0.5-7\,keV ({\it Chandra}),  2-7\,keV (XMM-XXL) and 0.2-2\,keV (XMMSL)
assuming a spectral index $\Gamma=1.9$}\label{fig:lxz}
\end{figure}

Figure    \ref{fig:lxz}    presents    the   distribution    on    the
luminosity-redshift plane of the X-ray sources detected in the Chandra
and  XMM-XXL survey  fields described  above.   Table \ref{table_data}
provides further  information on the  number of X-ray sources  in each
field  as  well  as  the corresponding  spectroscopic  or  photometric
redshift   completeness.  Figure   \ref{fig:area_slew}   presents  the
sensitivity curves of the {\it Chandra} and XMM-XXL surveys. 

\subsection{XMM Slew  Survey}

The  data  from the  {\it  Chandra}  and  {\it XMM}  dedicated  survey
programmes are  supplemented by  the wide area  and shallow  {\it XMM}
Slew Survey \citep[XMMSL,][]{Saxton2008}.  Spectroscopic redshifts for
the  XMMSL  sources  are  primarily from  the  SPIDERS  \citep[SPectroscopic
IDentification of eROSITA Sources,][]{Dwelly2017}  programme
of the  SDSS-IV \citep{Smee2013, Dawson2016} using  the SDSS telescope
\citep{Gunn2006}  and  spectrographs \citep{Smee2013}.   Our  analysis
uses   X-ray   sources  in   the   ``clean''   version  1.6   of   the
XXMSL\footnote{\url{http://www.cosmos.esa.int/web/xmm-newton/xmmsl1d-ug}}
catalogue that overlap with a  total of 3159 SDSS spectroscopic plates
observed up  to 11  May 2016  (inlcuding 392 observed  as part  of the
SDSS-IV). Spectroscopic data obtained up to that date will be included
in  the SDSS  Data Release  14 (DR14),  which will  be made  public in
summer 2017.   The total overlap between  the XMMSL v1.6 and  the DR14
plates  is  $\rm  1165\,deg^2$.   Within   that  area  there  are  942
detections  which  correspond to  816  unique  XMMSL sources.   Figure
\ref{fig:xmmsl} shows  the sky distribution of  the DR14 spectroscopic
plates within  the SDSS-IV footprint  and the spatial  distribution of
the XMMSL exposures within these plates.

The  XMMSL catalogue  lists  sources detected  independently in  three
energy bands, $0.2-2$, $0.2-12$ and  $2-12$\,keV. In this paper we use
$0.2-2$\,keV detected  sources from the  `clean' version of  the XMMSL
catalogue. The  choice of energy  interval is motivated by  the higher
sensitivity   and   lower   expected   spurious   detection   fraction
\citep[0.7\%][]{Saxton2008} compared  to e.g., the  2-12\,keV selected
sample.  The  soft-band selection  introduces a bias  against obscured
sources. The XMMSL sample however,  is dominated by luminous AGN ($L_X
(  \rm 2  - 10  \, keV  ) \ga  10^{44} \,  erg \,  s^{-1}$l;  see Fig.
\ref{fig:lxz}),  which  are  expected  to  include  a  large,  if  not
dominant,  fraction of unobscured  sources \citep[e.g.,][]{Akylas2006,
Ueda2014}.

Because of the  small exposure time per pixel  (typically few seconds)
the  XMMSL sources  typically consist  of  a small  number of  photons
detected against an almost  zero background. The estimated count rates
and fluxes are  therefore severely affected by the  Eddington bias. We
account for this effect using the source and background counts as well
as the exposure times listed in the XMMSL source catalogue.  Under the
assumption of Poisson statistics the  probability of a source having a
observed count rate $CR$ is

\begin{equation}\label{eq:slew_pcr}
P(CR, T) = \frac{E^T \, e^{-E}}{T!}.
\end{equation}

\noindent  where $T$  is the  observed  number of  counts (source  and
background) and  $E$ is  the expected number of  counts assuming  that the
source  count  rate  on  the  EPIC-PN  detector  is  $CR$\footnote{The
EPIC-MOS detectors are  not used in slew mode.},  the exposure time is
$t$, and the level of background at the source position is $B$

\begin{equation}
E = CR \times t + B.
\end{equation} 

\noindent The  conversion of 0.2-2\,keV  count rates to fluxes  in the
same energy interval assumes  a power-law spectral energy distribution
with  index $\Gamma=1.9$  modified  by the  Galactic neutral  hydrogen
column density,  $N_H$, in the direction  of each source\footnote{This
  is different from the XMMSL  v1.6 catalogue, where the listed fluxes
  are  determined assuming  a fixed  Galactic hydrogen  column density
  $N_H=  \rm 3.0\times10^{20}  cm^{-2}$.}.  For  the 0.2-2\,keV  XMMSL
sample  that overlaps  with the  SDSS  DR14 plates  the mean  Galactic
neutral hydrogen column density is $\log [ N_H/\rm cm^{-2}] =20.3$ and
the standard deviation of the distribution is 0.26.

\begin{figure}
\begin{center}
\includegraphics[height=0.85\columnwidth]{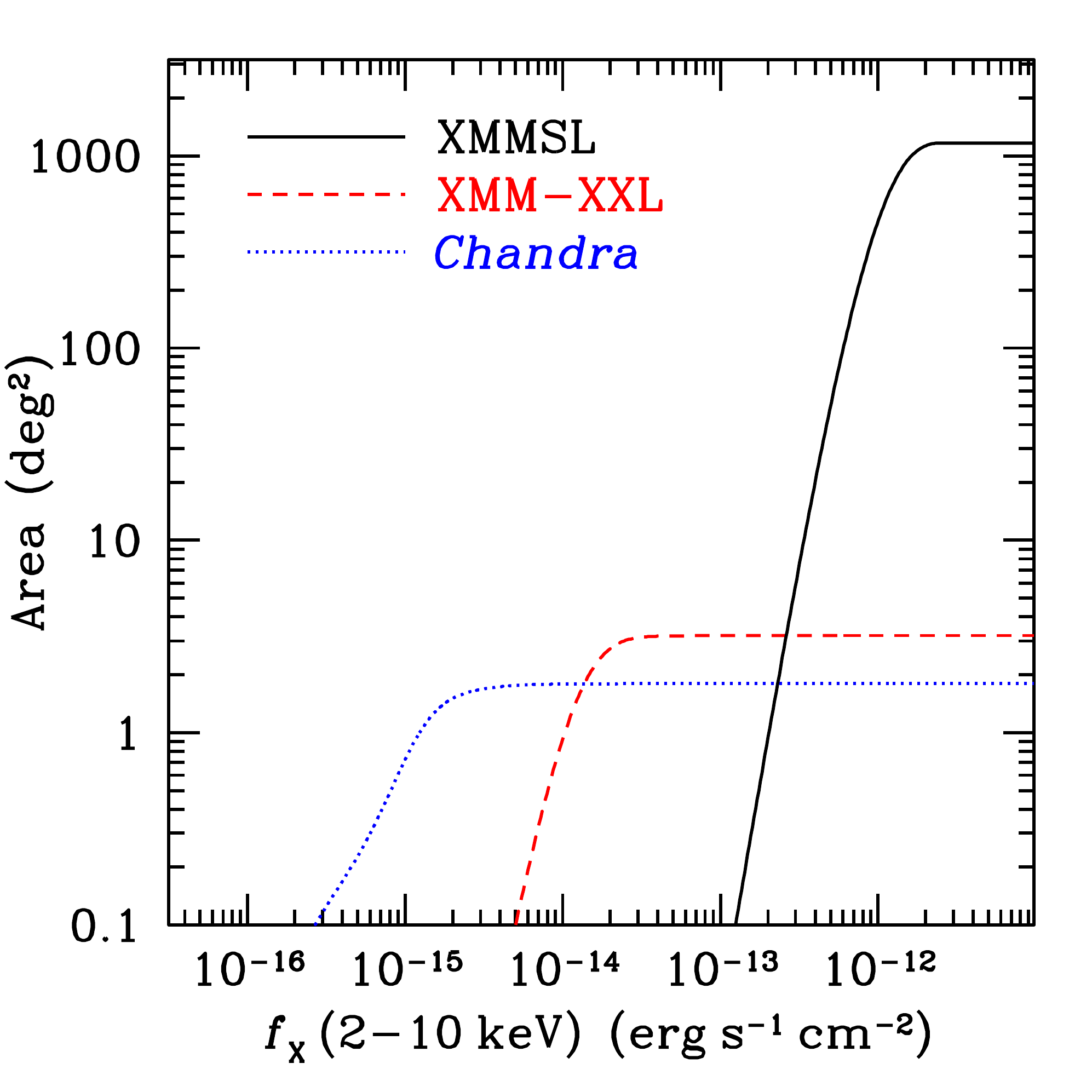}
\end{center}
\caption{Survey  area  (y-axis)  sensitive   to  X-ray  sources  of  a
particular  flux in  the  2-10\,keV energy  band (x-axis).   Different
colours and line types correspond  to the different X-ray surveys used
in our analysis, XMMSL  (black continuous), XMM-XXL (red dashed), {\it
Chandra} survey  fields (blue  dotted). For illustration  purposes the
count  rates from these  surveys have  been converted  to flux  in the
2-10\,keV band assuming an  X-ray spectral index of $\Gamma=1.9$.  The
XMMSL sensitivity curve is normalised  to the total overlap area ($\rm
1165\, deg^2$)  between the XMMSL  v1.6 source catalogue and  the DR14
SDSS-IV    plates,    accounting     for    the    SDSS    bright-star
mask.}\label{fig:area_slew}
\end{figure}

\begin{figure*}
\begin{center}
\includegraphics[height=0.85\columnwidth]{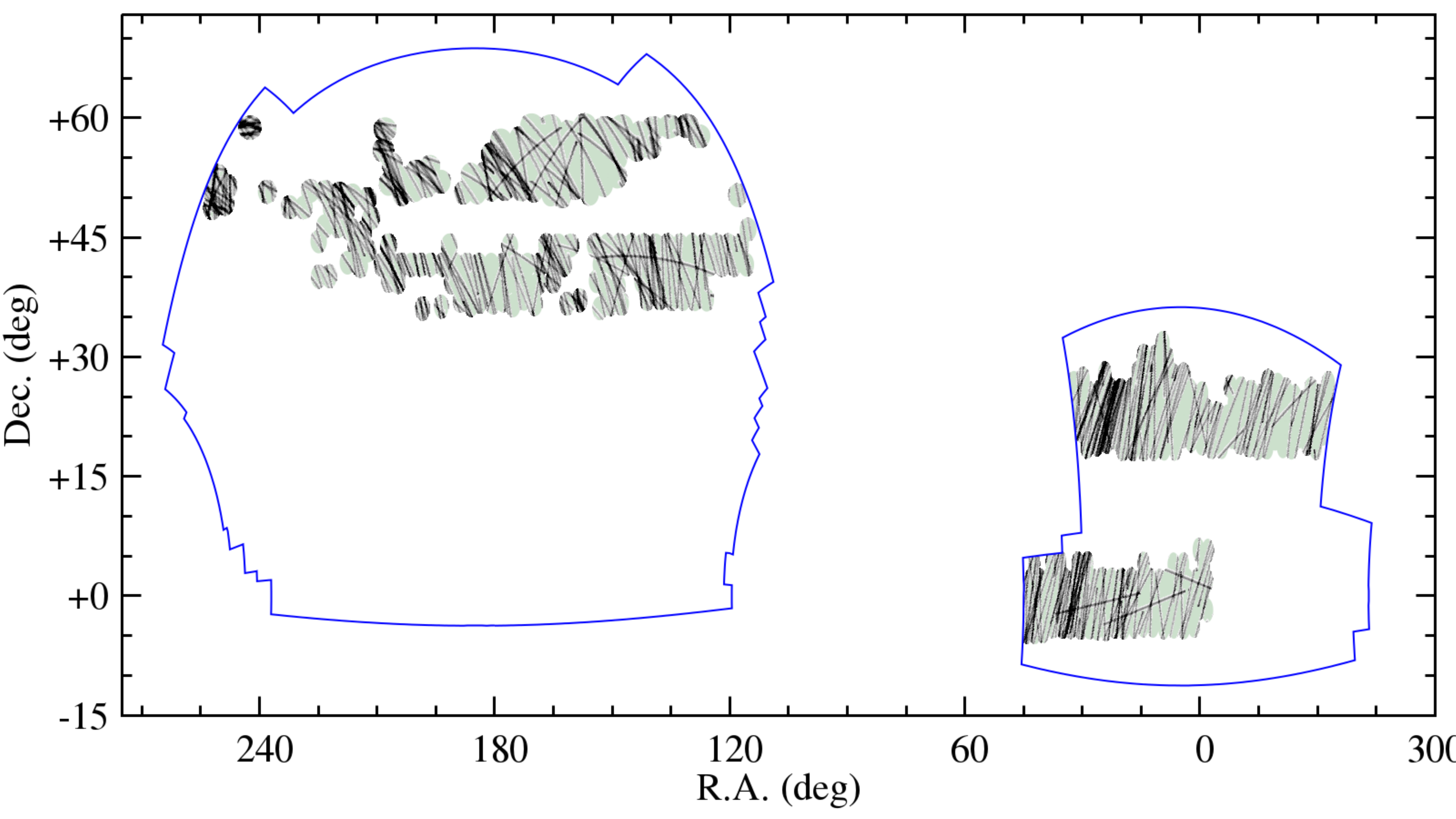}
\end{center}
\caption{Spatial  distribution of  the XMM  Slew sample  that overlaps
  with the SDSS-IV DR14 plates.  The blue lines mark the boundaries of
  the  BOSS  imaging  footprint,   the  shaded  regions  within  these
  boundaries show the positions in J2000 equatorial coordinates of the
  SDSS-DR14 spectroscopic plates and the dark tracks correspond to the
  XMM Slew exposure maps.  }\label{fig:xmmsl}
\end{figure*}

Although  sensitivity maps  are not  among  the data  products of  the
XMMSL,   it   is  possible   to   generate   the  selection   function
post-processing   \citep[e.g.,][]{Warwick2012}.    We  construct   the
0.2-2\,keV  sensitivity curve under  the assumption  that at  the flux
limit of the XMMSL \citep[$f_X(\rm 0.2-2\,keV)\ga 5.7\times10^{-13} \,
erg  \, s^{-1}  \, cm^{-2}$;  ][]{Saxton2008} the  differential number
count  distribution  of  the  extragalactic  component  of  the  XMMSL
catalogue is well represented by a power-law with the Euclidean slope,
i.e.,  $dN/dCR \propto  CR^{-2.5}$.  This  is a  reasonable assumption
given that the X-ray number counts in the 0.5-2\,keV band only deviate
from the Euclidean slope below the flux level of $f_X( \rm 0.5-2\,keV)
\approx     10^{-14}    \,     erg    \,     s^{-1}     \,    cm^{-2}$
\citep[e.g.,][]{Georgakakis2008_sense}.  The  sensitivity curve of the
sample  can  then   be  determined  as  the  ratio   of  the  observed
extragalactic  differential  number  counts,  $(dN/dCR)_{obs}$,  i.e.,
before accounting  for the  source detection limit  selection effects,
and the  expected true  differential number count  distribution, i.e.,
$dN/dCR  \propto CR^{-2.5}$.  The  $(dN/dCR)_{obs}$ is  constructed by
summing up the $P(CR,  T)$ probability density functions of individual
sources    estimated   from   Equation    \ref{eq:slew_pcr}.    Figure
\ref{fig:area_slew} presents the resulting sensitivity curve.

The XMMSL sources are associated with mid-infrared counterparts in the
AllWISE  catalogue   \citep{Cutri2013}  using   a  search   radius  of
60\,arcsec.   This   angle  should   be  compared  with   the  typical
$1\,\sigma$  rms  positional  uncertainty  of  the  XMMSL  sources  of
8\,arcsec \citep[][]{Saxton2008}.   Full details  of the  targeting of
the XMMSL sources are presented in \cite{Dwelly2017}. In brief,
a novel Bayesian cross-matching algorithm  is used (Salvato et al.  in
prep.),   which   is   based   on   the   techniques   introduced   by
\cite{Budavari_Szalay2008} and  \cite{Rots_Budavari2011}.    In   addition   to
astrometric  uncertainties,  the  identification is  guided  by  prior
information  on the  expected  distribution of  X-ray  sources in  the
mid-infrared  colour-magnitude  space  defined   by  the  WISE  $[W2]$
magnitude and  the $[W1-W2]$  colour. The  prior is  constructed using
bright  X-ray  sources [$f_X(\rm  0.2-2\,keV)  \ga10^{-13}  \, erg  \,
s^{-1}\, cm^{-2}$] from the 3XMM survey \citep{Watson2009, Rosen2016}.
The output of  the Bayesian cross-matching algorithm  is the posterior
probability, $P_{post}$, of  an XMMSL source being  associated with an
AllWISE counterpart.   In the case of  multiple potential associations
the one with the highest  $P_{post}$ is identified as the counterpart.
XMMSL associations  with $P_{post}>0.01$ are among  the candidates for
follow-up spectroscopy as part of the SPIDERS programme of the SDSS-IV
\citep{Dwelly2017}.  At this $P_{post}$  threshold the expected
spurious identification  rate is  about 12\%.   This value  is derived
empirically  using a  sample of  bright  X-ray sources  from the  3XMM
\cite[]{Watson2009},   the   Swift    XRT   Point   Source   Catalogue
\citep[1SXPS;][]{Evans2014}  and  the   Chandra  Source  Catalog  v1.1
\citep[CSC;][]{Evans2010}.  Optical  counterparts are searched  for in
the  SDSS photometric  catalogue  using the  AllWISE  positions and  a
search  radius of  1.5\,arcsec.  The  spectroscopic redshifts  for the
XMMSL sample  are from the SDSS-IV  SPIDERS \citep{Dwelly2017}
and    eBOSS   \cite[extended    Baryon   Oscillation    Spectroscopic
Survey,][]{Myers2015}   programmes,   as   well   as   previous   SDSS
spectroscopic  surveys, i.e.,  SDSS-III BOSS  \cite[Baryon Oscillation
Spectroscopic      Survey,][]{Dawson2013}      and     the      SDSS-I
\citep{Abazajian2009}.

The extragalactic sample used in this work is selected from a total of
595 XMMSL  sources that are  detected in the 0.2-2\,keV  spectral band
(B5 in XMMSL conventions) above  the maximum likelihood threshold {\sc
  det\_ml}=10 and overlap with the  SDSS-DR14 plates.  As a first step
we exclude  sources that lie  within the bright-star masks  defined as
part of the tiling process of the SDSS-III BOSS \citep{Dawson2013} and
SDSS-IV eBOSS  \citep{Dawson2016}.  This filtering reduces  the number
of 0.2-2\,keV selected XMMSL sources to 479.

Next  we  apply  a  cut   in  the  Bayesian  identification  posterior
probability,  $P_{post}>0.01$,  i.e.,  the  same  threshold  used  for
defining  the targets  for follow-up  spectroscopy with  SPIDERS. This
criterion  reduces  the  XMMSL  sample to  440  sources  with  AllWISE
counterparts.   The 39  sources with  $P_{post}<0.01$ include  AllWISE
associations  without optical  photometric  counterparts  in the  SDSS
(total of 16), stars (total of 3), galaxy cluster candidates (total of
2). These 39 sources are excluded from the analysis.

Among the 440  sources with $P_{post}>0.01$ there is  a large fraction
of non-AGN.  Stars are identified either as bright and saturated point
sources on the SDSS images  (33), as objects with significant non-zero
proper motions  \citep[total of  13;][]{Munn2004}, or as  sources with
stellar optical  spectra observed  as part  of the  SDSS spectroscopic
surveys (9).  X-ray sources that  lie within 1\,arcmin of a catalogued
galaxy cluster or group are also  marked as non-AGN (18). We also flag
a total of  9 X-ray sources that are associated  with BL Lacertae type
objects (BL  Lac) based on  spectroscopic information from  either the
SDSS or  the literature.   The spectral  energy distribution  of these
radio-loud sources is  dominated by emission from  a relativistic jet,
believed to be aligned to the line-of-sight of the observer. The X-ray
properties of these sources cannot be used as a proxy of the intrinsic
accretion  luminosity of  the active  nucleus.  The  above classes  of
sources (candidate stars, candidate clusters/groups and BL Lacs; total
of 82) are excluded from the analysis.  This selection process reduced
the 0.2-2\,keV selected XMMSL sample to 358 sources.

A total  of 337 of this  sample have redshift information  from either
the SDSS spectroscopic surveys up to DR14 (301) or the literature (36;
NASA  Extragalactic Database).   The majority  of these  redshifts are
spectroscopic  (333)  although  a  small  number  (4)  of  photometric
redshift  estimates  \citep{Csabai2007}   are  also  included.   XMMSL
sources  without spectroscopic  redshift  measurements  (total of  21)
include failed redshifts (total of  3), optically bright X-ray sources
($r\la17$\,mag;  total  of  7),  which  are  not  targeted  to  reduce
cross-talk  between   neighbouring  fibres,   sources  with   no  SDSS
photometric  counterparts (total  of  6), which  may include  spurious
associations, and optically  fainter X-ray sources that  have not been
targeted  because of  fibre-collision  problems,  the faint  magnitude
limit  of SPIDERS  target  selection ($r=22.5$)  or  because they  are
scheduled to be observed after 11  May 2016 and are therefore not part
of DR14.

In summary, there are potentially  397 0.2-2\,keV selected XMMSL X-ray
sources likely to be associated with AGN (358 + 39 sources; the latter
being those with $P_{post}<0.01$). We use in the analysis a total 337,
for which redshift information  is available.  These objects represent
85\% of the  parent sample of candidate AGN.   The 15\% incompleteness
also  includes  residual  Galactic star  contamination,  X-ray  galaxy
clusters,  as  well as  spurious  X-ray  detections and  associations.
Figure \ref{fig:lxz} presents the distribution on the X-ray luminosity
vs redshift space of all X-ray  sources used in this paper, i.e., from
the Chandra, XMM-XXL and XMMSL  surveys.  This figure demonstrates the
complementarity of the different samples.

\subsection{Stellar mass estimates for AGN host galaxies}\label{sec_mstar}

Stellar masses are  estimated using the {\sc cigale}  code version 0.8
\cite[Code   Investigating   GALaxy  Emission,][]{Noll2009}   to   fit
templates   to  the   observed  multiwavelength   photometry  (UV   to
mid-infrared)  of X-ray  selected AGN.   {\sc cigale}  convolves input
star-formation histories  with Single Stellar Population  (SSP) models
to  synthesise the  Spectral  Energy Distribution  (SED) of  galaxies.
Attenuation of  stellar radiation by  dust and the re-emission  of the
absorbed  energy in  the infrared  are also  accounted for  in a  self
consistent manner.  The emission associated with AGN is modelled using
the  \cite{Fritz2006}  library  of   template  SEDs  as  described  in
\cite{Ciesla2015}.   These  templates  correspond to  the  transmitted
radiation of a central source through obscuring material with toroidal
geometry. Free parameters  of the model include the  ratio between the
outer and inner  radius of the torus, the opening  angle of torus, the
optical depth  and density profile  of the obscuring material  and the
viewing angle  of the observer  relative to the equatorial  axis.  The
choice   of   model   parameters   for  the   AGN   templates   follow
\cite{Ciesla2015}. Table \ref{tab:cigale} presents the assumptions and
range  of {\sc  cigale} parameters  adopted in  this work  to generate
galaxy  and  AGN/galaxy  composite  templates.   We  use  the  delayed
star-formation history  model, in which the  star-formation rate (SFR)
as a function of time, $t$, is parametrised as $SFR(t)\propto t \times
\exp^{-t/\tau_1}$, where $\tau_1$  is the e-folding time  of the young
stellar population.   This parametric  form provides a  more realistic
description of  the time-averaged  star-formation rate of  galaxies in
semi-analytic  simulations  \citep{Ciesla2015}.   We  also  adopt  the
\cite{BC03}  SSPs  with  solar  metallicity,  the  \cite{Chabrier2003}
Initial  Mass   Function  (IMF)   and  the   \cite{Calzetti2000}  dust
extinction law.  The  UV/optical stellar emission absorbed  by dust is
remitted in the IR assuming the \cite{Dale2014} templates.

The input photometry  to {\sc cigale} is different for  each field. In
the  case   of  the   CDFS  we  use   the  photometric   catalogue  of
\cite{Hsu2014} that  combines photometry from the  the far-ultraviolet
($\rm \approx1500$\AA) to the mid-infrared ($\rm 8.0\mu m$).  Only the
broad-band filters listed in Table 1 of  Hsu et al. (2014) are used to
determine stellar masses.  For  the AEGIS-XD the multiwavelength ($\rm
\approx1500$\AA\, to  $\rm 8.0\mu m$) photometric  catalogue presented
by  \citet[][see  their  Table   7]{Nandra2015}  was  used  in  the
calculation of stellar masses.  In  the COSMOS-Legacy field the fluxes
in the CFHT $u^\star$, SUBARU $Vg^+r^+i^+z^+$ \citep{Capak2007}, UKIRT
WFCAM  $J$ \citep{McCracken2010},  CFHT WIRCAM  $Ks$ \citep{Capak2007}
and the four IRAC (3.6, 4.5,  5.8 and $8.0\,\mu m$) filters were used.
For  X-ray sources  detected in  the  XMM-XXL survey  we used  optical
photometry  ($ugriz$) from  the  CFHTLenS \citep[Canada-France  Hawaii
  Telescope Lensing  Survey;][]{Heymans2012, Erben2013}, near-infrared
fluxes ($YJHK$) from the VISTA  (Visible and Infrared Survey Telescope
for    Astronomy)     Deep    Extragalactic     Observations    survey
\citep[VIDEO;][]{Jarvis2013} based  on the  4th Data Release  (DR4) of
the Vista  Science Archive (VSA),  and the mid-infrared  photometry in
the     WISE     \citep[Wide-field    Infrared     Survey     Explorer
  mission;][]{Wright2010}  W1,  W2,  W3,  and W4  bands  with  central
wavelengths of 3.4,  4.6, 12, and $\rm 22\, \mu  m$, respectively from
the AllWISE data  release. The photometry for the  XMMSL Survey sample
is from the  GALEX (Galaxy Evolution Explorer)  All-Sky Imaging Survey
Data Release  5 \citep[GALEX-GR5;][]{Bianchi2011}, the  Sloan Digitial
Sky Survey  DR13 \citep{Albareti2016}, the  Two Micron All  Sky Survey
point   source  catalogue   \citep{Skrutskie2006}   and  the   AllWISE
catalogue.

\begin{figure}
\begin{center}
\includegraphics[height=0.7\columnwidth]{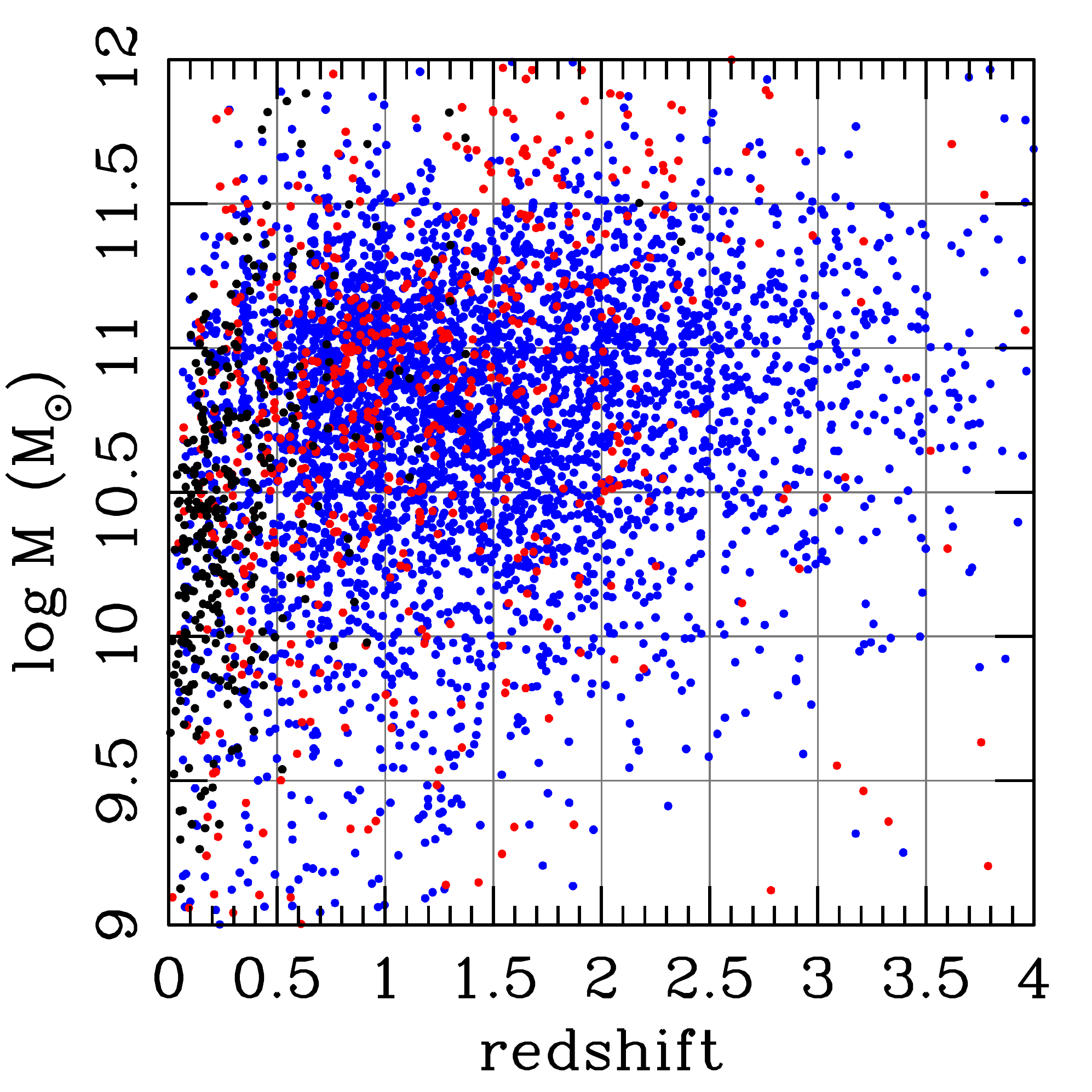}
\end{center}
\caption{Stellar mass  as a function redshift for  the sample 4821
X-ray selected AGN  used in this work. The blue  data points are X-ray
sources  detected  in  the   Chandra  surveys,  the  red  data  points
correspond to  the XMM-XXL sample and  the black ones to  the XMM Slew
survey.   Each data  point is  randomly drawn  from  the 2-dimensional
stellar mass and  redshift probability  distribution  function of
individual sources.}\label{fig:mz}
\end{figure}

{\sc  cigale}  provides  probability distribution  functions  for  the
output  parameters  determined from  the  SED  fits.  For  each  model
template the $\chi^2$  is determined based on the  input photometry of
each  galaxy.  The  probability distribution  function of  the derived
parameters of  interest (e.g., stellar mass,  star-formation rate) for
each source is then estimated based on the $\chi^2$ values of the full
set of  model SEDs.   Uncertainties associated with  e.g., photometric
errors  or  degeneracies  among   different  templates  are  therefore
accounted  for,  at least  to  the  first  approximation, and  can  be
propagated  into  the analysis.  Figure  \ref{fig:mz} shows  the
distribution  of the  AGN sample  on the  stellar mass  vs redshift
parameter space.

\begin{figure}
\begin{center}
\includegraphics[height=0.7\columnwidth]{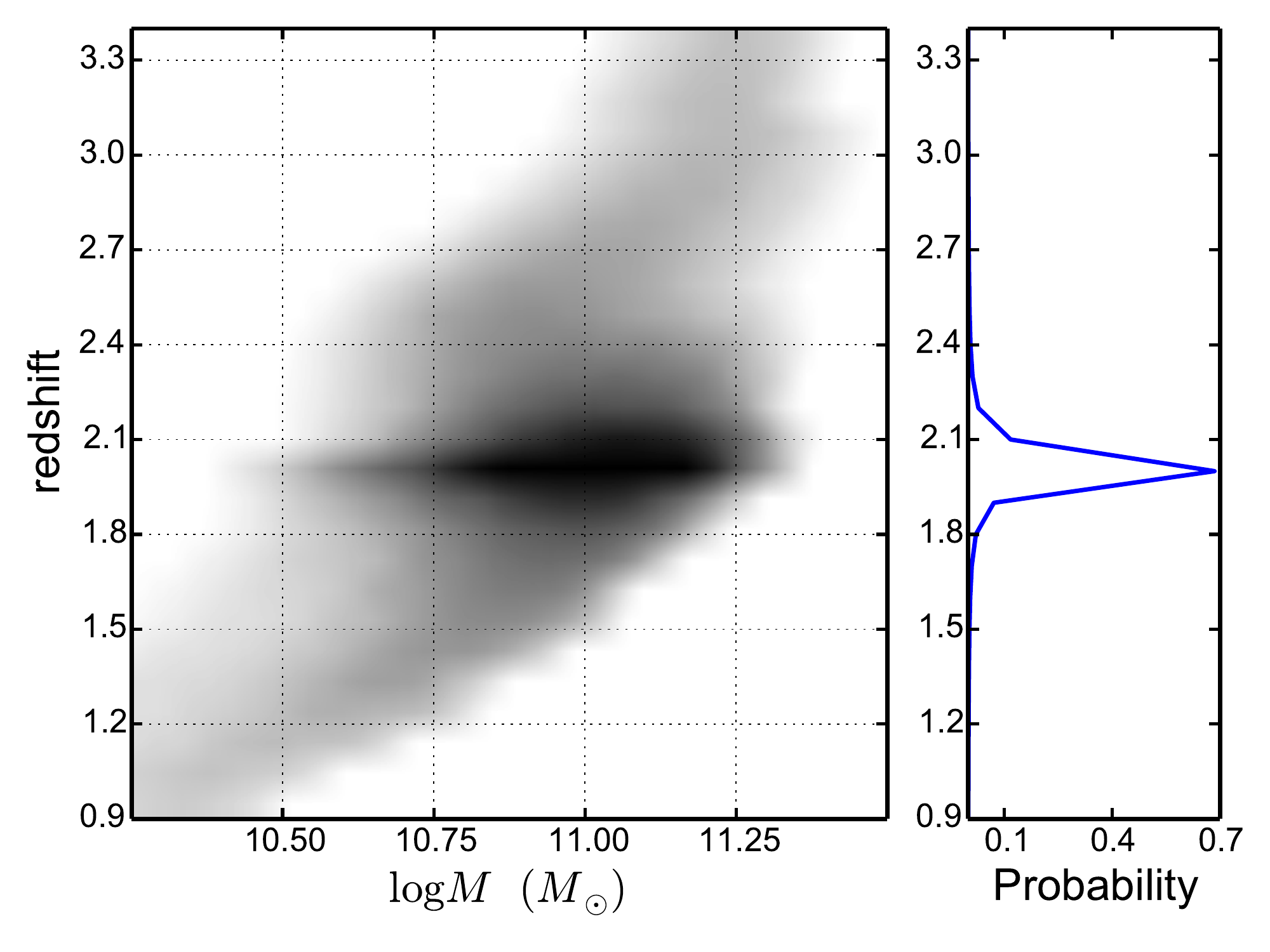}
\end{center}
\caption{The left panel shows the  probability density function in the
2-dimensional  space of  stellar mass  and redshift  for a  particular
X-ray  source   in  our  sample.   Darker   shading  signifies  higher
probability.  The  right panel presents the  corresponding photometric
redshift probability  density function,  estimated by  integrating the
2-dimensional   probability  map   along   the   stellar  mass   axis.
}\label{fig:mz_map}
\end{figure}

A fraction  of the  X-ray source  population has  photometric redshift
estimates  (40\%,  see   Table  \ref{table_data})  with  uncertainties
described  by  the  corresponding probability  distribution  functions
(PDZs).  When estimating  stellar masses we account  for these errors.
The PDZ of each  source is first sampled in steps of  $\Delta z = 0.1$
between redshifts $z=0$ and $z=6$. For  each bin $i$ at redshift $z_i$
the  {\sc cigale}  code is  applied  to the  observed photometry  with
redshift  fixed  to   $z_i$  to  yield  a   stellar  mass  probability
distribution  function.    The  normalisation  of  the   stellar  mass
probability distribution  function is not  unity but the value  of the
PDZ at  the corresponding  redshift bin  $i$.  This  approach produces
2-dimensional probability  distribution maps  in redshift  and stellar
mass, which provide a measure  of the uncertainty in the determination
of both quantities.   We use these maps in the  subsequent analysis to
determine the  specific accretion-rate distribution of  X-ray selected
AGN.  Figure \ref{fig:mz_map} shows an example of such a 2-dimensional
probability distribution map.

\subsection{Black hole mass estimates for broad line QSOs}\label{sec_mb}

The XMM-XXL and XMMSL survey  fields, because of their shallower depth
compared to  the {\it  Chandra} surveys used  in this work,  include a
large  fraction of type-1  AGN with  broad optical-emission  lines and
blue  continua.  The  determination  of stellar  masses  for the  host
galaxies of these sources is challenging because the AGN continuum and
emission lines  contaminate or  even dominate the  observed broad-band
fluxes. Stellar emission of the underlying host galaxy may however, be
significant    at   longer    wavelengths    \citep[i.e.,   rest-frame
near-infrared,][]{Bongiorno2012}   and  therefore  the   {\sc  cigale}
decomposition of the observed Spectral Energy Distribution into galaxy
and   AGN  contributions   can   at  least   partially  mitigate   AGN
contamination   issues.   Nevertheless,   some  level   of  systematic
uncertainties  should be  expected and  it is  therefore  desirable to
investigate  their  impact  on   the  results  and  conclusions.   For
broad-line AGN it is also  possible to have an independent estimate of
the mass  scale of  the system via  the determination  of single-epoch
virial   black-hole    masses   \citep[e.g.,][]{Shen_Liu2012}.    This
calculation comes with  its own set of systematics  and random errors,
which however  are likely to  be different from those  associated with
template fits to the  observed broad-band Spectral Energy Distribution
\citep[e.g.,][]{Ciesla2015}.    Virial   black-hole   mass   estimates
therefore  allow us  to explore  the  sensitivity of  our results  and
conclusions  to  the difficulties  in  determining  masses for  type-1
sources. Black-hole masses  for the XMM-XXL and the  XMMSL samples are
estimated for  broad-line AGN with SDSS spectroscopy  based on methods
similar     to    those     described    in     \cite{Shen2011}    and
\cite{Shen_Liu2012}. For sources in  the XMM-XXL the black hole masses
are presented in \cite{Liu2016}.  For the XMMSL AGN they are estimated
by one of us (AS).

\begin{figure}
\begin{center}
\includegraphics[height=0.7\columnwidth]{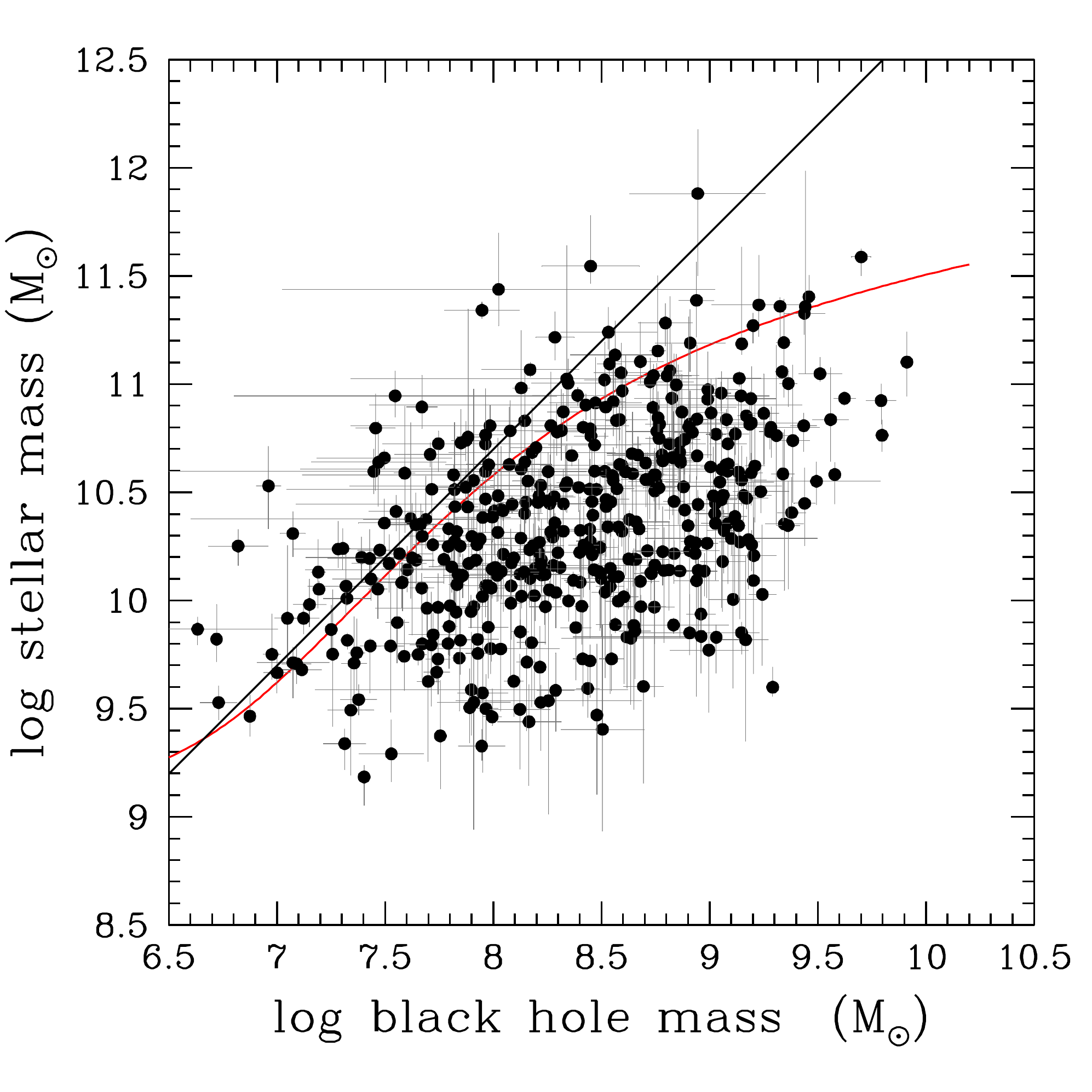}
\end{center}
\caption{Black-hole mass from single-epoch  optical spectra plotted as
a function of host galaxy stellar mass determined by fitting templates
to  the observed  broad-band  spectral energy  distribution. The  data
points  are broad-line  X-ray selected  AGN in  the XMM-XXL  and XMMSL
surveys.   The errorbars  correspond to  the 68th  percentiles of  the
stellar  and  black-hole mass  error  distributions.   The black  line
corresponds to the local relation  between stellar mass and black-hole
mass, $M_{BH}  = 0.002  M$ \protect\citep{Marconi_Hunt2003}.   The red
curve is the mean stellar mass  of galaxies at a given black-hole mass
assuming the  above intrinsic relation  between the two  quantities, a
scatter around this  relation that follows a  normal distribution with
$\sigma=0.3$\,dex,  and  a  galaxy   stellar  mass  function  with  an
exponentially decreasing space density at  the high-mass end (see text
for details).}\label{fig:bh_mstar}
\end{figure}

Figure \ref{fig:bh_mstar} displays stellar mass estimated via SED fits
against   black-hole  mass   determined   from  single-epoch   optical
spectroscopy  for   broad-line  sources  in  the   XMM-XXL  and  XMMSL
surveys. Also shown is the local relation between (bulge) stellar-mass
and      black-hole       mass      $M_{BH}      =       0.002      M$
\citep[e.g.,][]{Marconi_Hunt2003, Haring_Rix2004}.  Broad line AGN are
systematically  offset  by  a  factor of  about  three  toward  larger
black-hole masses compared  to the local relation.   The direction and
amplitude  of  this  offset  has been  reported  in  previous  studies
\citep[e.g][]{Lauer2007,  Matsuoka2014}.   It  may  indicate  redshift
evolution of the $M_{BH}-M$ relation \citep[e.g][]{Treu2004, Peng2006,
Canalizo2012},  differences in  the calibration  and normalisation  of
this    relation    compared    to   local    quiescent    black-holes
\citep[e.g][]{Matsuoka2014}   or  the   impact  of   selection  biases
\citep[][]{Lauer2007, Salviander2007, Schulze_Wisotzki2011}.  Although
the  redshift   evolution  of  the  $M_{BH}-M$   relation  is  debated
\citep[e.g.,][]{Shields2003, Shen2008,  Merloni2010}, it  is generally
accepted that selection  effects play an important  role. For example,
In Figure  \ref{fig:bh_mstar} we  plot the  relation between  the mean
stellar  mass of  galaxies for  a given  black-hole mass  estimated by
convolving  the  local  $M_{BH}-M$  relation  with  the  stellar  mass
function of galaxies  \citep[][]{Schulze_Wisotzki2011}.  The deviation
from the local  relation at the high-mass end is  a consequence of the
exponentially  decreasing space  density of  massive galaxies  and the
scatter    of    the   $M_{BH}-M$    relation    \citep[][]{Lauer2007,
Schulze_Wisotzki2011}.   At  fixed  black-hole  mass  there  are  more
galaxies  that   host  overmassive  black  holes   for  their  stellar
masses. Any  additional sample  selection effects, e.g.   flux limits,
luminosity cuts, will further impact the distribution of AGN in Figure
\ref{fig:bh_mstar}. It  also interesting that in  the recent work
of   \cite{Reines_Volonteri2015}   the   oposite  trend   from   Figure
\ref{fig:bh_mstar} is observed, i.e.  their local sample ($z < 0.055$)
of  low-luminosity broad-line  AGN selected  from the  SDSS has  lower
black  hole   masses  for  their   stellar  mass  compared   to  local
relations. This difference is also  likely related to sample selection
effects.

It is not the goal of this study to explore the $M_{BH}-M$ relation of
active galaxies, but rather to investigate systematics in the inferred
specific  accretion-rate distribution  of AGN  by testing  independent
methods of  measuring the mass scale  of the system.  In  the analysis
that  follows,  the  stellar  masses estimated  by  {\sc  cigale}  are
replaced by the virial black-hole mass estimates to explore systematic
differences in the  results.  Because our analysis  requires a measure
of the  stellar mass of  the AGN host (see  section \ref{sec_method}),
virial black-hole  masses are  converted to  stellar masses  using the
relation  $M_{BH} =  0.002 M$  \citep{Marconi_Hunt2003}.  We  mitigate
issues related  to the intrinsic  scatter of this scaling  relation or
possible evolution  with redshift by assigning  a Gaussian uncertainty
of  $\sigma=0.5$\,dex  \citep[e.g.][]{Shen2013}   to  the  logarithmic
stellar  mass  of  each  source   determined  by  scaling  the  virial
black-hole masses.

In the analysis that follows, broad-line  AGN in the XMM-XXL (total of
206 sources)  and XMMSL  (total of 253  sources) surveys  are assigned
stellar mass  probability distribution  functions that  are determined
indirectly  by  scaling  the   single-epoch  virial  black  hole  mass
estimates. We have tested that our main results and the conclusions do
not change if  instead we use stellar masses  determined from template
fits  to the  broad-band  spectral energy  distribution of  individual
sources. The  comparison of the specific  accretion-rate distributions
estimated  using  the  two  different approaches  to  approximate  the
stellar masses of  broad-line AGN is presented in  the Appendix. Small
differences are  observed at the  high specific-accretion rate  end of
the   distribution  in   the   redshift   interval  $z=0.0-0.5$   (see
Fig. \ref{fig:plz_bh_mstar}). This level  of difference indicates that
current uncertainties in the determination  of stellar masses for QSOs
(e.g.  factor of about 3 on the average in Fig. \ref{fig:bh_mstar}) is
adequate for  our purposes.   We also  emphasise the  relatively small
number of broad-line  quasars within our X-ray  selected sample, which
also plays  a role  in the  level of  difference between  the specific
accretion-rate distributions compared in Figure \ref{fig:plz_bh_mstar}.

\begin{table*}
\caption{Number of X-ray sources in the full-band selected sample}\label{table_data}

\begin{tabular}{l cc cc c c}

\hline
field & solid angle          & Number of      & Number of         & Number of                      & photometric     &  spectroscopic \\
      &  ($\rm deg^2$)    & X-ray sources  & optical/infrared IDs & sources used         & redshifts       &   redshift     \\

 (1) & (2) & (3) & (4) & (5) & (6) & (7) \\
\hline

4Ms CDFS  & $7.53\times10^{-2}$    & 417    & 413(1)  & 416 & 141  &  271  \\ 

AEGIS-XD  & $25.96\times10^{-2}$    &  818   & 793 (10) & 805 & 475  &  308  \\ 

COSMOS Legacy & 1.53  &  2676  & 2641 (18) & 2658 & 1087 (875)  & 1536 \\

XMM-XXL   & 3.19    & 682  & 605 (0) &  605  & 199  & 406 (206)\\ 

XMMSL & 1165 & 479  & 440 (82) & 337 & 4 & 333 (253)\\


\hline

\end{tabular}
\begin{list}{}{}
\item (1) Name of the X-ray survey fields used in this work. (2) Solid
angle  of each  sample in  square degrees  after excluding  regions of
poor/no photometry.  (3) Total number of X-ray sources detected in the
0.5-7\,keV energy band in the case of the {\it Chandra} survey fields,
the 2-8\,keV  band for the XMM-XXL  sample and the 0.2-2\,keV  band in
the case of the XMMSL.  Sources in areas of poor photometry associated
with e.g.  bright stars,  are excluded  in all  fields. (4)  Number of
X-ray  sources  with  optical/infrared associations.   The  number  in
brackets  indicates  X-ray  sources  associated  with  stars  in  each
sample. In  the case of the  XMMSL survey the number  in brackets also
includes BL Lacs and X-ray clusters (see text for details). (5) Number
of extragalactic  sources used in  the analysis.   In the case  of the
XMM-XXL  and {\it  Chandra}  field the  numbers  include sources  with
photometric or spectroscopic redshift but also sources without optical
counterparts.   (6)  Number  of   sources  with  photometric  redshift
estimates in  each survey  field.  In  the case  of the  COSMOS Legacy
field the number in brackets  corresponds to the number of photometric
redshift  estimates from  \protect\cite{Marchesi2016}.  (7)  Number of
sources  with   spectroscopic  redshift   estimates  in   each  survey
field.  For  the XMM-XXL  and  XMMSL  fields  the number  in  brackets
corresponds to  the number  of spectroscopically  confirmed broad-line
AGN in the sample, for which a black hole mass can be estimated.
\end{list}
\end{table*}

\begin{table*}
\caption{{\sc cigale} parameter settings for the determination of
  stellar masses of X-ray selected AGN.}\label{tab:cigale}

\begin{tabular}{l l}

\hline
Parameter description & values/range\\
\hline
\multicolumn{2}{c}{Stellar population parameters} \\
\\
Initial Mass Function & \protect\cite{Chabrier2003} \\
Metallicity & 0.02 (Solar) \\
Single Stellar Population Library  & \protect\cite{BC03} \\

\hline
\multicolumn{2}{c}{Delayed star-formation history law parameters} \\
\\
e-folding time, $\tau$, in Gyrs & 0.1, 0.3, 0.6, 1.0, 1.5, 2.0, 3.0, 4.0, 5.0, 7.0, 9.0, 13.0 \\
stellar population age, $t$, in Gyrs &  0.3, 0.5, 0.7, 1.0, 1.5, 2.0, 2.5, 3.0, 3.5, 4.0, 5.0, 6.0, 8.0, 10.0, 12.0 \\

\hline
\multicolumn{2}{c}{Dust extinction of stellar light} \\
\\
Dust attenuation law &  \protect\cite{Calzetti2000} \\
Reddening $E(B-V)$  & 0.01, 0.1, 0.2, 0.3, 0.4, 0.5, 0.6, 0.7, 0.8, 0.9, 1.2 \\
$E(B - V)$ reduction factor between old and young stellar populations 	& 0.44 \\

\hline
\multicolumn{2}{c}{\protect\cite{Fritz2006} library parameters}\\
\\
Ratio between outer and inner dust torus radii & 	30, 100 \\
$\rm 9.7\, \mu m$ equatorial optical depth & 0.3, 3.0, 6.0, 10.0 \\
\protect\cite{Fritz2006} density profile parameter $\beta^{\star}$ & -0.5 \\
\protect\cite{Fritz2006} density profile parameter $\gamma^{\star}$ & 0.0 \\
Viewing angle relative to  equatorial axis (deg) &  0.001, 50.100, 89.990 \\
AGN emission relative to infrared luminosity ($\rm 1-1000 \, \mu m$) & 0.0, 0.05, 0.1, 0.15, 0.2, 0.3, 0.4, 0.5, 0.6, 0.8, 0.9\\

\hline

\end{tabular}
\begin{list}{}{}
\item  $^\star$ The spatial distribution of the torus density in the
  \protect\cite{Fritz2006} models follows the functional form $\rho
  (r,\theta) \propto r^{\beta}\,e^{-\gamma \, |\cos\theta|}$, where $r$
  is the radial distance and $\theta$ is the polar distance. 
\end{list}
\end{table*}

\section{Methodology}\label{sec_method}

\subsection{The specific accretion-rate  distribution}

The determination  of the specific accretion-rate  distribution of AGN
is based on a modified version of the standard Poisson likelihood used
for the calculation of  the X-ray luminosity function \citep{Aird2010,
Aird2015, Buchner2015, Georgakakis2015}

\begin{equation}\label{eq:xlf_likeli}
\begin{split}  \mathcal{L}(  d_i \;|\;  \theta)  =  & e^{-\mu}  \times
\prod_{i=1}^{N} \int \mathrm{d}\log L_\mathrm{X} \,\, \frac{\mathrm{d}
V}{\mathrm{d} z}  \mathrm{d} z  \; \\ &  p(L_\mathrm{X},z |\; d_i) \;
\phi_{\rm T}(L_\mathrm{X},z \;|\; \mathbf{\theta_{\rm T}}),
\end{split}
\end{equation}

\noindent where  ${\rm d}V/{\rm d}z$  is the comoving volume  per unit
solid angle  and unit  redshift at redshift  $z$, $d_i$  signifies the
data available for source $i$ (e.g. X-ray counts, redshift and stellar
mass estimates)  and the summation  is over  all sources, $N$,  in the
sample.    The   quantity    $\phi_{\rm   T}(L_\mathrm{X},   z   \;|\;
\mathbf{\theta_{\rm  T}})$  is  the  luminosity  function  model  with
parameters  $\theta_{\rm T}$  that describes  the space  density as  a
function of redshift  and X-ray luminosity of  all extragalactic X-ray
populations.   The latter  includes mostly  AGN but  also a  small but
non-negligible contribution  from normal galaxies with  X-ray emission
dominated   by  stellar   processes   and/or   hot  interstellar   gas
\citep{Hornschemeier2002,     Georgantopoulos2005,    Georgakakis2007,
Lehmer2016}.   The  total  luminosity function  of  all  extragalactic
populations  can therefore  be expressed  as the  sum of  the AGN  and
normal galaxy luminosity functions

\begin{equation}\label{eq:xlf_total}  \phi_{\rm  T}(L_\mathrm{X},z  \;|\;
\mathbf{\theta_{\rm T}}) = \phi (L_\mathrm{X},z \;|\; \mathbf{\theta}) +
\phi_{\rm gal} (L_\mathrm{X},z \;|\; \mathbf{\theta_{\rm gal}}),
\end{equation}

\noindent where  $\phi (L_\mathrm{X},z \;|\; \mathbf{\theta})$  is the
AGN luminosity  function with parameters $\theta$  and $\phi_{\rm gal}
(L_\mathrm{X},z \;|\; \mathbf{\theta_{\rm gal}})$ is the normal galaxy
luminosity  function  described  by  a  different  set  of  parameters
$\theta_{\rm gal}$. In equation~\ref{eq:xlf_likeli} the multiplication
is over all  sources, $N$, in the sample, and  the integration is over
redshift  and X-ray  luminosity.  The  quantity $p(L_\mathrm{X},z  |\;
d_i)$ is  the probability of  a particular source having  redshift $z$
and X-ray luminosity $L_{\rm X}$  given the observational data, $d_i$.
This  approach captures  uncertainties  in the  determination of  both
redshifts (e.g.,  photometric redshifts)  and X-ray fluxes  because of
Poisson     statistics     and     the     Eddington     bias.      In
equation~\ref{eq:xlf_likeli}  the  parameter  $\mu$  is  the  expected
number of detected sources in the  survey area for a particular set of
luminosity function model parameters $\theta_T$

\begin{equation}\label{eq:xlf_mu} \mu =  \int \mathrm{d} \log L_\mathrm{X}
\,\,   \frac{\mathrm{d}   V}{    \mathrm{d}   z}   \mathrm{d}   z   \;
A(L_\mathrm{X},z)       \;       \phi_{\rm      T}(L_\mathrm{X},z\;|\;
\mathbf{\theta_{\rm T}}),
\end{equation}

\noindent where,  $A(L_\mathrm{X},z)$ is the  X-ray selection function
that  measures  the  probability  of  detecting a  source  with  X-ray
luminosity $L_X$ and redshift $z$ within the survey area.

The equations above can be modified by introducing the Eddington ratio
distribution  of  AGN,  $P(\lambda,  z)$,  which  is  defined  as  the
probability of  a galaxy  at redshift  $z$ and  with stellar  mass $M$
hosting an AGN accreting at  an Eddington ratio $\lambda$.  Given this
definition the AGN luminosity function can be written

\begin{equation}\label{eq:lambda2phi}  \phi(L_\mathrm{X},  z)  =  \int
\psi(M, z) \, P(\lambda(L_\mathrm{X}, M), z) \, \mathrm{d} \log M,
\end{equation}

\noindent where $\psi(M, z)$ is  the stellar mass function of galaxies
at redshift  $z$. The Eddington ratio  $\lambda$ is a  function of the
X-ray luminosity and stellar mass, under the assumption that the black
hole mass  is related to the stellar  mass of the AGN  host galaxy and
that the X-ray  luminosity can be used as proxy  of the bolometric AGN
luminosity.   The  quantity $P(\lambda,  z)$  as  defined  above is  a
probability density function, i.e., integrates to unity at any guiven
redshift 

\begin{equation}
\int  P(\lambda(L_\mathrm{X}, M), z) \, \mathrm{d}  \log M =1.
\end{equation}

\noindent  Equation \ref{eq:lambda2phi}  is  used to  express the  AGN
luminosity  function $\phi(L_X,  z)$ in  Equations \ref{eq:xlf_likeli}
and  \ref{eq:xlf_mu} in  terms of  the  galaxy mass  function and  the
Eddington ratio distribution of AGN

\begin{equation}\label{eq:plz_likeli}
\begin{split}
 \mathcal{L}( d_i \;|\; \omega, \theta_{\rm gal})  = & e^{-\mu} \times 
                                \prod_{i=1}^{N} \Biggl[ \int \mathrm{d}\log L_\mathrm{X} \,\, \frac{\mathrm{d}
V}{\mathrm{d}   z}  \mathrm{d}  z  \,\,  \mathrm{d} \log M  \\
                                 & p(L_\mathrm{X}, M, z \,|
                                 \; d_i)  \;
\psi(M, z) \, P(\lambda(L_\mathrm{X}, M), z \;|\; \mathbf{\omega})\\
                               & + \int \mathrm{d}\log L_\mathrm{X} \,\, \frac{\mathrm{d}
V}{\mathrm{d}   z}  \mathrm{d}  z  \\
                                 & p(L_\mathrm{X}, z | \; d_i)  \;
\phi_{\rm gal} (L_\mathrm{X},z \;|\; \mathbf{\theta_{\rm gal}})\Biggr],
\end{split}
\end{equation}

\noindent where the likelihood is explicitly split into AGN and normal
galaxy components. It  is also assumed that each  source in the survey
has an equal probability of being  a galaxy or an AGN, i.e., no prior
knowledge such as morphology  or X-ray--to--optical flux ratio is used
to    inform   the   nature    of   individual    sources   \citep[see
also][]{Aird2015}. In  the equation above $\omega$  represents the set
of parameter  related to the Eddington  ratio probability distribution
function model that  are to be determined from  the observations.  The
quantity  $p( L_\mathrm{X}, z,  M | \; d_i)$ is  the probability  of a
particular source  having redshift  $z$, X-ray luminosity  $L_{\rm X}$
and hosted by a galaxy with  stellar mass $M$.  The expected number of
detected sources in a survey  for a particular set of model parameters
$\omega$

\begin{equation}\label{eq:plz_mu}
\begin{split}  \mu  = &  \int  \mathrm{d}  \log L_\mathrm{X}  \,\, \frac{\mathrm{d} V}{  \mathrm{d} z} \mathrm{d} z  \,\, \mathrm{d} \log M \times   \\   
                      &    A(L_\mathrm{X},z)   \; \psi(M,z) \,  P(\lambda(L_\mathrm{X}, M), z \;|\; \mathbf{\omega})  \\
                      &  +  \int  \mathrm{d}  \log L_\mathrm{X}  \,\, \frac{\mathrm{d} V}{  \mathrm{d} z} \mathrm{d} z \; A(L_\mathrm{X},z)   \; \phi_{\rm gal} (L_\mathrm{X},z \;|\; \mathbf{\theta_{\rm gal}}).
\end{split}
\end{equation}

\noindent In the  equations above it is assumed  that the stellar mass
function of  galaxies is known to  a reasonable level  of accuracy. We
adopt the parametrisation presented by \cite{Ilbert2013} for the total
galaxy   mass   function.    They   used   two   Schechter   functions
\citep{Press1974} with parameters  evolving with redshift to represent
the mass function of galaxies in the redshift interval $z=0-4$.  Using
other   galaxy   mass    functions   available   in   the   literature
\citep[e.g.,][]{Muzzin2013} does not change our main results.

Assumptions also must  be made to estimate the  Eddington ratio of AGN
at redshift $z$, with X-ray  luminosity $L_X$, hosted by galaxies with
stellar mass  $M$. Ideally direct  estimates of black hole  masses are
needed  for  the  determination   of  Eddington  ratios.   This  is  a
challenging  task  however,  particularly  in  the  absence  of  broad
emission  lines in  the optical  spectra  of large  fraction of  X-ray
selected AGN  because of  e.g., obscuration or  AGN light  dilution by
stellar emission.  In this  paper a Magorrian-type relation is adopted
\cite{Magorrian1998} to scale the stellar  mass of the galaxy to black
hole mass.   There are  lengthy discussions in  the literature  on the
relation between  galaxy bulge mass  proxies and black hole  mass, the
level of scatter  of such correlations \citep[e.g.,][]{Kormendy_Ho2013}
or possible evolution with redshift \citep[e.g.,][]{Treu2004, Peng2006,
Canalizo2012}.   For the  shake  of simplicity,  the baseline  results
presented  in  this  paper  adopt a  single  and  redshift-independent
scaling relation  between black  hole mass and  bulge mass,  $M_{BH} =
0.002  M_{bulge}$  \citep{Marconi_Hunt2003}.   We  further  assume  no
scatter in  that relation and that  the AGN hosts  are bulge dominated,
i.e., that the  bulge mass can be approximated by  the stellar mass of
the  AGN host  galaxy, $M  =  M_{bulge}$.  A  single X-ray  bolometric
conversion factor is further adopted, $L_{bol}  = 25 \, L_X(\rm 2 - 10
\,  keV)$ \citep{Elvis1994}.   Under these  assumptions  the Eddington
ratio of AGN is estimated as

\begin{equation}\label{eq:lambda}
\lambda = \frac{25 \,\,  L_X(\rm 2-10\,keV)}{ 1.26 \times 10^{38} \,\, 0.002  \, M},
\end{equation}

\noindent where the X-ray luminosity  is in units of $\rm erg\,s^{-1}$
and the  stellar mass is  in solar  units. The $\lambda$  parameter is
essentially a scaled version of the specific accretion rate defined as
$L_X   /   M$   \citep[e.g.,][]{Bongiorno2012,   Aird2012,   Aird2013,
Bongiorno2016} expressed  in Eddington-ratio  units.  From here  on we
refer to $\lambda$ as the  specific accretion rate and $P(\lambda, z)$
as the specific accretion-rate distribution  of AGN, to emphasize that
these quantities  are derived  using the  stellar mass  as a  proxy of
black-hole mass  for the bulk of  the X-ray selected sample.   We also
underline  that the  only purpose  of Equation  \ref{eq:lambda} is  to
express the  specific accretion rate  in Eddington ratio  units, which
can be interpreted in a physical context.  Our results and conclusions
are  insensitive  to  the   assumptions  adopted  to  derive  Equation
\ref{eq:lambda}, e.g. bolometric correction. Nevertheless we have
also explored the  impact on the results and  conclusions of different
definitions  of the  $\lambda$  quantity. We  adopt  for example,  the
luminosity-dependent  bolometric  corrections  of  \cite{Marconi2004},
instead of  the fixed  conversion factor of  Equation \ref{eq:lambda}.
We also introduce scatter of $\sigma=0.38$ in the relation between the
black hole  mass and bulge mass  of \cite{Marconi_Hunt2003}.  Although
these  different  definitions  of   $\lambda$  yield  $P(\lambda,  z)$
distributions  that   differ  in   detail  from  our   baseline  model
assumptions, the  main characteristics of these  distributions as well
as our  results and conclusions on  the redshift evolution of  the AGN
duty-cycle (see Results and Discussion sections) remain unchanged.

The equations above ignore the impact of obscuration on the estimation
of  the luminosity  of individual  AGN  and the  determination on  the
specific accretion-rate distribution. A statistically robust treatment
of the obscuration requires  the estimation of neutral hydrogen column
densities, $N_H$,  from the  X-ray spectra of  individual AGN  and the
inclusion of model terms that  describe the dependence of the specific
accretion-rate distribution of AGN on $N_H$, i.e., $P(\lambda(L_X, M),
z, N_H)$.   We defer this  multi-dimensional analysis to  future work.
Accounting  for  obscuration  effects will  increase  the  space
density  of   AGN  we  recover   as  by-product  from   our  analysis,
particularly   at  intermediate   X-ray   luminosities.   The   higher
normalisation of the recovered X-ray luminosity function translates to 
an increase of the AGN duty cycle, i.e. the probability of a galaxy to
host  an AGN,  and  hence  a higher  amplitude  for  the the  inferred
$P(\lambda, z)$ at a given redshift. The impact of obscuration effects
on our results will also  decrease with increasing redshift because of
the nature of the X-ray k-corrections.

Poisson statistics are used to determine the flux distribution that is
consistent  with  the  extracted  source  and  background  counts.   A
power-law  X-ray spectrum  with  $\Gamma  = 1.9$  is  adopted in  this
calculation.   The resulting  flux distribution  is combined  with the
redshift  information, spectroscopic or  photometric redshift  PDZ, to
estimate  the luminosity  distribution  of each  source at  rest-frame
energies 2-10\,keV. The relevant k-corrections also assume a power-law
X-ray spectrum with $\Gamma  = 1.9$.  Equation \ref{eq:lambda} is used
to determine  the $\lambda$  distribution of individual  sources given
their  probability density distributions  in X-ray  luminosity stellar
mass, and redshift (see also section \ref{sec_mstar}).

A non-parametric approach for the determination of $P(\lambda(L_X, M),
z)$  is  adopted.   Our  baseline  model  assumes  that  the  specific
accretion-rate distribution of AGN  is independent of the stellar mass
of   the   host   galaxy   \citep[e.g.,][]{Aird2012,   Aird2013}.    A
two-dimensional  grid  in  redshift  and specific  accretion  rate  is
defined and $P(\lambda(L_X, M), z)$ is assumed to be a constant within
each  grid  pixel  with  dimensions ($\log  \lambda\pm  \mathrm{d}\log
\lambda$,  $z\pm  \mathrm{d}z$).   The   value  of  the  AGN  specific
accretion-rate  distribution  in each  grid  pixel  is determined  via
equation~\ref{eq:plz_likeli}. The edges of  the grid pixels in each of
the two dimensions are  $\log \lambda =$~($-10$, $-6$, $-5.5$, $-5.0$,
$-4.5$,  $-4.0$,  $-3.5$,  $-3.0$,  $-2.5$,  $-2.0$,  $-1.5$,  $-1.0$,
$-0.5$,  $0.0$, $0.5$, $1.0$,  $1.5$, $2.0$)  and $z=$(0.0,  0.5, 1.0,
1.5, 2.0, 3.0,  4.0, 6.0).  In Section \ref{sec_results}  we will also
discuss   modifications  to   the  baseline   model  to   explore  the
stellar-mass dependence of the $P(\lambda(L_X, M), z)$.

Importance sampling  \citep{NR1992} is used to  evaluate the integrals
in  equation~\ref{eq:plz_likeli}.   For  each source  we  draw  random
($\log  \lambda$, $z$,  $M$)  points from  the full  three-dimensional
probability  distribution.  The  specific accretion-rate  distribution
model  is then  evaluated  for  each sample  point.   The integral  of
equation~\ref{eq:plz_likeli} is simply the average $P(\lambda(L_X, M),
z)$  of the  sample.  The  Hamiltonian Markov  Chain Monte  Carlo code
Stan\footnote{\url{http://mc-stan.org/shop/}} \citep{Carpenter2016} is
used for Bayesian statistical inference.

The  Appendix  compares our  results  on  the specific  accretion-rate
distribution of  AGN with those of  Aird et al. (in  prep.). The later
work  developed  parallel  to   our  study  and  uses  an  independent
methodology  to infer  the  $P(\lambda,M,z)$ and  a  different set  of
multiwavelength photometric catalogues  to estimate galaxy properties,
such  as  stellar  mass.   The  agreement  between  the  two  distinct
approaches compared in Figure \ref{fig:plz_aird}  is remarkable.

\subsection{The X-ray luminosity function of normal galaxies}

The normal  galaxy (non-AGN) X-ray luminosity  function parameters can
in   principle  be   determined  simultaneously   with   the  specific
accretion-rate  distribution  parameters.    For  the  sake  of  speed
however, we  follow the alternative approach of  adopting a parametric
model for  the X-ray luminosity  function of galaxies  with parameters
fixed to  values from the  literature.  This approach is  motivated by
the  fact  that  the  X-ray  luminosity function  and  evolution  with
redshift of this class of sources  have been studied to some detail in
the    past    \citep[e.g.,][]{Georgantopoulos2005,   Georgakakis2006,
Ptak2007,  Tzanavaris2008,  Aird2015}.   These  previous  studies  can
therefore  inform  our expectations  for  the  contribution of  normal
galaxies to  the current sample.  Moreover,  normal galaxies represent
only  a small  component of  current X-ray  survey  source populations
\cite[e.g.,][]{Lehmer2016}.    We   adopt   the   parametrisation   of
\cite{Aird2015}  and use  a  Schecter form  for  the X-ray  luminosity
function of galaxies 

\begin{equation}
\phi_{\rm gal} (L_X, z) = K \;
\Bigg(\frac{L_X}{L_X^{\star}} \Bigg)^{-\alpha}\;e^{-\frac{L_X}{L_X^{\star}}}. 
\end{equation}

\noindent where the characteristic luminosity, $L_X^{\star}$, evolves as

\begin{equation}  
\log L_X^{\star}(z) =  \left\{\begin{array}{ll}   \log L_0 + \beta \, \log (1 + z), & z<z_c,  \\ 
                                                                     &   \\ 
 \log L_0 + \beta \, \log (1 + z_c), & z \ge z_c. \\
\end{array} \right.
\end{equation}

\noindent The  parameters are  fixed to the  maximum-likelihood values
determined  by \citet[][their  Table 10]{Aird2015}.   For completeness
they  are  also  listed  in  Table  \ref{tab:xlf_results}.   Equations
\ref{eq:plz_likeli} and \ref{eq:plz_mu} are then used to determine the
specific  accretion-rate distribution  of  AGN by  keeping the  normal
galaxy luminosity  function parameters  fixed.  This  approach ignores
uncertainties  in the  determination  of the  galaxy X-ray  luminosity
function.   The  advantage is  that  the  required  CPU time  for  the
determination of the $P(\lambda, z)$ is significantly reduced.

\subsection{The non-parametric luminosity function of X-ray selected AGN}

A by-product of our analysis is the reconstructed AGN X-ray luminosity
function derived  from Equation \ref{eq:lambda2phi}  by convolving the
$P(\lambda,z)$  distribution   with  the  stellar   mass  function  of
galaxies. It is desirable to assess the quality of this reconstruction
by comparing with independent estimates of the AGN space density as a
function of redshift and X-ray luminosity. 

For  this exercise  we derive  non-parametric estimates  of  the X-ray
luminosity function  of AGN. A 2-dimensional grid  in X-ray luminosity
and   redshift   is   defined   with   pixel   edges   $\log   L_X(\rm
2-10\,keV)=$~($38$,  $39$,  $40$,  $41$,  $42$,  $43$,  $43.5$,  $44$,
$44.5$,  $45$,  $46$),  $z=$(0.0,   0.5,  1.0,  1.5,  2.0,  3.0,  4.0,
6.0). Equation~\ref{eq:plz_likeli}  is then used to  determine the AGN
space density, $\phi(L_X, z)$, assuming that this quantity is constant
within each grid pixel.  The Hamiltonian Markov Chain Monte Carlo code
Stan is used  to sample the likelihood function  described by Equation
\ref{eq:xlf_likeli} and estimate the model parameters.  We account for
the contribution of normal galaxies  to the X-ray source population by
adopting  the  Schechter  model  described above  for  the  luminosity
function  of these  sources and  fixing the  parameters to  the values
listed   in  Table  \ref{tab:xlf_results}.   Appendix  \ref{sec:agnLF}
compares the  reconstructed and non-parametric  determinations of the
AGN X-ray luminosity function.

\begin{table} 
\caption{Galaxy X-ray luminosity function parameters}\label{tab:xlf_results} 
\begin{center} 
\begin{tabular}{l  c}
\hline 
parameter & value \\
 (1)      &  (2)         \\
\hline
$\log K$   &   $-3.59$ \\
$\log L_0$ &   $41.12$ \\
$\alpha$   &   $0.81$  \\
$\beta$    &   $2.66$  \\
$z_c$      &   $0.82$  \\
\hline
\end{tabular} 
\end{center}
\end{table}

\begin{figure*}
\begin{center}
\includegraphics[height=0.85\columnwidth]{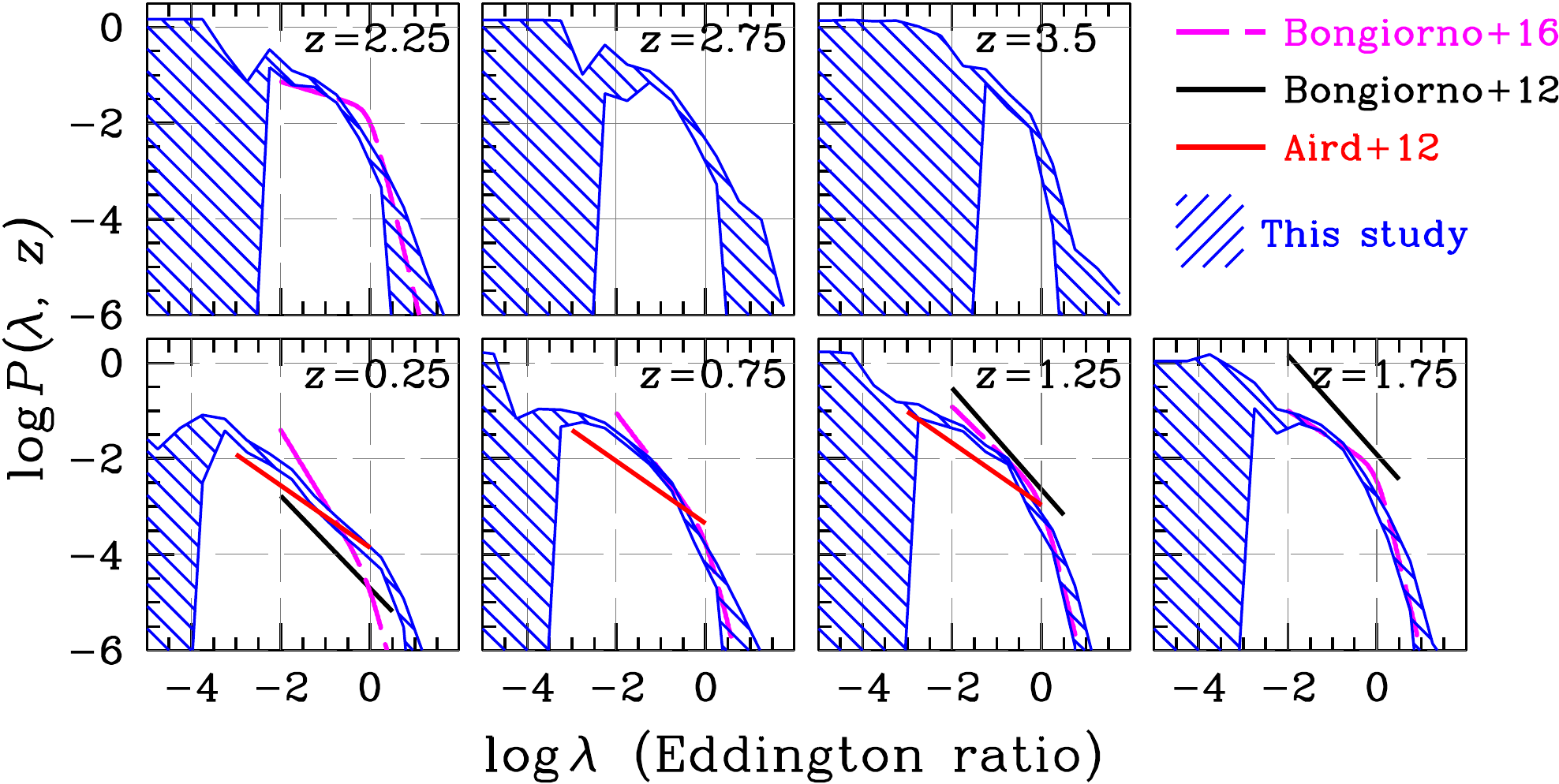}
\end{center}
\caption{Specific accretion-rate distribution,  $P(\lambda, z)$ of AGN
across  the full  stellar mass  range considered  in the  paper, $\log
M/M_{\odot}=8-13$.   Each panel  corresponds to  a different  redshift
interval,  $z=0.0-0.5$,  $0.5-1.0$, $1.0-1.5$,  $1.5-2.0$,  $2.0-3.0$,
$3.0-4.0$.  The mean  redshift of each bin is labelled  on each panel.
The  blue   hatched  regions  are  the   non-parametric  observational
constraints on  $P(\lambda, z)$  from the  analysis presented  in this
paper.  The extent  of the blue hatched regions correspond  to the 5th
and 95th  percentiles of the $P(\lambda,  z)$ probability distribution
function.   The   parametric  specific   accretion-rate  distributions
estimated by  \protect\cite{Aird2012} and \protect\cite{Bongiorno2016}
are shown  with the red-solid and  magneta-dashed lines, respectively.
The \protect\cite{Aird2012} relation applies  to $z\la1$ and therefore
we choose not not extrapolate beyond $z>1.5$. The black line shows the
power-law fits of \protect\cite{Bongiorno2012} to their AGN samples in
the redshift  intervals $0.3-0.8$, $0.8-1.5$ and  $1.5-2.5$.}\label{fig_plz}
\end{figure*}

\begin{figure*}
\begin{center}
\includegraphics[height=0.85\columnwidth]{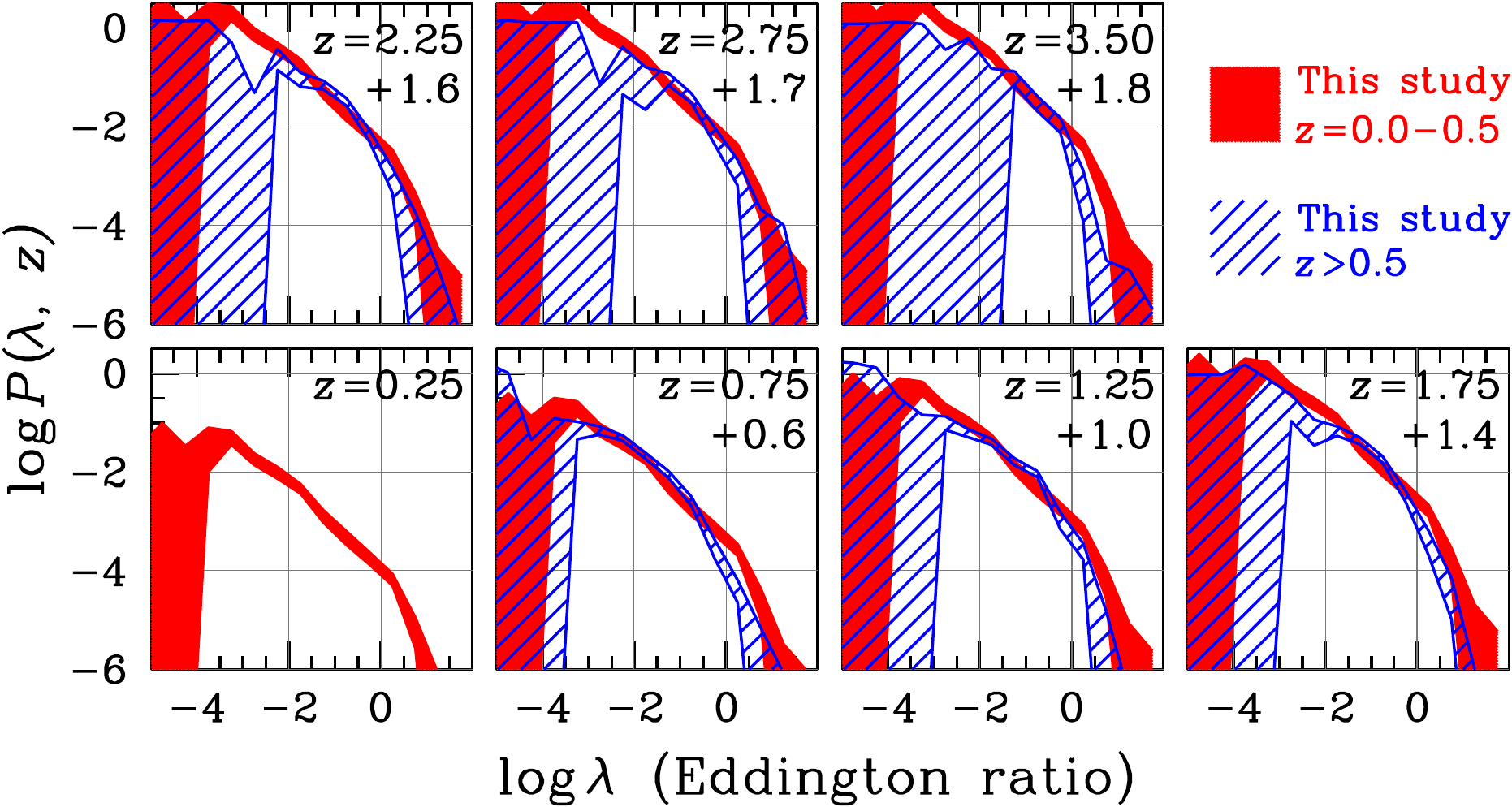}
\end{center}
\caption{Same  format   as  Figure  \ref{fig_plz}.   The  red  hatched
histogram in each panel is the specific accretion-rate distribution of
the  lowest  redshift  bin,  $z=0.0-0.5$.  This  relation  is  shifted
upwards  by the  logarithmic offset  marked  in each  panel under  the
redshift label.   The offset  is not estimated  by a  fitting process,
instead it  is empirically visually  determined to demonstrate  that a
simple re-normalisation of the $P(\lambda, z=0.0-0.5$) approximates in
a rough  manner the redshift evolution of  the specific accretion-rate
distribution of AGN.  }\label{fig:plz_over}
\end{figure*}
\begin{figure*}

\begin{center}
\includegraphics[height=0.85\columnwidth]{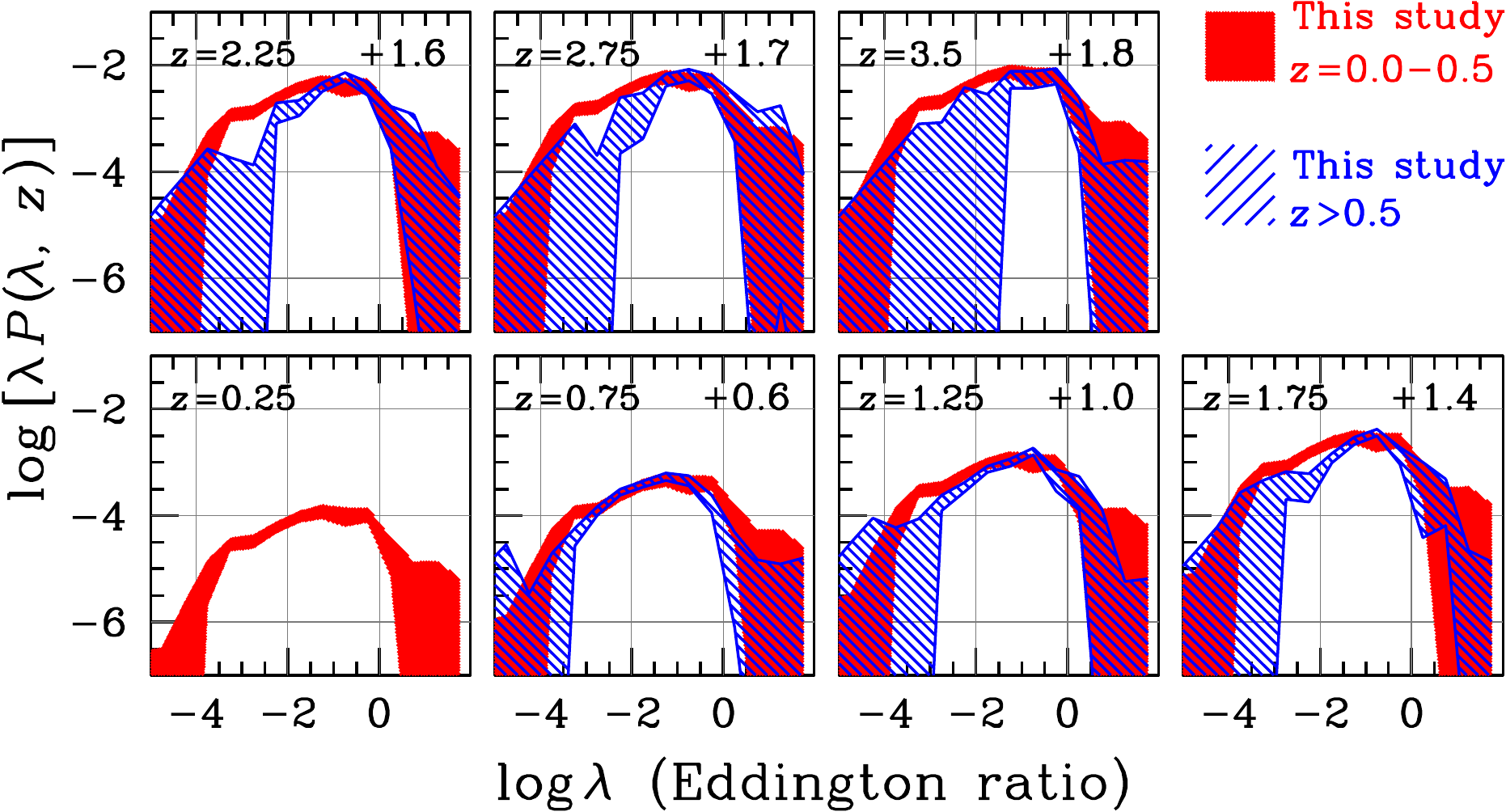}
\end{center}
\caption{The distribution  of the quantity $\lambda  \, P(\lambda, z)$
  as  a function  of the  specific  accretion rate.   The red  hatched
  histogram  in each  panel  corresponds to  the  distribution of  the
  lowest redshift  bin, $z=0.0-0.5$.  This  is shifted upwards  in the
  y-direction by the  logarithmic offset marked in each  panel next to
  the redshift label.  
}\label{fig:lplz}
\end{figure*}

\begin{figure}
\begin{center}
\includegraphics[height=0.9\columnwidth]{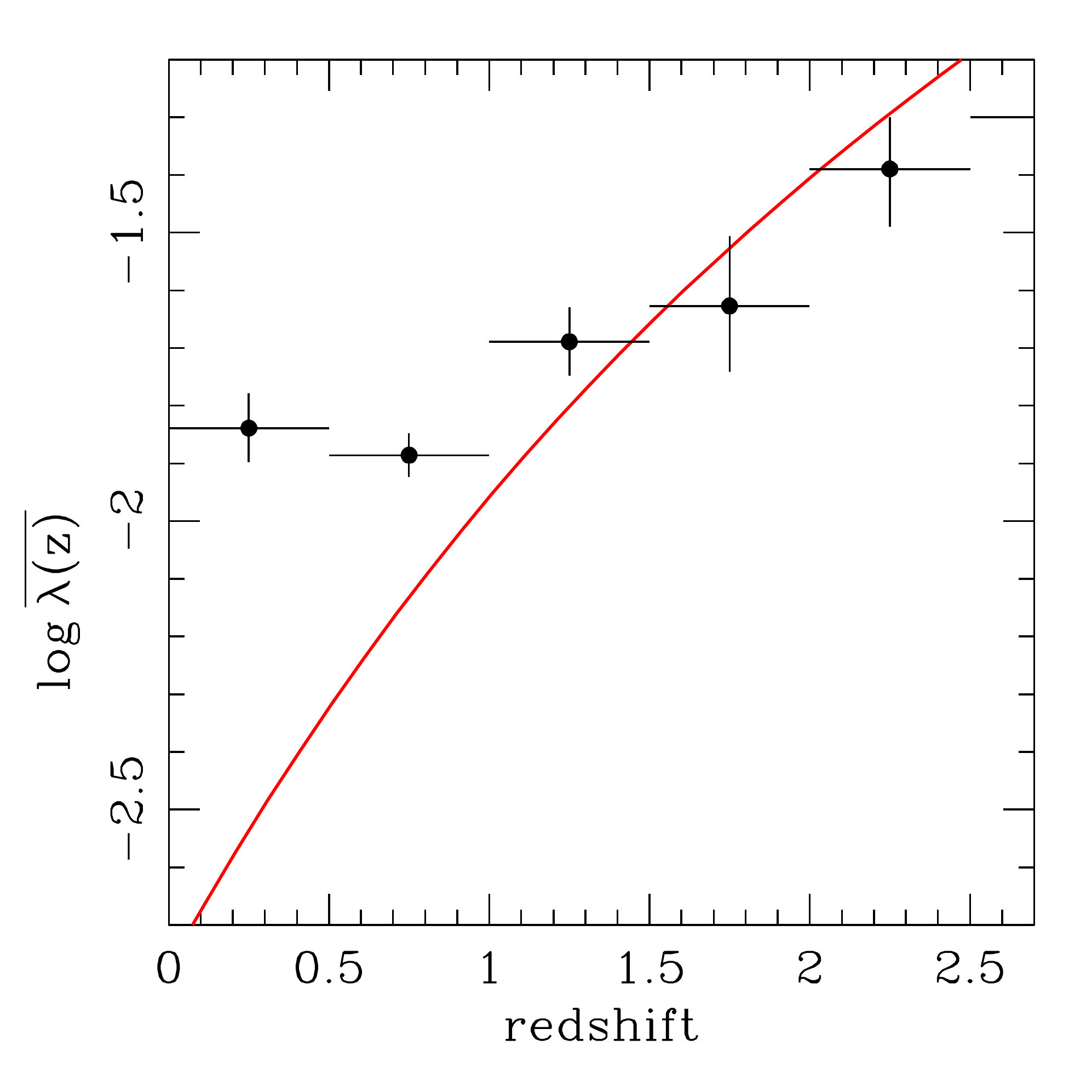}
\end{center}
\caption{Mean specific accretion rate of AGN as a function of
redshift.  The  mean Eddington ratio  is estimated by  integrating the
distributions    of    Figure    \ref{fig_plz}   above    the    limit
$\lambda=10^{-3}$.   Above   this  threshold  and  for   $z<2.5$,  the
$P(\lambda, z)$  distributions presented  in Figure  \ref{fig_plz} are
reasonably well  constrained. The  red curve  shows the  mean specific
star-formation   rate   of   main-sequence  galaxies   determined   by
\protect\cite{Elbaz2011}  (their   equation  13)  and  scaled   by  an
arbitrary logarithmic factor of -1.7 to facilitate the comparison with
the mean specific accretion rate.}\label{fig:mean_lambda}
\end{figure}

\section{Results}\label{sec_results}

\subsection{The specific accretion-rate distribution of AGN}
 
As  a first  approximation  we  ignore a  possible  dependence of  the
specific  accretion-rate distribution of  AGN on  the stellar  mass of
their host galaxies; such a  dependence will be explored later in this
section. Equation \ref{eq:plz_likeli} is applied to the full sample of
X-ray selected  AGN without  a division into  stellar mass  bins.  The
resulting constraints  on the  $P(\lambda,z)$ are displayed  in Figure
\ref{fig_plz}, where  the different panels correspond  to the redshift
intervals  defined  in section  \ref{sec_method}.   The  width of  the
hatched  regions  in  Figure  \ref{fig_plz} corresponds  to  the  90\%
confidence interval and shows the  region of the parameter space where
the data provide meaningful constraints to the specific accretion-rate
distribution of  AGN.  As expected, the sample  becomes insensitive to
low  specific  accretion-rate  AGN  with  increasing  redshift.   This
behaviour is manifested by  a threshold in the specific accretion-rate
below which the width of the hatched regions is comparable to the full
range of  the $P(\lambda,z)$ prior  (i.e., many orders  of magnitude).
This  specific accretion-rate  threshold moves  to higher  values with
increasing redshift.

In  each redshift  panel  of Figure  \ref{fig_plz} the  $P(\lambda,z)$
distributions, i.e. the  probability of a galaxy  hosting an accretion
event, increase with  decreasing specific accretion rate,  at least to
the limit $\lambda\approx10^{-4}  - 10^{-3}$.  Also, the  slopes of the
distributions    steepen    close    to    the    Eddington    limit,
$\lambda\approx0$.   These results  are consistent  with studies  that
approximate  the specific  accretion-rate distribution  of AGN  with a
single power-law  and find evidence  for a break at  approximately the
Eddington  limit \citep{Aird2012,  Bongiorno2012}. This  break can  be
parametrised    by   a    second   steeper    power-law   distribution
\citep{Aird2013}.    At    very   low   specific    accretion   rates,
$\lambda\la10^{-4} - 10^{-3}$ there is evidence for a flattening of the
$P(\lambda,z)$  distribution in  all panels  of Figure  \ref{fig_plz}.
This trend is driven by the data  and is only partially imposed by the
requirement that the $P(\lambda,z)$ distribution integrates to unity.

Figure  \ref{fig_plz} also  compares our  results with  the parametric
specific accretion-rate  distribution of \cite{Aird2012}.   They model
the $P(\lambda,z)$  distribution with  a single  power-law that  has a
normalisation    that    evolves    with    redshift    as    $\propto
(1+z)^\gamma$. These results are formally  valid for AGN with specific
accretion-rate in  the range $\lambda  = 10^{-4}-1$ and  for redshifts
$z\la1$.   We choose  not to  extrapolate the  \cite{Aird2012} results
beyond  $z>1.5$  and  outside  the  specific  accretion-rate  interval
$\lambda =  10^{-4}-1$.  Our analysis suggests  a $P(\lambda,z)$ shape
that  is  more complex  than  a  single power-law.   Nevertheless  the
functional form  derived by  \cite{Aird2012} roughly  approximates the
slope and overall redshift evolution of the low Eddington-ratio end of
the distributions plotted in Figure \ref{fig_plz}.  Also shown in this
figure  are   the  power-law  fits  to   the  specific  accretion-rate
distributions estimated  by \cite{Bongiorno2012} for their  samples at
redshift  intervals $z=0.3-0.8$,  $0.8-1.5$  and  $1.5-2.5$.   The
specific accretion rates in the Bongiorno et al. work are converted to
the  Eddington-ratio  units  adopted   in  our  paper  using  equation
\ref{eq:lambda}.  At the lowest redshift  bin there is fair agreement
between the slope  found by \cite{Bongiorno2012} and  our results.  At
higher redshift their slope appears to be in better agreement with the
high specific  accretion-rate end of  our results.   The specific
accretion-rate  distributions  estimated by  \cite{Bongiorno2016}  are
also  plotted in  Figure \ref{fig_plz},  after scaling  their specific
accretion  rate  to the  units  adopted  in  our work  using  Equation
\protect\ref{eq:lambda}.  We plot the  specific accretion-rate term of
\cite{Bongiorno2016} (their equation 11) estimated at the average mass
$\log  M/M_{\odot}=10.5$  of our  sample.   Because  of the  different
methodology adopted  by \cite{Bongiorno2016} their  specific accretion
rate  distribution  has  no  normalisation.  Their  curves  in  Figure
\ref{fig_plz}  have been  arbitrarily normalised  to roughly  match the
amplitude of  the $P(\lambda, z)$  distributions derived in  our work.
Nevertheless, the overall shape  of the distributions, particularly at
$z\ga1$,  is similar.  We also  caution that  the \cite{Bongiorno2016}
results are an extrapolation at redshifts below $z\approx0.5$.

The  redshift dependence  of  the $P(\lambda,z)$  distribution can  be
explored by  comparing the  different panels of  Figure \ref{fig_plz}.
It  can  be  argued  that   the  fundamental  characteristics  of  the
$P(\lambda,z)$ do not change drastically with redshift.  Despite small
differences  in the  slopes, which  will  be discussed  later in  this
section, the general  shape of the $P(\lambda,z)$  remains roughly the
same  across  redshift.  The  most  striking change  is  that  of  the
normalisation  of the  $P(\lambda,z)$, which  increases toward  higher
redshift.  This trend  appears  to  saturate at  $z>2$,  i.e., in  the
redshift   bins  $z=2.0-2.5$,   $z=2.5-3.0$  and   $z=3.0-4.0$.   This
behaviour is further demonstrated  in Figure \ref{fig:plz_over}, which
compares  the  $P(\lambda,z)$  distribution derived  in  the  redshift
interval  $z=0.0-0.5$  to  the   distributions  determined  at  higher
redshifts.  In  each panel of  this figure  a vertical shift  has been
applied to  the $P(\lambda,z=0.0-0.5)$ distribution to  facilitate the
comparison.  The amplitude of this shift is indicated in each panel of
Figure  \ref{fig:plz_over}.   We  emphasise   that  these  shifts  are
empirically  visually  estimated  to  demonstrate  that  the  observed
redshift evolution of the  specific accretion-rate distribution of AGN
can be represented in a rough  manner by a change of the normalisation
of the $P(\lambda,z=0.0-0.5)$.

An implication  of the  evolutionary scheme  in which  the fundamental
shape  of  the  $P(\lambda,z)$ is  roughly  redshift
independent is  that the  typical specific-accretion  rate of  the AGN
population does  not evolve strongly  with redshift. We show
this point  using  the the quantity  $\lambda \,  P(\lambda,z)$, in
which  the  linear  power-law  slope of  the  specific  accretion-rate
distribution  is removed  to  highlight the  value  of $\lambda$  that
dominates the AGN population at a  given redshift.  This is plotted in
Figure \ref{fig:lplz}.  For comparison in that figure the distribution
derived in  the redshift  interval $z=0.0-0.5$  is overplotted  on the
distributions determined at  higher redshifts.  In each  panel of this
figure the  same vertical  shift as  in Figure  \ref{fig:plz_over} has
been applied to the  $\lambda \, P(\lambda,z=0.0-0.5)$ distribution to
facilitate the  comparison. In  Figure \ref{fig:lplz} the  $\lambda \,
P(\lambda,z)$ shows a  plateau or a peak at $\log  \lambda \approx -1$
that does  not change significantly with  increasing redshift. Another
way  to  demonstrate this  point  is  to  estimate the  mean  specific
accretion-rate of the distributions in Figure \ref{fig_plz}, i.e., the
integral of the quantity $\lambda \, P(\lambda,z)$

\begin{equation}
\overline{\lambda(z)}  =   \frac{\int_{\lambda_{l}}^{\inf}  P(\lambda,   z)  \,
  \lambda \, \rm d\log\lambda}  {\int_{\lambda_{l}}^{\inf} P(\lambda, z) \, \rm
  d\log\lambda},
\end{equation}

\noindent  where   the  $\lambda_{l}$  is  the  lower   limit  of  the
integration.  For  this exercise the  integration is performed  in the
$\lambda$  interval over  which the  $P(\lambda,z)$  distributions are
well constrained.  We choose  as an integration lower limit $\lambda_l
=  10^{-3}$ and  restrict this  calculation to  redshift  $z<2.5$.  At
higher   redshift   our   $P(\lambda,z)$  constraint   suffers   large
uncertainties  for  $\lambda   <10^{-3}$.  Different  choices  of  the
integration lower  limit will yield different  mean specific accretion
rates but  will not  alter the redshift  dependence of  this quantity,
which  is the focus  of this  exercise.  This  is presented  in Figure
\ref{fig:mean_lambda} to redshift  $z=2.5$.  For comparison also shown
in  this figure  is the  expected amplitude  of the  evolution  of the
specific star-formation rate of galaxies in the same redshift interval
\citep{Elbaz2011}.   This   comparison  is  interesting   because  the
star-formation  rate density  and  the accretion density  of the  Universe
follow similar patterns \citep[e.g.,][]{Aird2010, Aird2015}, and links
are often proposed between the  formation of stars in galaxies and the
growth  of black holes  at their  centres \citep[e.g.,][]{Santini2012,
Rosario2012}.  Figure \ref{fig:mean_lambda} demonstrates than the mean
specific  accretion rate of  the X-ray  AGN population  evolves mildly
with  redshift  when compared  with  the  level  of evolution  of  the
specific  star-formation  rate  of   galaxies  in  the  same  redshift
range.  At  relatively  low  redshift,  $z\la  1$  the  mean  specific
accretion  rate  is nearly  constant,  while  at  higher redshifts  it
appears to track the  overall evolution of the specific star-formation
rate of galaxies.  This can be interpreted as  evidence that accretion
events onto  supermassive black holes  decouple from the  formation of
new stars in galaxies below redshift $z\approx1$.

Next,  the  zero-order approximation  of  a  universal shape  for  the
specific accretion-rate  distribution is scrutinised  by investigating
changes  with  redshift of  the  basic  shape characteristics  of  the
$P(\lambda,z)$.    This   is   easier    done   using   the   quantity
$\lambda\,P(\lambda,z)$  plotted in  Figure \ref{fig:lplz},  which has
the   linear   power-law   slope  of   the   specific   accretion-rate
distributions removed to highlight  subtle differences in slopes among
redshift bins. In Figure \ref{fig:lplz} there is evidence that at high
redshift,   $z\ga1$,  the   slope  of   the  specific-accretion   rate
distributions below $\log \lambda \approx  -2$ is flatter than that at
low redshifts, $z=0.0-0.5$.   This trends is manifested  by the offset
between the red (lowest redshift bin; $\lambda\,P(\lambda,z=0.0-0.5)$)
and the  blue (high redshift  bin) distributions at $\log  \lambda \la
-2$.   This  difference becomes  more  pronounced  at  $z >  1.5$  and
suggests that the value of the specific accretion rate below which the
distributions of  Figure \ref{fig_plz} turn over  depends on redshift,
i.e., shifts  to higher  $\lambda$ with increasing  redshift.  Similar
trends  are  claimed by  \cite{Bongiorno2016}  and  Aird et  al.   (in
prep.).

\begin{figure*}
\begin{center}
\includegraphics[height=0.6\columnwidth]{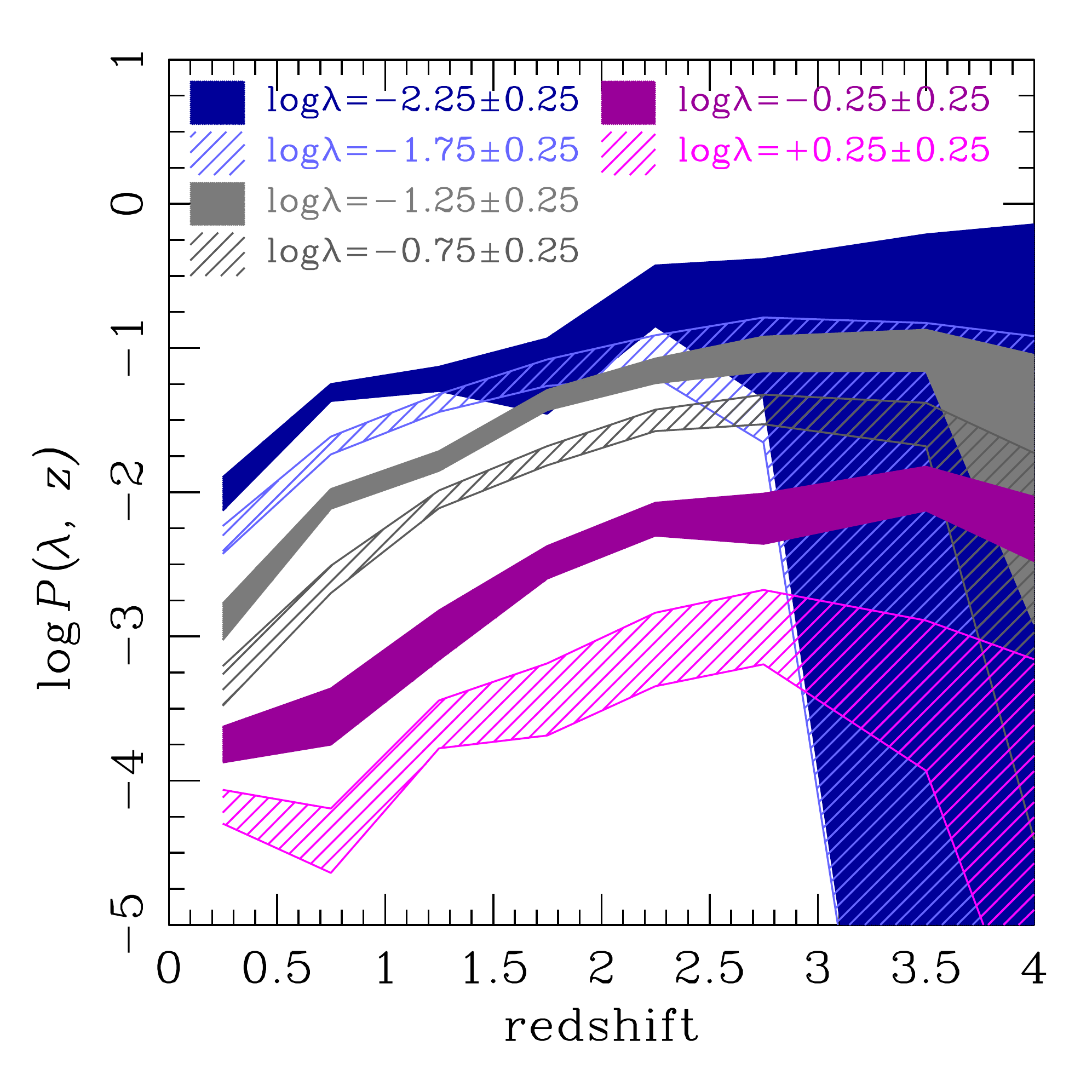}
\includegraphics[height=0.6\columnwidth]{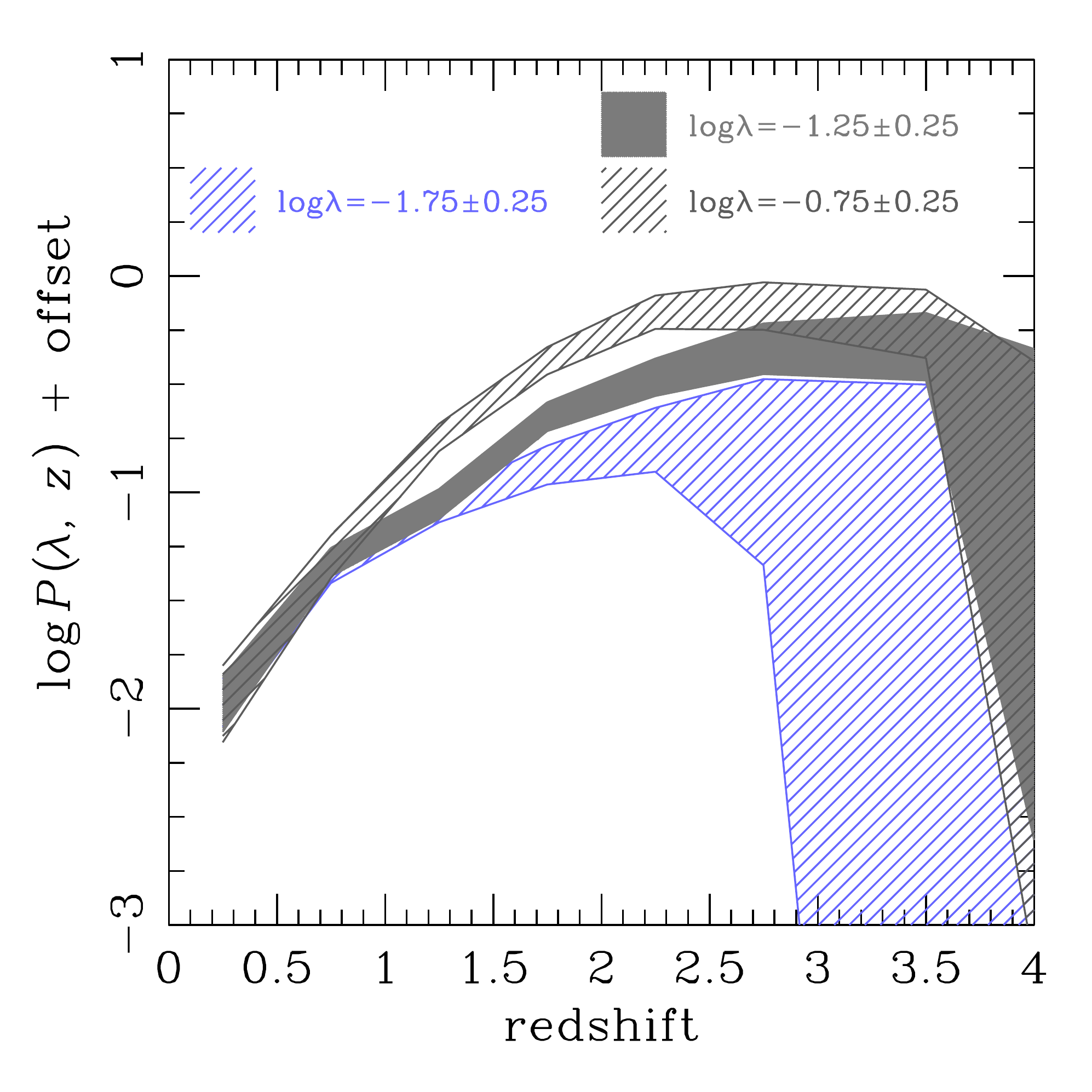}
\end{center}
\caption{{\bf Left:}  $P(\lambda, z)$ as  a function of  redshift. The
hatched  regions with  different  colours and  shadings correspond  to
different $\lambda$-bins  as indicated in  the panels.  The  extent of
the shaded regions corresponds to  the 90\% confidence interval around
the median of the $P(\lambda, z)$ distributions. {\bf Right:} same as
the  left panel  but  the $P(\lambda,  z)$ are  shifted  upward by  an
arbitrary  offset to  force them  to overlap  at low  redshifts.  This
superposition  of the  specific  accretion  rate distributions  allows
direct  comparison and  emphasises the  differential evolution  of the
curves.   For clarity  only three  $\lambda$-bins are  displayed. 
}\label{fig:pl_vs_z}
\end{figure*}

Figure \ref{fig:pl_vs_z}  explores subtle differences in  the redshift
evolution of the probability of a galaxy experiencing accretion events
of a  given specific accretion  rate.  The  left panel of  that figure
plots the $P(\lambda, z)$ in different $\lambda$-bins as a function of
redshift. The right panel  faciliates the comparison between different
$\lambda$-bins by normalising  them to a fixed value  at low redshift.
Low  and  moderate specific  accretion-rate  bins  ($\log \lambda  \la
-0.5$) evolve  at a  similarly high rate  between redshifts  $z=0$ and
$z\approx1$.   At  higher redshift  however,  $z\approx  1 -  3$,  the
evolution rate of low specific  accretion-rate bins ($\log \lambda \la
-1.5$)  slows compared  to  higher  specific accretion-rate  intervals
($-1.5  \la  \log\lambda  \la  0$).  This  effect  is  manifested  for
example, in  the right panel  of Figure \ref{fig:pl_vs_z} by  the fact
that the  (normalised) probability  of an  accretion event  with $\log
\lambda  =  -0.75\pm0.25$   lies  above  that  for   $\log  \lambda  =
-1.25\pm0.25$  or  $\log  \lambda  =  -1.75\pm0.25$  in  the  redshift
interval $z\approx  1 - 3$. It  is also interesting that  close to the
Eddington limit,  i.e., for  $\lambda =  -0.25\pm0.25$ and  $\lambda =
+0.25\pm0.25$,  the evolution  rate of  the $P(\lambda,z)$  starts off
slow  at low  redshift,  contrary to  other  $\lambda$-bins, but  then
increases  and broadly  follows  the evolution  of  the high  specific
accretion-rate  intervals. This  type  of evolution  implies that  the
probability   (or   duty-cycle)   of  moderate   and   high   specific
accretion-rate  events  increases relative  to  that  of low  specific
accretion-rate episodes toward higher  redshift. Therefore one expects
an increasing  fraction of  moderate and high  specific accretion-rate
AGN  with   increasing  redshift.   It  therefore   appears  that  the
evolution,   in  which   only   the  overall   normalisation  of   the
$P(\lambda,z)$  distributions  changes  with  redshift,  needs  to  be
revised to  include a  $\lambda$-dependent differential  evolution. We
emphasise  neverthless,  that this  a  subtle  effect, with  the  main
evolution pattern being that of an overall increase in normalisation.

\begin{figure*}
\begin{center}
\includegraphics[height=0.9\columnwidth]{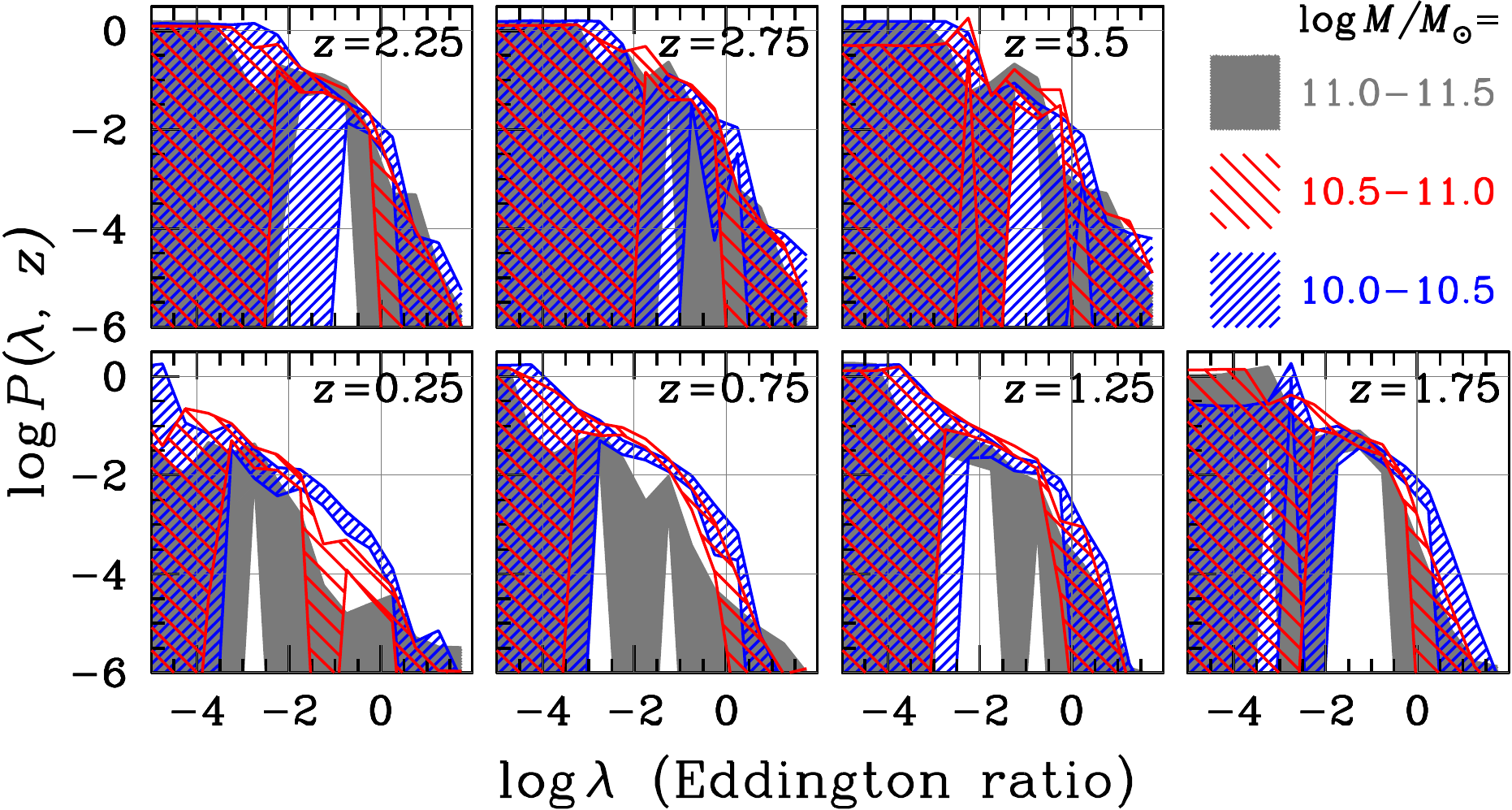}
\end{center}
\caption{Comparison of the $P(\lambda,  z)$ distributions of X-ray AGN
samples split by host-galaxy stellar mass. Each panel corresponds to a
different  redshift   interval,  $z=0.0-0.5$,   $0.5-1.0$,  $1.0-1.5$,
$1.5-2.0$, $2.0-3.0$,  $3.0-4.0$. The  hatched regions  with different
colours  and shadings  correspond to  different stellar  mass bins  as
indicated in the legend. The  extent of the shaded regions corresponds
to the 90\%  confidence interval around the median  of the $P(\lambda,
z)$ distributions.  }\label{fig:pl_vs_mass}
\end{figure*}

We further  explore the stellar-mass dependence  of the $P(\lambda,z)$
distribution  by  splitting the  sample  into  stellar mass  bins  and
determining   the  specific   accretion-rate  distribution   for  each
separately.   The results  are shown  in Figure  \ref{fig:pl_vs_mass},
which compares  the $P(\lambda,z)$  distributions in three  mass bins:
$\log M  / M_{\odot}= 10.0 -  10.5$, $\log M /M_{\odot}=  10.5 - 11.0$
and $\log  M /M_{\odot}= 11.0  - 11.5$.   There is a  systematic trend
whereby more  massive galaxies are less  likely to host AGN  with high
specific  accretion  rates  approaching  the  Eddington  limit,  $\log
\lambda \approx 0$.  This trend  is more evident at redshifts $z\la2$,
where  the  uncertainties  of   the  $P(\lambda,z)$  distributions  in
individual mass bins are smaller.   At the low specific accretion-rate
end of the $P(\lambda,z)$  distributions ($\log\lambda\la-1$) there is
fair agreement  between the different  mass samples, at least  for the
redshift   panels   of    Figure   \ref{fig:pl_vs_mass},   where   the
uncertainties of each mass sample are reasonably small.

In  the  analysis  that  follows   we present results using the
specific accretion-rate distributions that do not depend on the stellar
mass. This because   of  the relatively  large uncertainties  of the
$P(\lambda,z)$  constraints for  individual mass bins.  Nevertheless,
where necessary we will also discuss the impact on the results and
conclusions of the mass-dependent specific accretion-rate distributions
plotted in Figure \ref{fig:pl_vs_mass}. 

\subsection{X-ray AGN mass function and duty cycle}

The redshift  evolution of the $P(\lambda,z)$  distributions of Figure
\ref{fig_plz},  which  can  be  approximated by  an  increase  of  the
normalisation with  increasing redshift,  indicates that  galaxies are
more likely to host an AGN at earlier cosmic times.

\begin{figure*}
\begin{center}
\includegraphics[height=0.85\columnwidth]{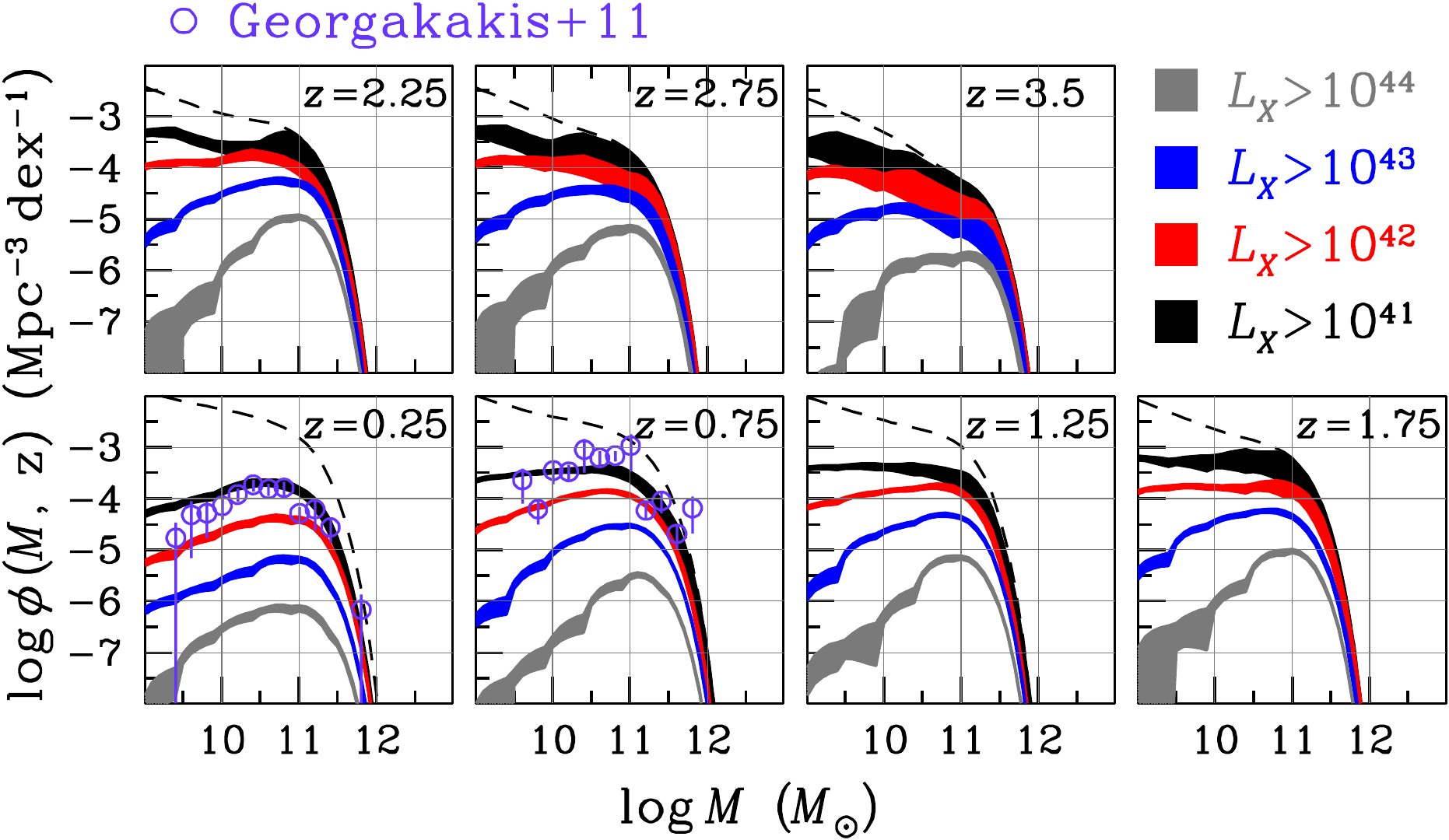}
\end{center}
\caption{Stellar  mass  function  of  AGN host  galaxies  derived  by
convolving  the  $P(\lambda,  z)$  of  Fig.   \ref{fig_plz}  with  the
\protect\cite{Ilbert2013}  galaxy stellar  mass function.   Each panel
corresponds to a  different redshift interval, $z=0.0-0.5$, $0.5-1.0$,
$1.0-1.5$, $1.5-2.0$, $2.0-3.0$, $3.0-4.0$.  The mean redshift of each
bin  is  marked  on  each  panel. The  shaded  regions  correspond  to
different  X-ray luminosity cuts:  $L_X( \rm  2-10\,keV)>10^{41}\, erg
\,s^{-1}$ (black), $L_X( \rm 2-10\,keV)>10^{42}\, erg \,s^{-1}$ (red),
$L_X(  \rm  2-10\,keV)>10^{43}\,   erg  \,s^{-1}$  (blue),  $L_X(  \rm
2-10\,keV)>10^{44}\, erg  \,s^{-1}$ (grey). The  black-dashed curve in
each  panel  is  the  \protect\cite{Ilbert2013}  galaxy  stellar  mass
function. The extent  of the shaded regions correspond  to the 5th and
95th percentiles of the AGN  host galaxy stellar mass function.  Also
plotted with the  purple open circles are the  AGN host galaxy stellar
mass functions  independently derived by \protect\cite{Georgakakis2011}
for systems with $L_X(  \rm 2-10 \, keV ) > 10^{41}  \, erg \, s^{-1}$
at mean redshifts $z=0.1$ and  $z=0.8$. The purple open circles should
be compared with the black shaded curve.}\label{fig:mfz}
\end{figure*}

This  result  can  be  demonstrated by  estimating  the  stellar  mass
function of  AGN hosts  above a given  X-ray luminosity  cut following
relation

\begin{equation}\label{eq:lambda2phi}  \phi(M, z)  =  \int_{L_X>L_{cut}}
\psi(M, z) \, P(\lambda(L_\mathrm{X}, M), z) \, \mathrm{d} \log \lambda,
\end{equation}

\noindent  where  the  integration  is  over  the  range  of  specific
accretion rates  that yield  X-ray luminosities  (for a  given stellar
mass) larger than the cut $L_{cut}$. Figure \ref{fig:mfz} presents the
stellar mass  function of  AGN at different  X-ray luminosity  cuts in
comparison with the mass function  of the galaxy population determined
by \cite{Ilbert2013}.  Also displayed  in Figure \ref{fig:mfz} are the
stellar mass  functions of X-ray selected  AGN with $L_X( \rm  2-10 \,
keV) > 10^{41} \, erg  \, s^{-1}$ derived by \cite{Georgakakis2011} at
mean redshifts  $z\approx0.1$ and $z\approx0.8$.  Although  that study
shares  some of  the  data  used in  the  present  analysis, the  mass
functions  have  been estimated  using  a  very different  methodology
compared to  the one  presented here.   The reasonably  good agreement
between the mass functions of \cite{Georgakakis2011} and those derived
in this  paper further  underlines the robustness  of the  results.  A
feature  of  Figure  \ref{fig:mfz}   is  that  with  decreasing  X-ray
luminosity cut the mass function  of AGN becomes typically broader and
extends to lower stellar masses.   Another important point to be taken
from  Figure \ref{fig:mfz}  is that  at fixed  stellar mass  and X-ray
luminosity cut  the fraction  of AGN  among galaxies  increases toward
higher redshift and  even approaches unity for certain  regions of the
stellar mass, redshift and X-ray luminosity threshold parameter space.

\begin{figure*}
\begin{center}
\includegraphics[height=0.85\columnwidth]{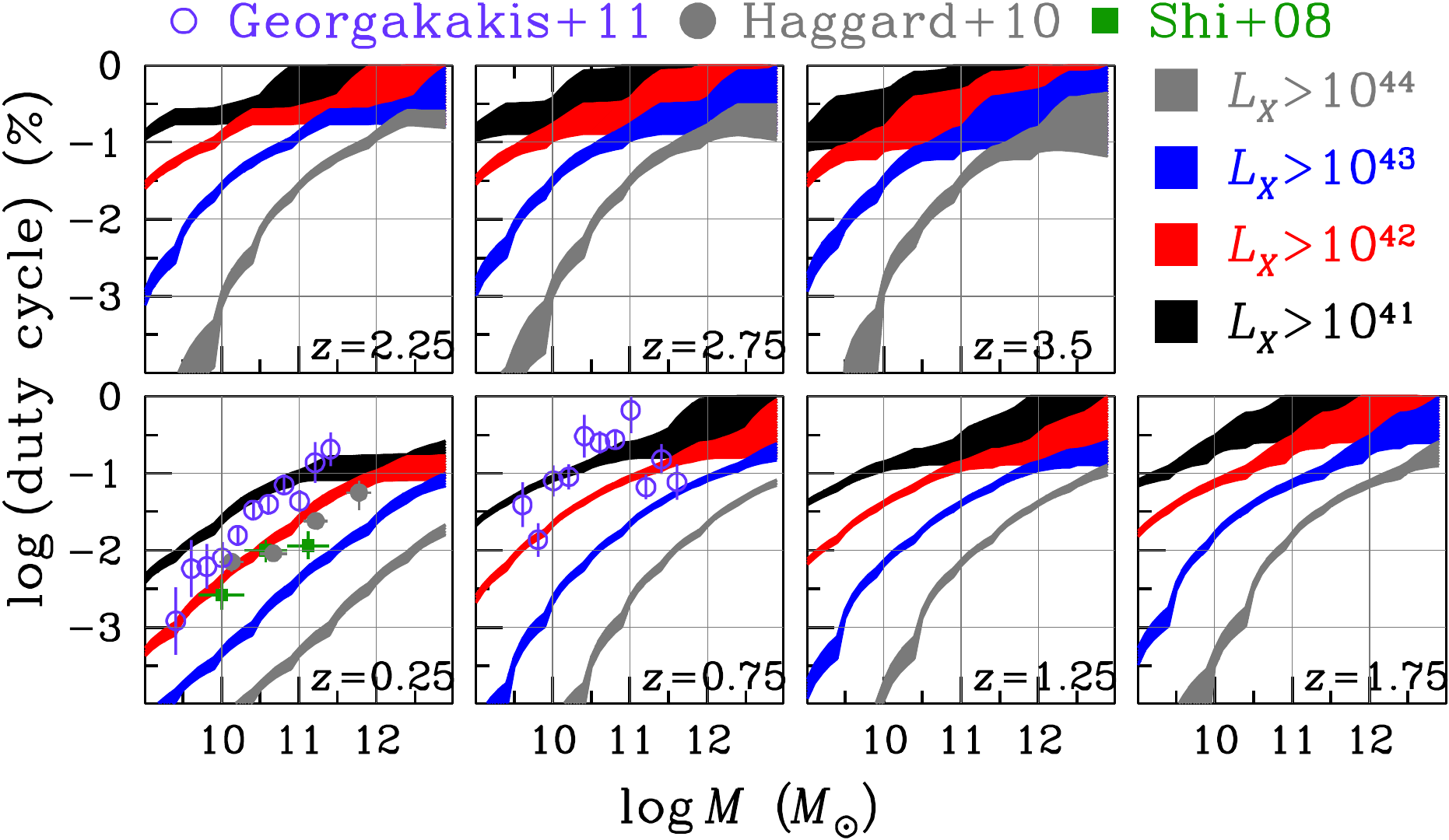}
\end{center}
\caption{AGN duty  cycle as  a function of  stellar mass  estimated by
dividing the stellar mass function  of AGN in Fig.  \ref{fig:mfz} with
that of  galaxies \protect\citep{Ilbert2013}.  Each  panel corresponds
to a  different redshift interval,  $z=0.0-0.5$, $0.5-1.0$, $1.0-1.5$,
$1.5-2.0$,  $2.0-3.0$, $3.0-4.0$.  The  mean redshift  of each  bin is
marked on  each panel.  The  different colours correspond  to different
X-ray  luminosity cuts: $L_X(  \rm 2-10\,keV)>10^{41}\,  erg \,s^{-1}$
(black), $L_X( \rm 2-10\,keV)>10^{42}\, erg \,s^{-1}$ (red), $L_X( \rm
2-10\,keV)>10^{43}\,     erg    \,s^{-1}$     (blue),     $L_X(    \rm
2-10\,keV)>10^{44}\, erg  \,s^{-1}$ (grey).  The extent  of the shaded
regions  correspond  to the  5th  and  95th  percentiles of  the  AGN
duty-cycle at a  given stellar mass.  Also plotted  are the independent
constraints  on  AGN   duty-cycle  from  \protect\citet[][purple  open
circles;  $L_X(\rm  2-10\,keV)>10^{41}\, erg  \,  s^{-1}$, $z=0.1$  and
$z=0.8$]{Georgakakis2011},   \protect\citet[][grey   filled   circles;
$L_X(\rm     2-10\,keV)    \ga     10^{42}\,     erg    \,     s^{-1}$,
$z=0.125-0.275$]{Haggard2010}, and \protect\citet[][green squares;
$L_X(\rm     2-10\,keV)    \ga     10^{42}\,     erg    \,     s^{-1}$,
$z=0.1-0.4$]{Shi2008}. 
}\label{fig:dcz}
\end{figure*}

Another way of presenting
this latter result  is via the AGN duty cycle, i.e.,  the ratio of the
mass functions of AGN and galaxies above a particular X-ray luminosity
cut.  This relation is presented as a function of stellar mass and for
different  X-ray   luminosity  cuts  in   Figure  \ref{fig:dcz}.   For
comparison, independent duty-cycle  observational constraints from the
literature  \citep{Shi2008,  Haggard2010,  Georgakakis2011}  are  also
shown   at  the   appropriate  redshift-interval   panels   of  Figure
\ref{fig:dcz}.  At  fixed stellar mass and  X-ray luminosity threshold
the duty cycle of AGN increases with increasing redshift.

\subsection{X-ray AGN downsizing}

\begin{figure}
\begin{center}
\includegraphics[height=0.85\columnwidth]{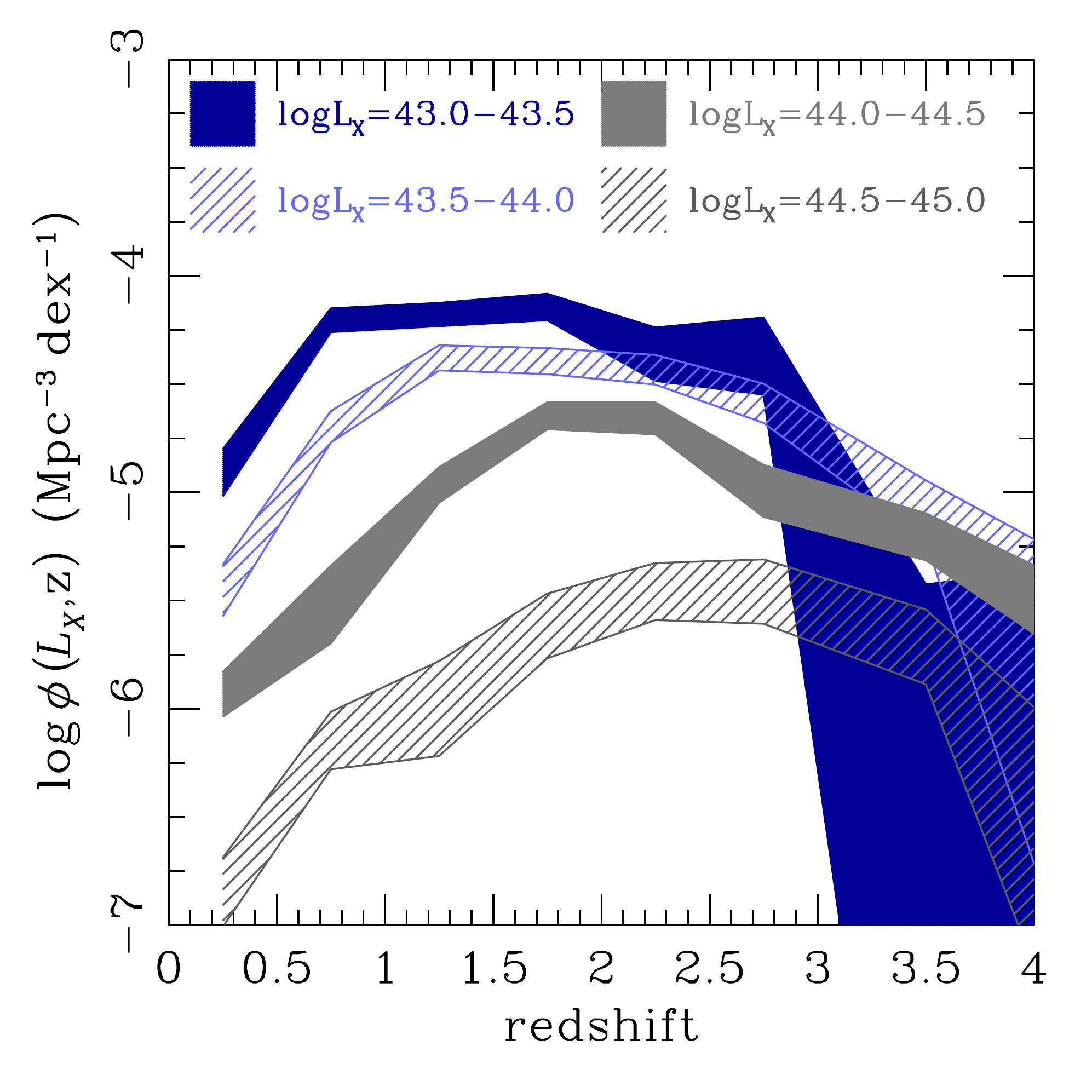}
\end{center}
\caption{X-ray AGN space density  in different luminosity intervals as
a  function  of  redshift.   The hatched  regions  are  non-parametric
constraints of the AGN space density in different X-ray luminosity and
redshift bins.  The colours and shadings correspond  to four different
X-ray luminosity  bins as  indicated on the  plot.  The extent  of the
shaded regions corresponds to  the 90\% confidence interval around the
median of the $\phi(L_X, z)$ distributions.  }\label{fig:xlf_dsize}
\end{figure}

AGN downsizing  is defined as the  trend whereby the space  density of
luminous AGN  peaks at earlier  cosmic times compared to  systems with
lower accretion  luminosities \citep{Barger2005,  Hasinger2005}.  This
differential  evolution  with  X-ray  luminosity is  shown  in  Figure
\ref{fig:xlf_dsize},  which  plots  as  a  function  of  redshift  our
non-parametric constraints  on the space  density of AGN  in different
X-ray luminosity bins.  In this  section we discuss this observational
trend  in the  context of  the specific  accretion rate  distributions
derived in our work.

\begin{figure*}
\begin{center}
\includegraphics[height=0.85\columnwidth]{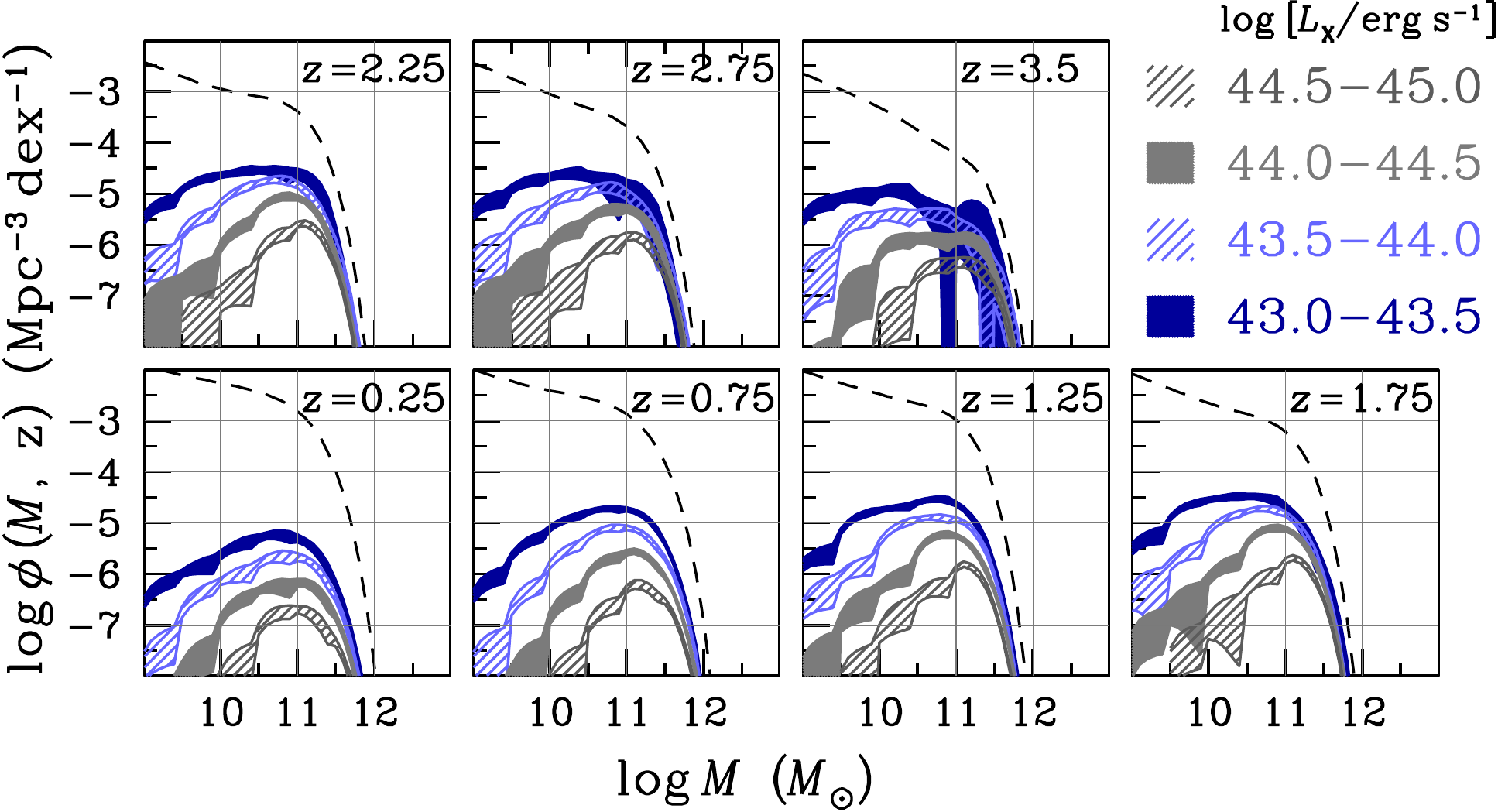}
\includegraphics[height=0.85\columnwidth]{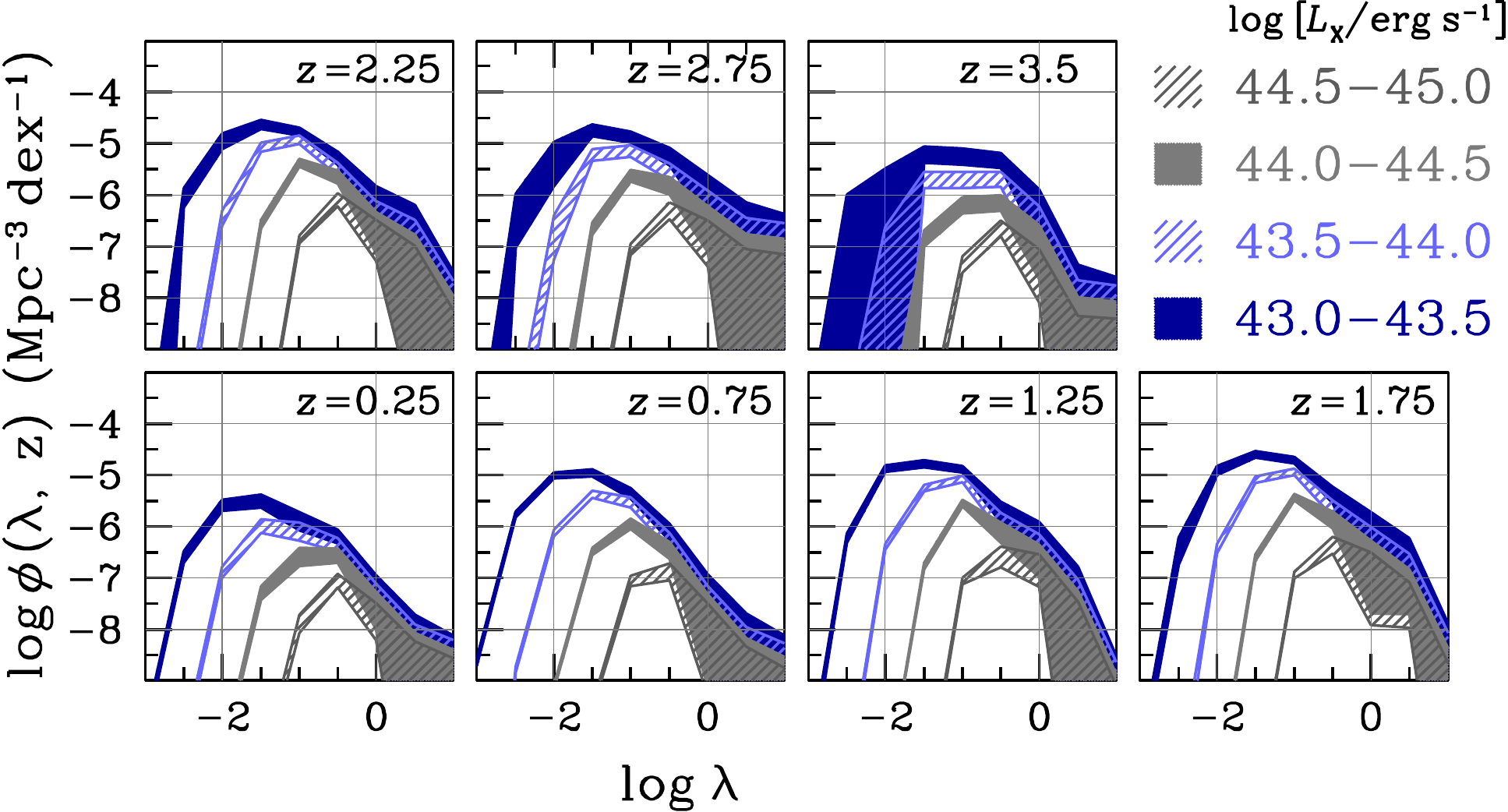}
\end{center}
\caption{{\bf Top  set:} Stellar mass  function of AGN  hosts galaxies
derived by convolving the  $P(\lambda, z)$ of Fig.  \ref{fig_plz} with
the  \protect\cite{Ilbert2013}  galaxy  stellar mass  function.   Each
panel  corresponds  to  a  different redshift  interval,  $z=0.0-0.5$,
$0.5-1.0$,  $1.0-1.5$,  $1.5-2.0$,  $2.0-3.0$,  $3.0-4.0$.   The  mean
redshift  of each bin  is marked  on each  panel.  The  shaded regions
correspond to different X-ray luminosity cuts: $\log \left[ L_X( \rm 2
- 10  \,keV )/erg \,s^{-1}\right]  =43-43.5$ (blue-solid), $43.5-44.0$
(light-blue   hatched),    $44.0-44.5$   (grey   solid),   $44.5-45.0$
(dark-grey  hatched).  The black-dashed  curve in  each panel  is the
\protect\cite{Ilbert2013} galaxy stellar mass function.  The extent of
the shaded regions correspond to  the 5th and 95the percentiles of the
AGN  host galaxy stellar  mass function.   {\bf Bottom  set:} Specific
accretion-rate function of AGN  for the same X-ray luminosity selected
sub-samples and  same redshift intervals  presented in the top  set of
panels.  }\label{fig:mf_dsize}
\end{figure*}

\begin{figure*}
\begin{center}
\includegraphics[height=0.65\columnwidth]{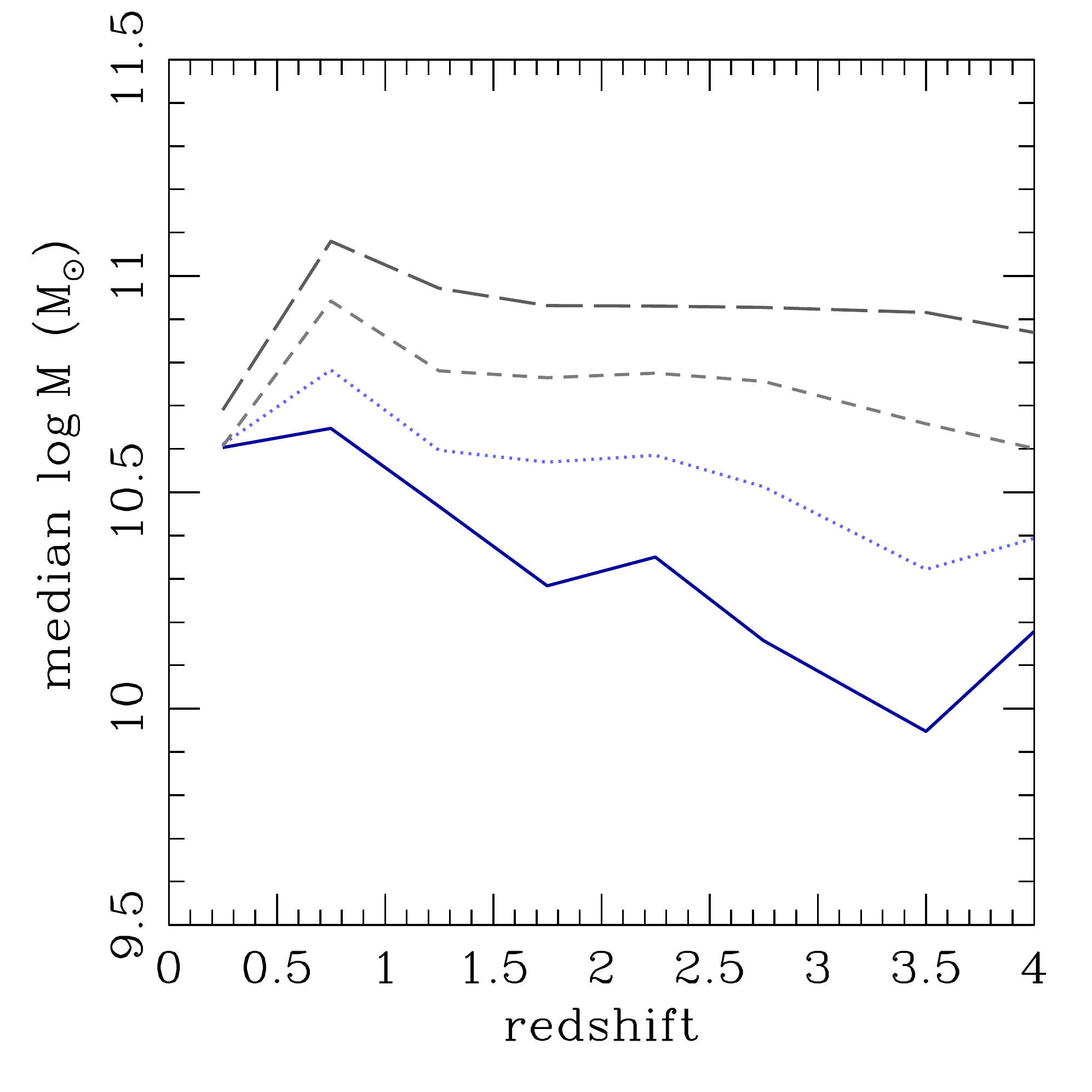}
\includegraphics[height=0.65\columnwidth]{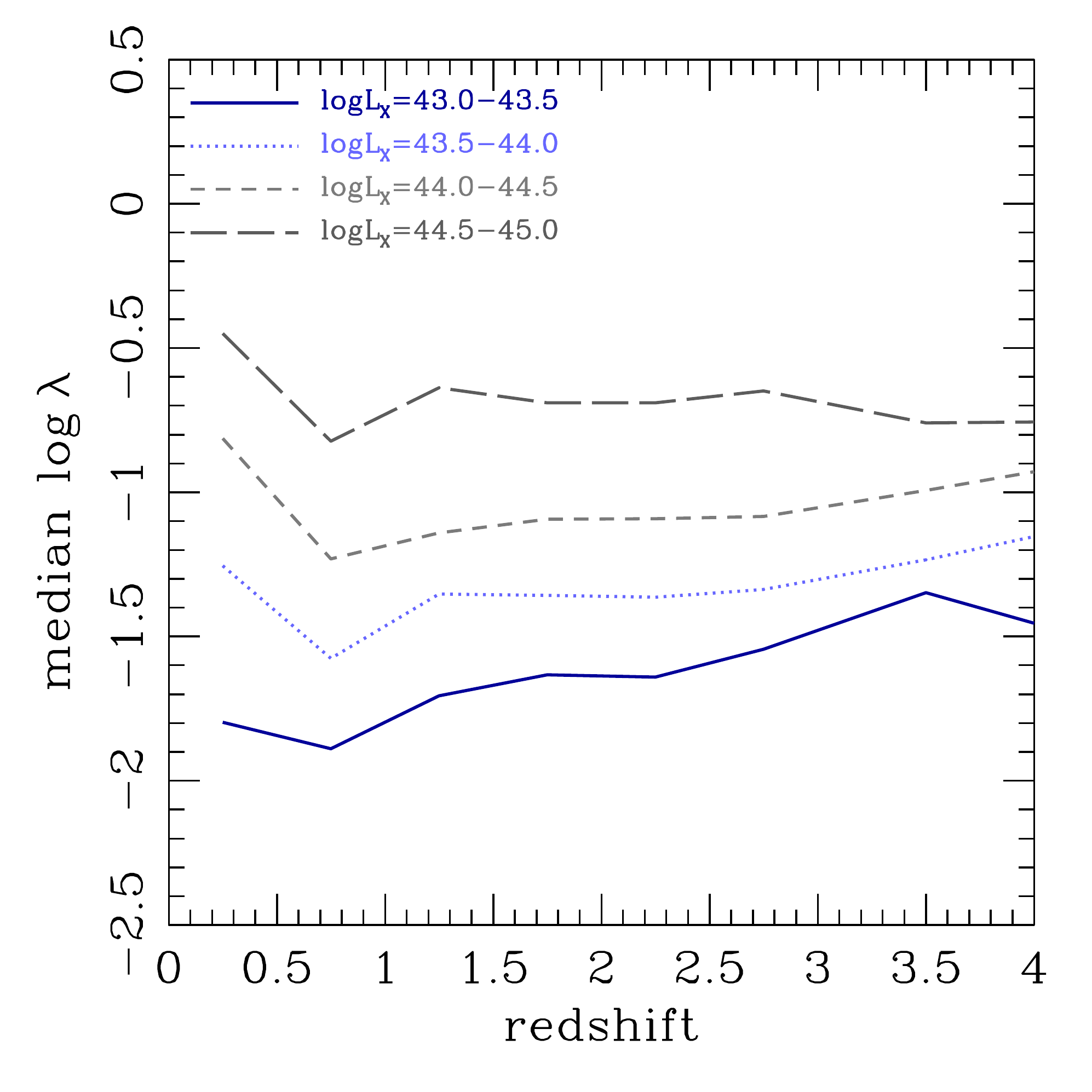}
\includegraphics[height=0.65\columnwidth]{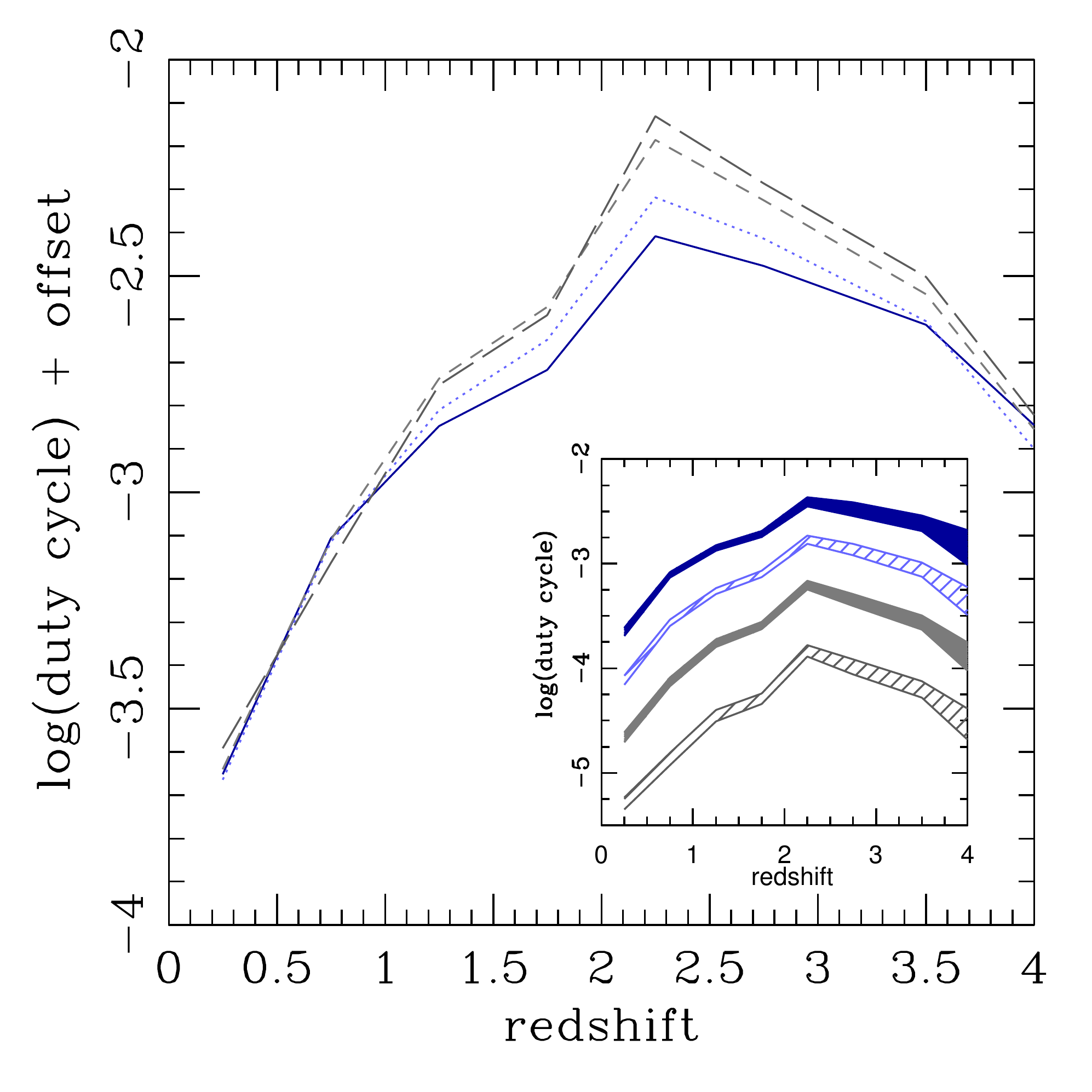}
\end{center}
\caption{{\bf Left:}  Median stellar  mass of  AGN hosts  in different
2-10\,keV X-ray  luminosity intervals as  a function of  redshift. The
median stellar mass is estimated  from the mass function distributions
presented in the top set of panels of Figure \ref{fig:mf_dsize}.  {\bf
Middle:} Median specific accretion rate  of AGN in different 2-10\,keV
X-ray luminosity intervals  as a function of redshift.   The median is
estimated from the specific accretion rate functions in the bottom set
of panels of Figure \ref{fig:mf_dsize}. {\bf Right:} Median duty cycle
of AGN in different 2-10\,keV X-ray luminosity intervals as a function
of redshift. The duty cycle is estimated by dividing the median of the
integrated  AGN mass  functions  of Figure  \protect\ref{fig:mf_dsize}
(i.e. split  by 2-10\,keV X-ray  luminosity) with the  integrated mass
function  of  the  galaxy  population. The  duty  cycle  measures  the
probability of  a galaxy  at redshift  $z$ hosting  an AGN  with X-ray
luminosity in the  intervals $\log \left[ L_X( \rm 2  - 10 \,keV )/erg
\,s^{-1}\right]  =43-43.5$   (blue  solid),   $43.5-44.0$  (light-blue
dotted),   $44.0-44.5$   (grey    short-dashed),   $44.5-45.0$   (grey
long-dashed).  An  arbitrary vertical offset  has been applied  to the
curves  to  force  them  to  overlap at  low  redshift  and  highlight
differences in the evolution of the duty-cycle curves.  The inset plot
shows the same  quantity, i.e.  duty cycle, as a  function of redshift
without any vertical offsets applied.   The widths of the shaded areas
also show  the 90\% confidence  interval around the median.  The color
coding        is       the        same       as        in       Figure
\protect\ref{fig:mf_dsize}. }\label{fig:median_dsize}
\end{figure*}

Figure \ref{fig:mf_dsize} displays in different redshift intervals the
space density of AGN as a function of stellar mass (mass function) and
specific accretion  rate ($\lambda$  function) in the  same luminosity
intervals  as in  Figure  \ref{fig:xlf_dsize}.   These quantities  are
estimated  by  convolving  the  mass function  of  galaxies  with  the
$P(\lambda,z)$ distribution and integrating over the appropriate X-ray
luminosity range.   The mass  functions in  the top  set of  panels of
Figure  \ref{fig:mf_dsize}   become  broader  with   decreasing  X-ray
luminosity and the peaks of  the distributions shift to somewhat lower
stellar mass  with increasing redshift.   The latter point  is further
demonstrated in the left panel of Figure \ref{fig:median_dsize}, which
presents  the median  stellar mass  of  the mass  functions of  Figure
\ref{fig:mf_dsize}.  The  decreasing trend of the  median stellar mass
with increasing redshift is more  pronounced for lower luminosity AGN.
Nevertheless, the variation in stellar  mass is about 0.5\,dex or less
over  a  wide range  of  redshifts,  even  for the  lowest  luminosity
interval $\log  L_X(\rm 2-10\,keV) = 43.0-43.5$.   A similar amplitude
of redshift variation is therefore expected for the specific accretion
rate of the X-ray luminosity  subsamples.  The specific accretion rate
function   is  shown   in  the   bottom  set   of  panels   of  Figure
\ref{fig:mf_dsize}.  There  is a clear trend  whereby lower luminosity
AGN  are systematically  shifted  to lower  specific accretion  rates,
albeit with  a tail  extending to high  $\lambda$ values.   The median
specific accretion  rate of  these distributions  is displayed  in the
middle  panel  of Figure  \ref{fig:median_dsize}.   There  is a  clear
offset  in  the  median  specific accretion  rate  of  the  subsamples
selected  by X-ray  luminosity.  AGN  with luminosities  $\log L_X(\rm
2-10\,keV)  = 44.5-45.0$  are  typically associated  with black  holes
accreting  with   $\log  \lambda  \approx  -0.7$,   while  those  with
luminosities  $\log L_X(\rm  2-10\,keV)  =  43.0-43.5$ typically  have
accretion  rates  $\log \lambda  \approx  -1.7$,  with some  level  of
variation across redshift.  The trend whereby X-ray selected AGN split
by luminosity  have distinct median  specific accretion rates  is also
evident when using the mass-dependent $P(\lambda, z)$ distributions to
convolve with the mass function of  galaxies and produce the curves in
Figure \ref{fig:median_dsize}. This result  is therefore robust to the
stellar mass dependence of the specific accretion-rate distribution of
AGN.

The  AGN downsizing  of  Figure \ref{fig:xlf_dsize}  can therefore  be
understood  as  the   result  of  two  effects.   The   first  is  the
stratification  in  median  specific  accretion rate  of  AGN  samples
selected    by    X-ray    luminosity,     as    shown    in    Figure
\ref{fig:median_dsize}.   The  second  effect is  that  high  specific
accretion-rate   events    are   more    likely   relative    to   low
specific-accretion ones  with increasing redshift. In  other words the
probability of  a galaxy hosting  an AGN with  a certain value  of the
specific  accretion rate  depends on  both $z$  and $\lambda$.   These
trends  are  indicated  in  Figure \ref{fig:pl_vs_z}. They are further
demonstrated in the right panel of 
Figure \ref{fig:median_dsize},  which plots  the duty-cycle of  AGN in
the luminosity  intervals defined  in Figure  \ref{fig:xlf_dsize}. The
duty-cycle  is the  probability of  galaxy with  a given  stellar mass
hosting an AGN with X-ray luminosity in a given range. This is estimated
by dividing the the AGN stellar  mass functions plotted in the top set
of panels of  Figure \ref{fig:mf_dsize} with the  total galaxy stellar
mass function  and then integrating  along the stellar mass  axis. AGN
with $\log L_X(\rm 2-10\,keV) = 43.0-43.5$ include a large fraction of
sources  with  $\lambda  \la  -1.5$  and  therefore  their  duty-cycle
evolution slows  at redshifts $z\ga0.5-1$  relative to AGN  with $\log
L_X(\rm  2-10\,keV) =  44.5-45.0$.  The  latter population  includes a
large fraction of accretion events  with $\lambda \ga -1.0$, which are
characterised by a higher rate of duty-cycle evolution to redshifts $z
\approx  2-2.5$.  The  duty-cycle  evolution of  all accretion  events
eventually  slows  at  redshift  beyond  $z\ga2.5$  resulting  in  the
observed decrease of the space density of AGN at early cosmic times.

\begin{figure}
\begin{center}
\includegraphics[height=0.8\columnwidth]{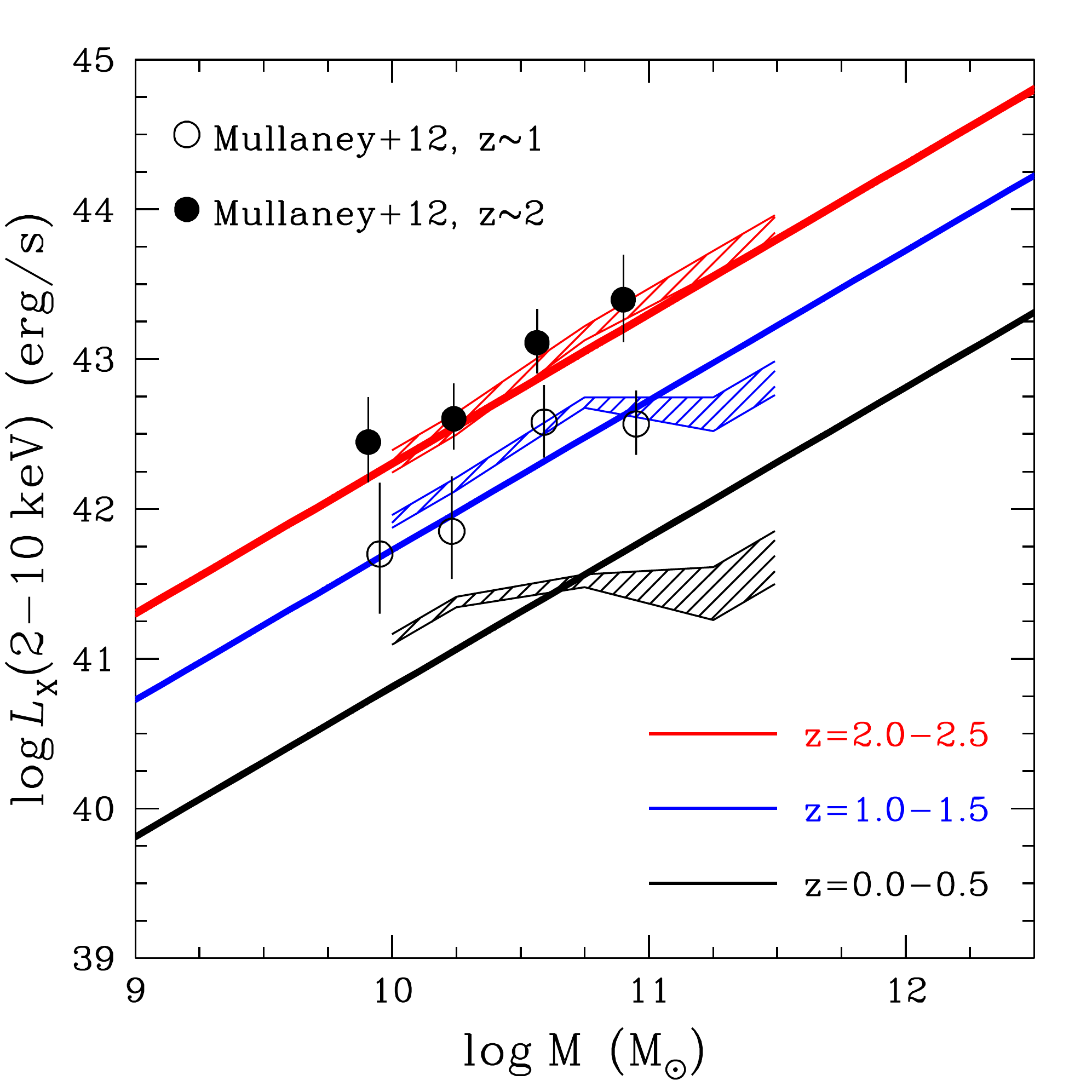}
\end{center}
\caption{Mean 2-10\,keV X-ray luminosity of  galaxies as a function of
  stellar  mass. The  shaded  bands are  estimated  by convolving  the
  $P(\lambda,     z)$    of     Fig.     \ref{fig_plz}     with    the
  \protect\cite{Ilbert2013}   galaxy   stellar   mass   function   and
  integrating  in  luminosity.   The expectation  in  three  different
  redshift  intervals is  indicated, $z=0.0-0.5$  (black), $z=1.0-1.5$
  (blue)  and $z=2.0-2.5$  (red).  The hatched  regions are  estimated
  using  the mass-dependent  specific accretion-rate  distributions of
  Figure \ref{fig:pl_vs_mass}. In this case, the mean X-ray luminosity
  is determined in the mass range $\log M/M_{\odot} = 10-11.5$, where
  the mass-dependent $P(\lambda, z)$ distributions are reasonably well
  constrained. The color coding of  the hatched regions corresponds to
  the redshift  intervals indicated  in the plot.  Also shown  are the
  X-ray          stacking         analysis          results         of
  \protect\cite{Mullaney2012_hidden}  at   redshifts  $z\approx1$  and
  $z\approx2$. We  caution that  these data  points correspond  to the
  average  X-ray luminosity  of  star-forming galaxies,  not the  full
  galaxy population.  }\label{fig:meanlx_mass_z}
\end{figure}

\begin{figure}
\begin{center}
\includegraphics[height=0.8\columnwidth]{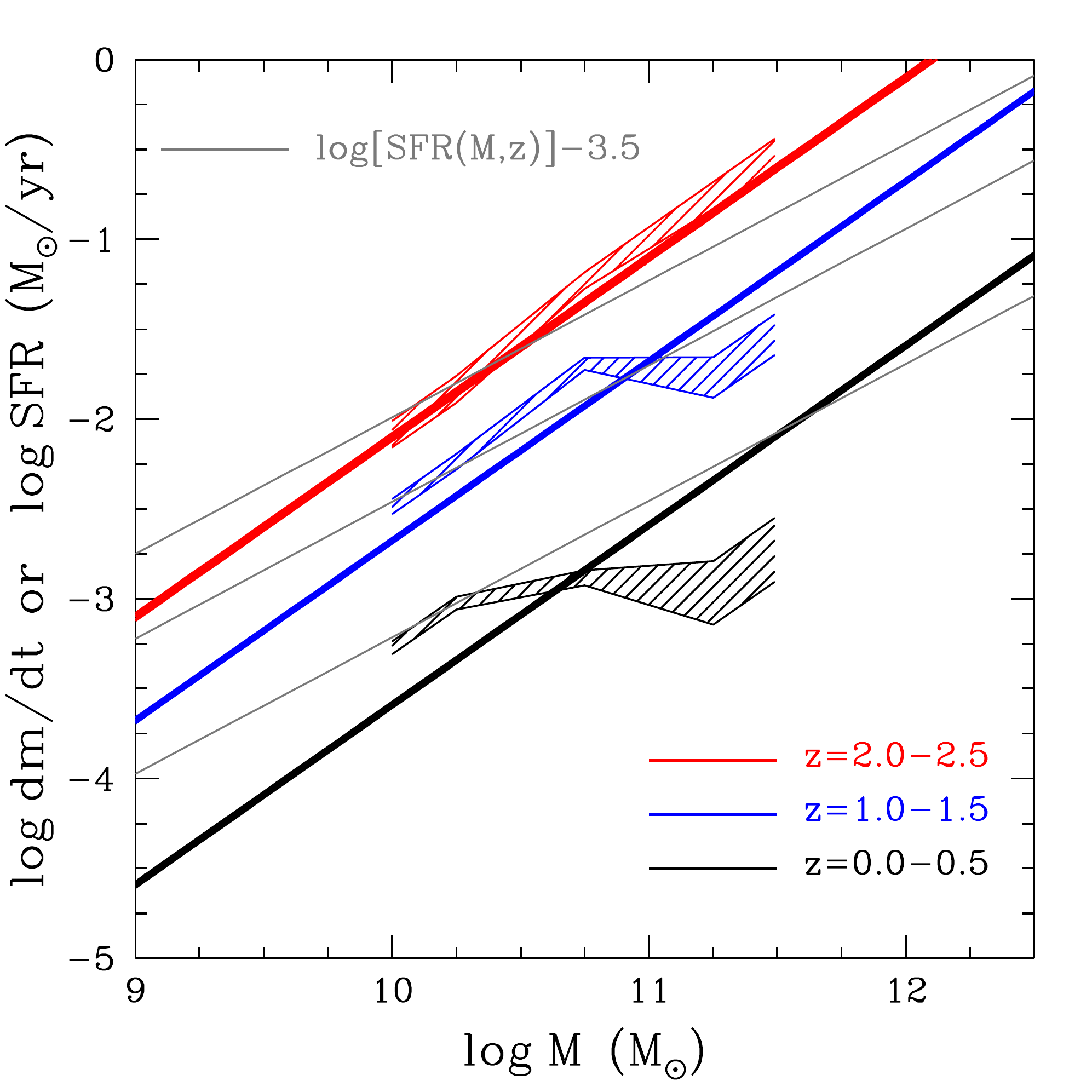}
\end{center}
\caption{Black hole accretion rate, $dm/dt$,  in units of solar masses
  per year as a function of stellar mass. The colored shaded bands and
  hatched   regions   correspond   to    those   plotted   in   Figure
  \ref{fig:meanlx_mass_z}.  Accretion  rates are estimated  from X-ray
  luminosities  assuming   a  radiative  efficiency  of   10\%  and  a
  bolometric  correction of  25. Also  shown  with grey  lines is  the
  star-formation rate (SFR) vs stellar  mass relation of Main Sequence
  galaxies at the  corresponding redshift \protect\cite{Aird2017}. The
  slope  of  the SFR  vs  stellar  mass  relation  is about  0.8.  The
  normalisation of  the star-formation  rate vs stellar  mass relation
  has  been  scaled  down  by  -3.5  in  logarithmic  units  to  allow
  comparison   with   the   black  hole   accretion   rate   relation.
}\label{fig:growth}
\end{figure}

\begin{figure}
\begin{center}
\includegraphics[height=0.8\columnwidth]{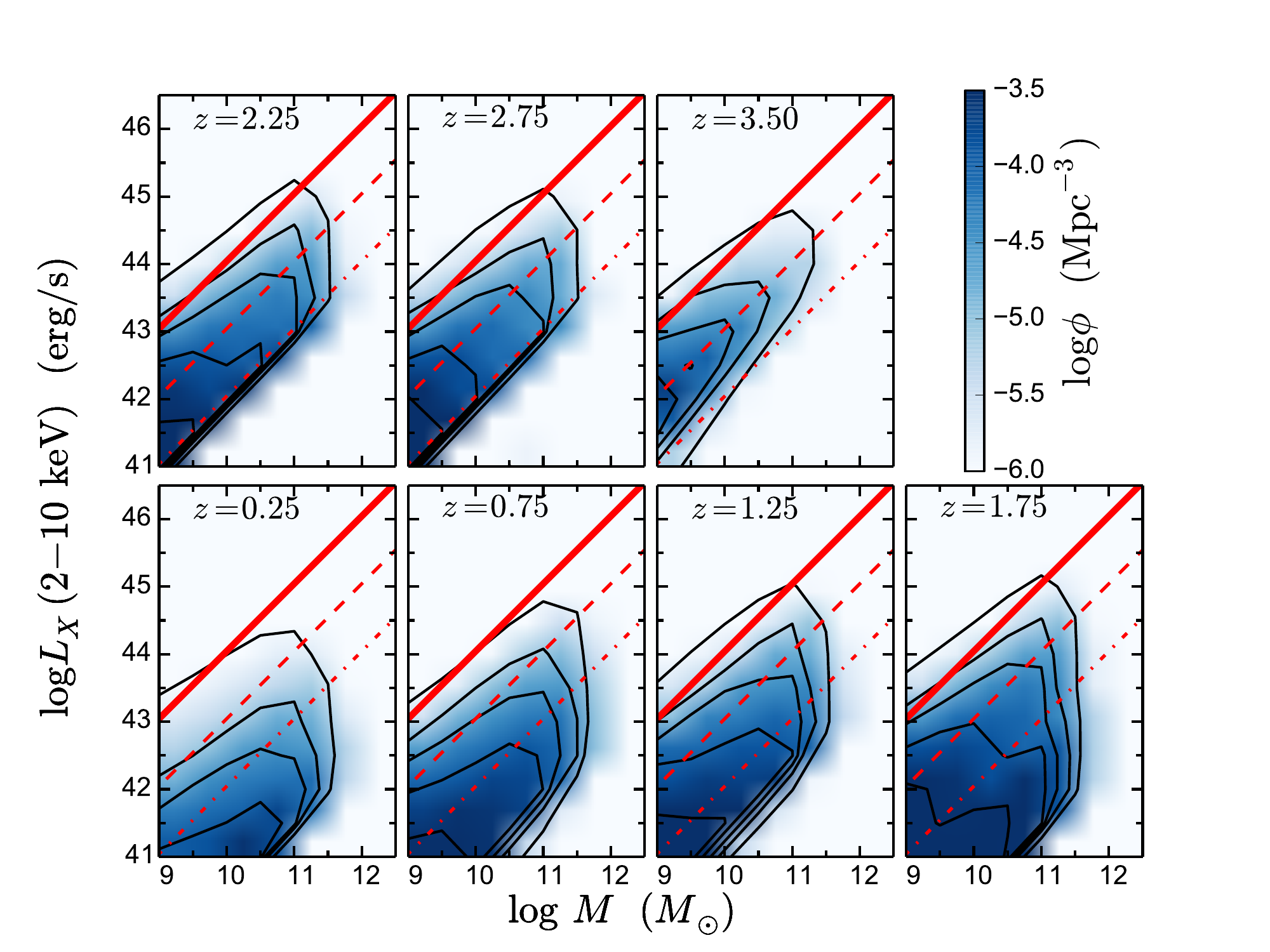}
\end{center}
\caption{Space  density  distribution of  X-ray  selected  AGN in  the
2-dimensional  space  of  stellar  mass  and  X-ray  luminosity.   The
different  panels correspond  to  different  redshift intervals.   The
spade density is  estimated by convolving the $P(\lambda,  z)$ of Fig.
\ref{fig_plz} with  the \protect\cite{Ilbert2013} galaxy  stellar mass
function.   The shading  corresponds  to different  AGN space  density
levels in logarithmic units as  indicated in the colour-bar. The black
contours  mark logarithmic  space  densities of  $\log  \phi =  -6.0$,
--5.5, --5.0, --4.5,  --4.0 and --3.5.  The thick  continuous red line
marks the specific accretion rate  $\lambda=1$. The thinner dashed and
dashed-dotted  red  lines  indicate $\lambda=10^{-1}$  and  $10^{-2}$,
respectively.  We caution that at $z>2$ and $\lambda\la-2$ the
distributions   are   affected   by    the   lack   of   observational
constraints on the $P(\lambda, z)$ distribution.}\label{fig:lx_m_z} 
\end{figure}

\section{Discussion}

In this paper we  combine wide-area/shallow and pencil-beam/deep X-ray
surveys to compile one of the largest samples of X-ray selected AGN to
date, with a  total of 4821 sources.  The  X-ray data are supplemented
by multiwavelength observations  to derive the specific accretion-rate
distribution of  AGN, $P(\lambda,z)$. This quantity is  defined as the
probability of a galaxy at a  given redshift, $z$, hosting an AGN with
specific  accretion  rate  $\lambda$.   Our  approach  uses  the  mass
function of  galaxies as  a boundary condition  and has  the advantage
that is non-parametric in nature, i.e., no assumptions are made on the
underlying   shape  or   model  that   describes   the  $P(\lambda,z)$
distribution. Additionally, uncertainties  in the determination of the
properties  of individual  AGN  in  our sample,  such  as host  galaxy
stellar masses,  X-ray luminosities and redshifts  are propagated into
the analysis.

\subsection{AGN evolution and
  relation to star-formation}\label{sec:discussion1} 

Our   non-parametric  method   recovers   a  specific   accretion-rate
distribution of  AGN (see  Fig.  \ref{fig_plz}) that  increases toward
low specific  accretion rates to  $\log \lambda\approx-3$ and  shows a
break close to  the Eddington limit, above which the  probability of a
galaxy hosting an accretion event  drops steeply.  Below $\log \lambda
\approx  -3$  there is  evidence  that  the  increasing trend  of  the
$P(\lambda, z)$  is inverting  and the  distribution drops.   There is
also evidence that the turnover point at low specific accretion rates,
$\log \lambda \la  -3$, is a function of redshift,  i.e., it shifts to
higher   $\lambda$    values   with   increasing    redshift.    These
characteristics  are broadly  consistent  with previous  observational
studies that either derive the specific accretion-rate distribution of
AGN via non-parametric methods \citep[][]{Aird2012, Bongiorno2012}, or
adopt an analytic  model for the specific  accretion rate distribution
and recover the relevant parameters  by requiring that the convolution
of this  model with the  AGN or galaxy  mass functions yields  the AGN
luminosity   function  \citep[][]{Aird2013,   Bongiorno2016}.   Recent
hydrodynamical  cosmological  simulations   of  galaxy  formation  and
black-hole growth also produce  Eddington ratio distributions that are
qualitatively similar,  in terms of  shape and redshift  evolution, to
the    observational   results    shown    in   Fig.     \ref{fig_plz}
\citep{Sijacki2015}.

We find that  the specific accretion-rate distribution  of AGN evolves
strongly with cosmic time so that  the probability of a galaxy hosting
an AGN  increases toward higher  redshifts.  In other  words, galaxies
were on average more nuclear-active in the past. This trend translates
to a  higher AGN  duty cycle  at earlier epochs  and for  more massive
galaxies  (see  Fig.   \ref{fig:dcz}).   At   $z>1.5$  and  $\rm  M  >
10^{11}\,M_{\odot}$,  for example,  nearly 100\%  of the  galaxies are
predicted to host  an AGN with $L_X (\rm  2 - 10 \, keV)  > 10^{41} \,
erg \,  s^{-1}$.  It  is interesting to  compare the  high duty-cycles
inferred by  our analysis with independent  observational estimates of
the   incidence   of   AGN    among   galaxies   at   high   redshift.
\cite{Genzel2014} studied the kinematics of massive galaxies $\rm \log
M /M_{\odot}  > 10.9$ at  $z=1-3$ and found  that 2/3 of  their sample
show broad nuclear emission-line components $\approx \rm 450 - 5300 \,
km\,  s^{-1}$.   These  objects  may  be  associated  with  black-hole
accretion events and suggest a high  duty-cycle of AGN outflows at the
redshifts and stellar-mass  intervals above.  Such a  high fraction is
not inconsistent with the duty cycles implied by our analysis.

The  rapid increase  of  the  AGN duty-cycle  from  redshift $z=0$  to
$z\approx1-1.5$ is also driving the  strong evolution of the AGN space
and   luminosity   densities   between  these   redshifts   \citep[see
also][]{Aird2013}. In Figure \ref{fig:dcz}  the change in amplitude of
the AGN duty-cycle  at fixed stellar mass and luminosity  cut from low
redshift  to $z\approx1$  is  about  1\,dex. This  is  similar to  the
increase  of  the corresponding  integrated  AGN  space or  luminosity
densities \citep[e.g.][]{Ueda2014,Aird2015}.

The observed increase of the AGN duty-cycle toward higher redshift may
be related  to the evolution  of the star-formation rate  in galaxies.
As  a  first  step  we  explore  such  a  relation  in  a  qualitative
manner. Star-forming galaxies  are known to occupy  a relatively tight
sequence    on   the    star-formation   vs    stellar   mass    plane
\citep[][]{Noeske2007,  Speagle2014, Whitaker2012,  Tomczak2016}.  The
dominant evolution pattern  is a shift of this  sequence toward higher
star-formation rates with  increasing redshift, i.e., a  change of the
sequence's     normalisation    \citep[e.g.,][]{Speagle2014}.      Put
differently, at  fixed stellar  mass galaxies form  stars at  a higher
rate    with    increasing    redshift    \citep[e.g.,][]{Speagle2014,
Ilbert2015}.  Figure  \ref{fig:meanlx_mass_z} shows  our observational
results on  the mean  X-ray luminosity  of galaxies  as a  function of
stellar mass  and for  different redshift intervals.   Our constraints
are compared with X-ray stacking  analysis results from the literature
\citep[][]{Mullaney2012_hidden}. At  fixed stellar mass  galaxies have
systematically  higher X-ray  luminosities  with increasing  redshift,
i.e.,  qualitatively similar  behaviour  as  the star-formation  rate.
Similar  trends are  evident  when using  the mass-dependent  specific
accretion-rate   distributions  of   Figure  \ref{fig:pl_vs_mass}   to
determine the mean X-ray luminosity as  a function of stellar mass and
redshift. In this case however, the  slopes of the $L_X - M$ relations
at $z\la2$ are shallower than unity, i.e.  the slope derived using the
mass-independent $P(\lambda, z)$ distribution.  The difference is more
striking  at the  lowest  redshift-bin, $z=0.0-0.5$  and becomes  less
pronounced with increasing redshift.

Next  we explore  quantitatively the  relation between  the growth  of
black  holes   and  the  star-formation  rate   in  galaxies.   Figure
\ref{fig:growth}  compares the  accretion  rate onto  the black  hole,
$dm/dt$, with the star-formation rate  (SFR) of Main Sequence galaxies
at the  corresponding redshift. For  the conversion of the  mean X-ray
luminosities  of  Fig.  \ref{fig:pl_vs_mass} to  black-hole  accretion
rates we  assume radiative efficiency  of 10\% and a  fixed bolometric
correction of 25.  The relation between SFR, stellar mass and redshift
is  from  \cite{Aird2017}.   Figure \ref{fig:growth}  shows  that  the
overall normalisations of the Main Sequence of star-formation and mean
black-hole accretion rate evolve similarly with redshift. However, the
mass dependence  of the Main  Sequence SFR and $dm/dt$  have different
slopes.    Adopting  a   stellar-mass   independent  $P(\lambda,   z)$
distribution to derive the mean  X-ray luminosity of galaxies yields a
slope  of unity,  compared  to slopes  in the  range  $0.6-1$ for  the
star-formation rate \citep[e.g.][]{Speagle2014}.   Taking into account
the mass-dependence of the $P(\lambda,  z)$ introduces a tilt into the
$dm/dt$ vs stellar mass relation, the amplitude of which is a function
of  redshift.  At  $z=0.0-0.5$ the  mass-dependence of  the black-hole
accretion rate is flatter than  that of the star-formation rate, while
at $z=1.5-2$ it  is steeper.  This suggests that  the relation between
star-formation and black-hole accretion events is redshift dependent. 

A  different  manifestation of  this  effect  is  presented in  Figure
\ref{fig:mean_lambda}, which plots the redshift dependence of the mean
specific accretion rate,  $\overline{\lambda}$, of the $P(\lambda, z)$
distribution.  We find that  this quantity remains nearly constant out
to  redshift $z\approx1$,  and then  at higher  redshifts  follows the
evolution  of  the  specific  star-formation  rate  of  Main  Sequence
galaxies.  There  is therefore evidence for  two distinct evolutionary
regimes of the mean specific  accretion rate of AGN.  At high redshift
the  quantity $\overline{\lambda}$  couples  to the  Main Sequence  of
star-formation, while  at lower redshift  it appears to  separate from
the  evolution  of the  specific  star-formation  rate  of the  galaxy
population.   This   differential  behaviour  with   redshift  can  be
interpreted as  evidence for multiple  black-hole accretion mechanisms
that dominate at different cosmic epochs.  This scenario is consistent
with     some     recent     cosmological     Semi-Analytic     Models
\citep[][]{Fanidakis2012,  Hirschmann2012},   which  invoke  different
black-hole  fuelling  modes to  reproduce  the  evolution  of AGN  and
galaxies.  In these models the different accretion modes take place in
distinct environments with respect  to the level of star-formation and
become important at different redshifts. 

Finally,  we  note  that  the  different  stellar-mass  dependence  of
black-hole accretion  rate and star-formation rate  contradicts recent
results  by \cite{Mullaney2012_hidden}.   These authors  estimated the
mean X-ray and  infrared luminosities of galaxies and  find that these
quantities have  remarkably similar dependence on  galaxy stellar mass
and  redshift.  This  was interpreted  in the  context of  a universal
ratio between  mean black hole  growth rate and SFR.   The discrepancy
between  Figure  \ref{fig:growth} and  the  \cite{Mullaney2012_hidden}
results may be related to sample  selection effects (e.g. X-ray AGN vs
infrared  galaxies) or  biases associated  with the  stacking analysis
(e.g., Eddington  bias, see  Mullaney et  al. 2012).  It  is also
noted  that the  trends observed  in Figure  \ref{fig:growth} are  not
sensitive  to   the  adopted  conversion  from   X-ray  to  bolometric
luminosity. The  use of the \cite{Marconi2004}  bolometric corrections
yields similar results.

\subsection{Constraints on black-hole fuelling modes}

Our  results  suggest  that  the  probability  of  galaxy  hosting  an
accretion event onto  its supermassive black hole depends  on both the
redshift  and  the  specific  accretion-rate  of  the  event.   Figure
\ref{fig:pl_vs_z}  shows  that  high  specific  accretion-rate  events
($\lambda  \ga -1.5$)  become  more likely  relative  to low  specific
accretion-rate  ones with  increasing redshift.   \cite{Bongiorno2016}
parametrise  the  specific accretion-rate  distribution  of  AGN by  a
double power-law with a slope at the low-$\lambda$ end that depends on
redshift.  They  then infer a behaviour  similar to that found  by our
non-parametric  approach, i.e.,  a low-$\lambda$  slope that  flattens
with increasing redshift resulting in  a differential evolution of the
specific accretion-rate distribution.  \cite{Merloni_Heinz2008} solved
the  continuity equation  of the  black-hole mass  function using  the
locally determined  one as  boundary condition  and assuming  that the
evolution of the X-ray luminosity function of AGN traces the growth of
black-holes  as   a  function  of   cosmic  time.   They   recover  an
Eddington-ratio distribution for AGN that flattens at higher redshift,
i.e., indicating an increasing relative importance of highly accreting
objects at earlier cosmic epochs.  We emphasise that this differential
evolution is a second order effect in our analysis. The bulk effect is
an overall increase of the  normalisation of the $P(\lambda,z)$ toward
higher redshift. 

The  differential  evolution of  the  $P(\lambda,z)$  with respect  to
$\lambda$ may be related to the  fact that galaxies at higher redshift
are, on average, more gas-rich compared to lower redshift counterparts
having the same  stellar mass \citep[e.g.,][]{Genzel2010, Tacconi2010,
Tacconi2013,  Magdis2012,  Santini2014}.   This  description  however,
cannot be the only effect.  If  a larger reservoir of gas is available
to the supermassive black hole,  and the same black-hole fuelling mode
or AGN triggering mechanism is in operation across all redshifts, then
one  might  expect  a  higher   frequency  of  accretion  events  with
increasing redshift over  the full range of  specific accretion rates.
Our analysis  suggests the existence of  additional physical processes
that suppress  lower specific  accretion-rate events relative  to high
specific accretion-rate  ones at higher redshift.   One possibility is
that there are multiple AGN fuelling or triggering modes that dominate
at  different cosmic  times and  specific accretion-rate  regimes (see
also Section  \ref{sec:discussion1}).  For example, one  may imagine a
simple  scenario, in  which  the  growth of  black  holes in  gas-rich
galaxies  proceeds  at  relatively  high  Eddington  rates,  while  in
gas-poor  early-type galaxies  black  holes grow  at typically  slower
rates        relative       to        the       Eddington        limit
\citep[e.g.,][]{Kauffmann_Heckman2009,  Fanidakis2012}.    The  latter
fuelling mode may  become less important with  increasing redshift, as
galaxies  become, on  average, more  gas-rich at  fixed stellar  mass,
thereby  leading  to  the   observed  differential  evolution  of  the
$P(\lambda,z)$   distribution    with   respect   to    the   specific
accretion-rate.  This scenario is  consistent with the distribution of
the AGN space density on  the 2-dimensional parameter space of stellar
mass and X-ray luminosity plotted  in Figure \ref{fig:lx_m_z}.  At low
redshift   the   AGN  population   is   dominated   by  low   specific
accretion-rate systems,  $\log\lambda\la-2$.  At  higher redshift
moderate  specific accretion-rate  events, $-2\la  \log\lambda \la-1$,
become increasingly  prominent toward $z\approx2$. At  higher redshift
the  AGN  population is  dominated  by  accretion events  with  $-2\la
\log\lambda  \la-1$.  This   also  partly  because  of   the  lack  of
observational constraints below $log\lambda\la -2$ at $z>2$.

Variations  of the specific  accretion-rate distribution  of AGN  as a
function of the star-formation rate of their host galaxies, a proxy to
gas availability,  is an  important diagnostic of  multiple black-hole
fuelling   modes   \citep[][]{Kauffmann_Heckman2009,  Georgakakis2014,
Azadi2015}. There is evidence that high specific accretion-rate events
are less common among  quiscent galaxies compared to star-forming ones
(Aird  et  al. in  prep).  Moreover,  recent  work suggests  that  the
signatures of multiple black-hole  fuelling modes are imprinted on the
large  scale  environment of  AGN  and  are  manifested as  systematic
variations  of the  clustering amplitude  with redshift  and accretion
luminosity          \citep[][]{Fanidakis2013,         Koutoulidis2013,
  Mountrichas2016}.  

Semi Analytic Models (SAM) that simulate the formation of galaxies and
the growth of  black holes at their centres  must also adopt different
AGN  triggering processes to  match the  observational results  on the
cosmological      evolution      and      demographics     of      AGN
\citep[e.g.,][]{Fanidakis2012,     Hirschmann2012,     Hirschmann2014,
Gutcke2015}.  AGN  triggering mechanisms in  these simulations include
galaxy  mergers, disk  instabilities  and hot-gas  accretion, each  of
which becomes  important at different  cosmic times, dark  matter halo
environments and accretion luminosity regimes.  For example, there are
striking  similarities  between  our  Figure  \ref{fig:lx_m_z},  which
displays the space density of AGN in the 2-dimensional stellar mass vs
X-ray luminosity space,  and the \cite{Hirschmann2014} SAM predictions
shown  in their  Figure 12,  which  presents the  simulated AGN  space
density distribution in the  black-hole mass and bolometric luminosity
plane. In that figure of Hirschmann  et al., there is a clear shift of
the   AGN  population,   in   terms  of   space   density,  from   low
Eddington-ratio  and  high black-hole  mass  systems  at low  redshift
(radio-mode  accretion) to high-Eddington  ratio and  lower black-hole
mass  sources (quasar-mode accretion)  with increasing  redshift, i.e.
similar to  the trends in Figure \ref{fig:lx_m_z}.   This behaviour of
the simulated AGN in the  \cite{Hirschmann2014} work is because of the
interplay of  the two black hole  fuelling modes adopted  in that SAM.
The   quasar-mode   corresponds   to  high   Eddington-ratio   events,
$\lambda_{Edd}>0.01$,    while   the    radio-mode    corresponds   to
$\lambda_{Edd}<0.01$.   The difference  between the  two modes  is the
level of feedback,  which is assumed to be more  efficient in the case
of the radio-mode black-hole fuelling.  This assumption represents the
impact  on the  interstellar medium  of  radio jets  from massive  and
slow-accreting black holes hosted  by early-type and massive galaxies.
In  that  model the  radio-mode  component  becomes significant  below
redshift  $z \approx  1$,  while at  higher  redshift the  quasar-mode
increasingly dominates  the growth of black holes.   Therefore in this
simulation  there is  differential  evolution of  the Eddington  ratio
distribution, qualitatively similar to the evolution of $P(\lambda,z)$
we  find  in  our  work  (e.g.  see  Fig.   \ref{fig:pl_vs_z}).   High
Eddington-ratio  AGN  evolve  faster  at moderate  and  high  redshift
compared to low Eddington-ratio ones.

Our analysis also suggests  differences in the specific accretion-rate
distribution of AGN split by stellar mass. There is evidence that more
massive galaxies  are less likely to  host AGN with  an accretion rate
approaching  the Eddington  limit (Fig.   \ref{fig:pl_vs_mass}).  This
trend is stronger at $z<2$ and disappears at higher redshift, although
the uncertainties  at $z>2$ are also larger.   The observed variations
in the shape of the $P(\lambda,z)$ with stellar mass may be related to
gas availability  in massive galaxies  that also live, on  average, in
denser   environments.    \cite{Bongiorno2016}   model  the   specific
accretion rate distribution of AGN in the COSMOS survey using a double
power-law and introduce a stellar  mass dependence for the break point
between the  two power-law slopes. They  find that the  break moves to
lower specific  accretion rates with increasing  stellar mass, similar
to our  non-parametric approach.  \cite{Schulze2015}  suggest that the
type-I quasar duty-cycle decreases  with increasing black-hole mass at
least  for redshift  $z<1$,  indicating a  suppression  of type-I  AGN
activity  at higher mass  systems, in  qualitative agreement  with our
results.

\subsection{Implications on the downsizing of X-ray selected AGN}

Our analysis  also provides observational  clues on the origin  of AGN
downsizing.   We  find  that  this observational  trend  is  primarily
related  to the  redshift evolution  of the  AGN duty-cycle,  i.e. the
probability  of a  galaxy hosting  an accretion  event within  a given
X-ray luminosity interval.

Firstly, it  is shown  that the  median stellar mass  of AGN  split by
accretion luminosity  differs by less than  a factor of  six for X-ray
luminosities  spanning two orders  of magnitude  between $L_X=10^{43}$
and  $\rm  10^{45}\, erg  \,  s^{-1}$ (Fig.   \ref{fig:median_dsize}).
There is also no strong redshift dependence of the median stellar mass
between  $z=0$  and  $z=3$  for the  luminosity  selected  sub-samples
(maximum factor  of three  for AGN with  $\log L_X  =43-43.5$).  These
findings argue against black hole mass being primarily responsible for
the  AGN downsizing,  under the  assumption that  stellar  mass tracks
black-hole  mass.  Moreover,  our analysis  suggests that  AGN samples
selected by luminosity differ in their median specific accretion rate,
i.e., at least  one order of magnitude between  $L_X=10^{43}$ and $\rm
10^{45}\,  erg   \,  s^{-1}$  (Fig.    \ref{fig:median_dsize}).   More
luminous  sources  are   associated  typically  with  higher  specific
accretion  rates  compared  to  their lower  luminosity  counterparts.
There  is  also little  redshift  dependence  of  the median  specific
accretion rate for the  luminosity selected AGN sub-samples. Accretion
events  within  a  given  luminosity  interval have  the  same  median
specific accretion  rate across redshift. The parameter  that is found
to evolve differentially with redshift  is the duty-cycle of AGN split
by X-ray luminosity. The  right panel of Figure \ref{fig:median_dsize}
shows  that the  probability  of  a galaxy  hosting  a moderate  X-ray
luminosity  event relative to  a high  X-ray luminosity  one decreases
with  increasing redshift  in the  interval $z\approx1-2.5$  (see also
Fig.  \ref{fig:pl_vs_z}).   We therefore find that  the AGN downsizing
trend is  primarily a duty-cycle  effect, i.e.  it is  associated with
the differential evolution of the probability of a galaxy to host an
accretion  event.  This  conclusion does  not change  if we  take into
account  the stellar-mass  dependence of  the  specific accretion-rate
distribution  shown in  Figure  \ref{fig:pl_vs_mass}, whereby  massive
galaxies are less likely to host AGN with high accretion rates.

The conclusions  above on  AGN downsizing are  compared to  the recent
work  of  \cite{Bongiorno2016}.  These  authors  argue  that the  main
effect in  AGN downisizing is  the differential redshift  evolution of
the  specific  accretion-rate  distribution  of AGN,  with  the  space
density  of high  specific  accretion-rate events  peaking at  earlier
epochs.  This is similar to  our results, although in detail there are
differences.  We do not find evidence that the probability of galaxies
hosting  accretion events  with  different $\lambda$  values peaks  at
different  redshifts  (see  Fig.   \ref{fig:pl_vs_z}).   Instead,  the
evolution of the $P(\lambda, z)$ for different specific accretion-rate
bins shows a  broad plateau in the redshift  interval $z\approx1 - 3$.
Also our conclusions on the origin of AGN downsizing differ from those
of \cite{Schulze2015}.  They  studied the Eddington-rate (not specific
accretion rate) distribution of type-I quasars and found evidence that
the black  hole mass is the  main driver of  downsizing. This apparent
discrepancy  may  indicate   observational  selection  effects,  i.e.,
broad-line QSOs vs X-ray detections, or the biases associated with the
use of stellar mass to approximate the Eddington ratio of AGN.

Our  results  on  the  origin  of  the AGN  downsizing  are  at  least
qualitatively     consistent     with     recent    simulations     by
\cite{Hirschmann2014}.   In  this  latter  work it  is  the  interplay
between  the quasar  and radio  mode accretion  as well  as  the heavy
black-hole seeding at early times  that drives the downsizing trend of
the AGN  luminosity function. Therefore at  least in that  SAM the AGN
downsizing is directly  related to the presence of  multiple modes for
fuelling the black holes of galaxies.

Finally, our  results can be used  to explore the origin  of the break
$L_X^{*}$ of the X-ray  luminosity function of AGN. Our non-parametric
analysis suggests a break luminosity  in the range $\log L_X^{*} / \rm
erg  \, s^{-1}  \approx43.5-44.5$  depending on  redshift (see  Fig.
\ref{fig:xlf_comp1}).    Such   AGN   are   associated   with   median
specific-accretion  rates $\lambda \approx  10^{-1.5} -  10^{-1}$ (see
Fig.  \ref{fig:median_dsize}).  These range  of values  brackets the
position where the slope  of the specific accretion-rate distributions
of  Figure  \ref{fig_plz}   steepens  toward  high  $\lambda$  values.
Additionally AGN with $\log L_X / \rm erg \, s^{-1} \approx 43.5-44.5$
are associated with median galaxy  stellar masses in the range $\log M
/ M_{\odot}  = 10.5-11.0$.   This interval brackets  the break  of the
stellar mass  function of  the galaxy population  $\log M  / M_{\odot}
\approx  10.7$   \citep{Ilbert2013}.   The  $L_X^{*}$   of  the  X-ray
luminosity function is therefore related  to the breaks of the stellar
mass   function   and   the  specific   accretion-rate   distributions
\citep{Aird2013}.  It is also possible to show analytically that these
quantities are related  under the assumptions of a  Schechter form for
the  stellar mass  function and  a double  power-law for  the specific
accretion rate distribution \citep[][]{Caplar2015}.

\section*{Conclusions}

The main conclusions from our work are summarised below

\begin{itemize}

\item  The  non-parametric  approach  for  the  determination  of  the
  specific accretion-rate distribution of AGN recovers a shape that is
  characterised by an increase toward low specific accretion rates, a
  break close to the Eddington  limit, above which 
  the probability  of a  galaxy hosting  an AGN  drops steeply,  and a
  turnover at  very low  specific accretion  rates, $\log  \lambda \la
  -3$.

\item   The  redshift   evolution  of   the  specific   accretion-rate
  distribution is  such that the fraction  of AGN among galaxies  at a
  given  accretion-luminosity limit  (duty-cycle) increases  with both
  increasing stellar mass and redshift.

\item  The  rapid  increase  of the  AGN  duty-cycle  with  increasing
  redshift is  primarily responsible for  the strong evolution  of the
  AGN space  and luminosity densities  between the local  Universe and
  $z=1-1.5$.

\item  There  is  evidence  for  differential  evolution  of  the  AGN
  duty-cycle  with X-ray  luminosity or  specific accretion  rate.  We
  argue  that  this  trend  is   primarily  responsible  for  the  AGN
  downsizing.

\item Our analysis suggests that the relation between accretion events
  onto  supermassive  black  holes   and  star-formation  episodes  is
  redshift dependent.   At redshift $z<1$  there is evidence  that the
  evolution of the specific accretion rate of AGN is disjoint from the
  evolution of the specific star-formation rate of galaxies. At higher
  redshift  the two  quantities evolve  similarly. This  is consistent
  with suggestions  that there are multiple  black-hole fuelling modes
  that dominate at different redshifts.

\item  The specific  accretion-rate  distribution of  AGN likely  also
  depends on  stellar-mass. More  massive galaxies  are systematically
  shifted to lower  specific accretion rates compared  to less massive
  systems.

\end{itemize}

\section*{Acknowledgements}

The authors thank  the anonymous referee for their  careful reading of
the paper  and their constructive  comments. This work  benefited from
the {\sc thales} project 383549 that is jointly funded by the European
Union  and the  Greek Government  in  the framework  of the  programme
``Education and lifelong learning''.  JA acknowledges support from ERC
Advanced  Grant FEEDBACK  340442.  A.S. acknowledges  support by  JSPS
KAKENHI  Grant Number  26800098.  Funding  for the  Sloan Digital  Sky
Survey IV  has been provided  by the  Alfred P. Sloan  Foundation, the
U.S.  Department  of Energy Office  of Science, and  the Participating
Institutions. SDSS acknowledges support  and resources from the Center
for High-Performance Computing at the University of Utah. The SDSS web
site is www.sdss.org.   SDSS is managed by  the Astrophysical Research
Consortium   for   the   Participating  Institutions   of   the   SDSS
Collaboration  including   the  Brazilian  Participation   Group,  the
Carnegie  Institution for  Science,  Carnegie  Mellon University,  the
Chilean   Participation  Group,   the   French  Participation   Group,
Harvard-Smithsonian    Center   for    Astrophysics,   Instituto    de
Astrof\'isica  de  Canarias,  The   Johns  Hopkins  University,  Kavli
Institute for  the Physics  and Mathematics of  the Universe  (IPMU) /
University of  Tokyo, Lawrence  Berkeley National  Laboratory, Leibniz
Institut  f\"ur Astrophysik  Potsdam (AIP),  Max-Planck-Institut f\"ur
Astronomie  (MPIA Heidelberg),  Max-Planck-Institut f\"ur  Astrophysik
(MPA  Garching), Max-Planck-Institut  f\"ur Extraterrestrische  Physik
(MPE), National  Astronomical Observatory  of China, New  Mexico State
University,   New  York   University,   University   of  Notre   Dame,
Observat\'orio   Nacional  /   MCTI,   The   Ohio  State   University,
Pennsylvania  State  University,  Shanghai  Astronomical  Observatory,
United Kingdom  Participation Group, Universidad Nacional  Autonoma de
Mexico,  University  of  Arizona,   University  of  Colorado  Boulder,
University of  Oxford, University  of Portsmouth, University  of Utah,
University  of  Virginia,  University  of  Washington,  University  of
Wisconsin, Vanderbilt University, and Yale University.

\appendix

\section{The impact of different approaches for assessing the  stellar mass of AGN hosts}\label{sec:BH_VS_MSTAR}

Two  methods are  adopted  in our  work  to estimate  the host  galaxy
stellar masses  of AGN with  broad optical-emission lines.   The first
fits hybrid  AGN/galaxy templates to the  observed broad-band spectral
energy distribution of X-ray sources  to decompose the emission of the
underlying   host   from  that   of   the   nuclear  source   (Section
\ref{sec_mstar}). The second approach approximates the stellar mass of
the galaxy by simply scaling the virial black-hole mass estimated from
single-epoch spectroscopy (Section  \ref{sec_mb}). The two methods are
affected  by different  systematics and  random errors  and  provide a
handle  to assess  the impact  of uncertainties  in  broad-line quasar
stellar  masses  to  the  results  and  conclusions  on  the  specific
accretion-rate distribution.

\begin{figure*}
\begin{center}
\includegraphics[height=0.85\columnwidth]{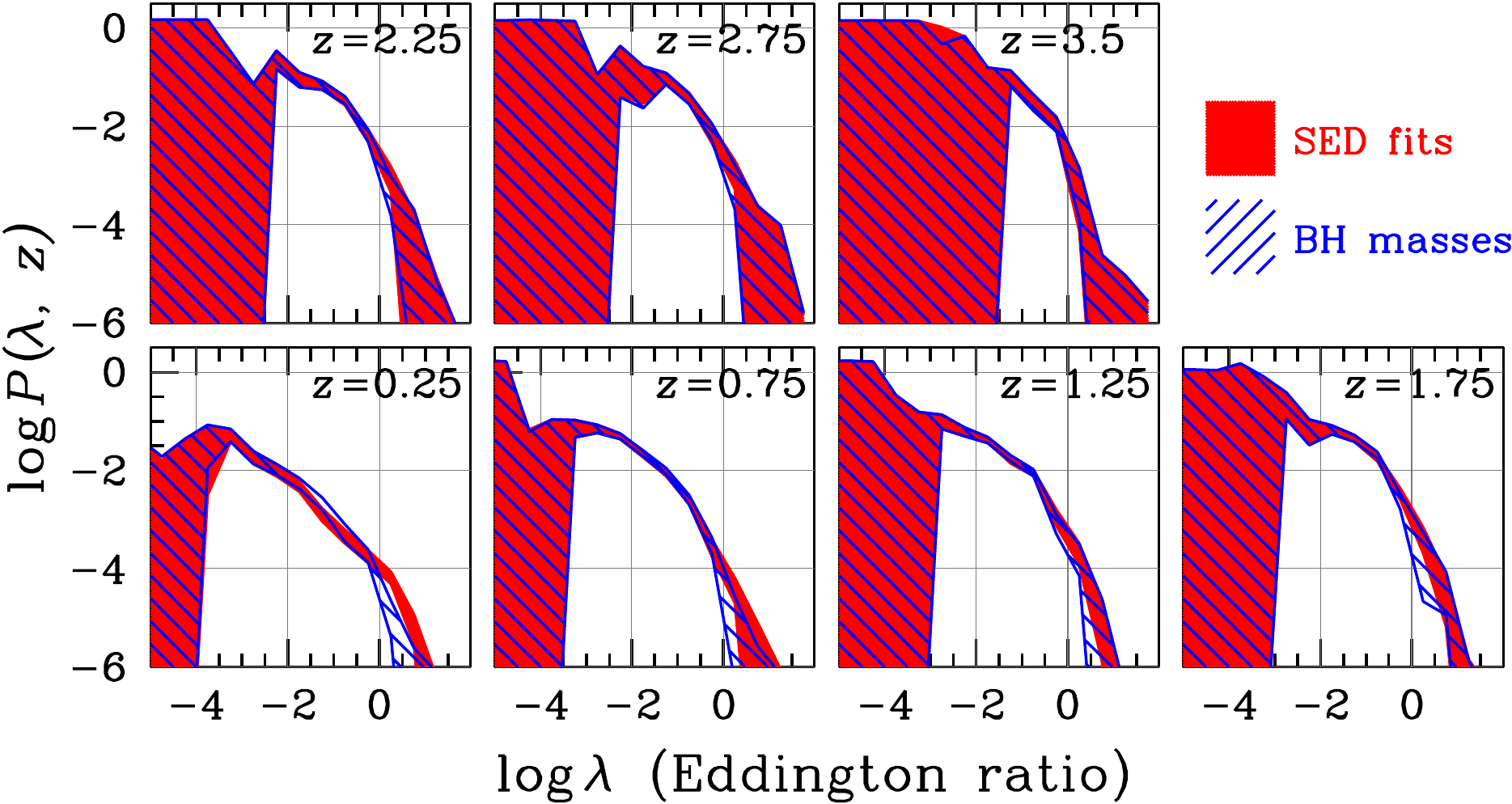}
\end{center}
\caption{Specific accretion-rate distribution for different methods of
approximating the host-galaxy stellar mass of broad-line quasars.. The
blue  hatched  regions are  the  same  as  those presented  in  Figure
\ref{fig_plz} and correspond to  the case of stellar masses determined
from virial black-hole masses.  The red shaded curves indicate stellar
masses determined using AGN/galaxy  decomposition via template fits to
broad-band photometry.  }\label{fig:plz_bh_mstar}
\end{figure*}

Figure \ref{fig:plz_bh_mstar} compares  the $P(\lambda,z)$ for the two
different approaches  of approximating the stellar  mass of broad-line
quasars. There is a steepening of the high specific accretion-rate end
of  the   distributions,  when  using  virial   black-hole  masses  to
approximate the  stellar mass of the underlying  galaxy. Most affected
are  the  specific  accretion-rate  distribution  in  the  two  lowest
redshift  bins.   This  behaviour  arises  because   these  X-ray  AGN
populations include a large number of luminous broad-line quasars from
the XMMSL sample. Nevertheless,  the overall effect is moderate and
has no impact on the any of our final results and conclusions.

\section{Comparison with Aird et al.}\label{sec:comp_aird}

\begin{figure*}
\begin{center}
\includegraphics[height=0.85\columnwidth]{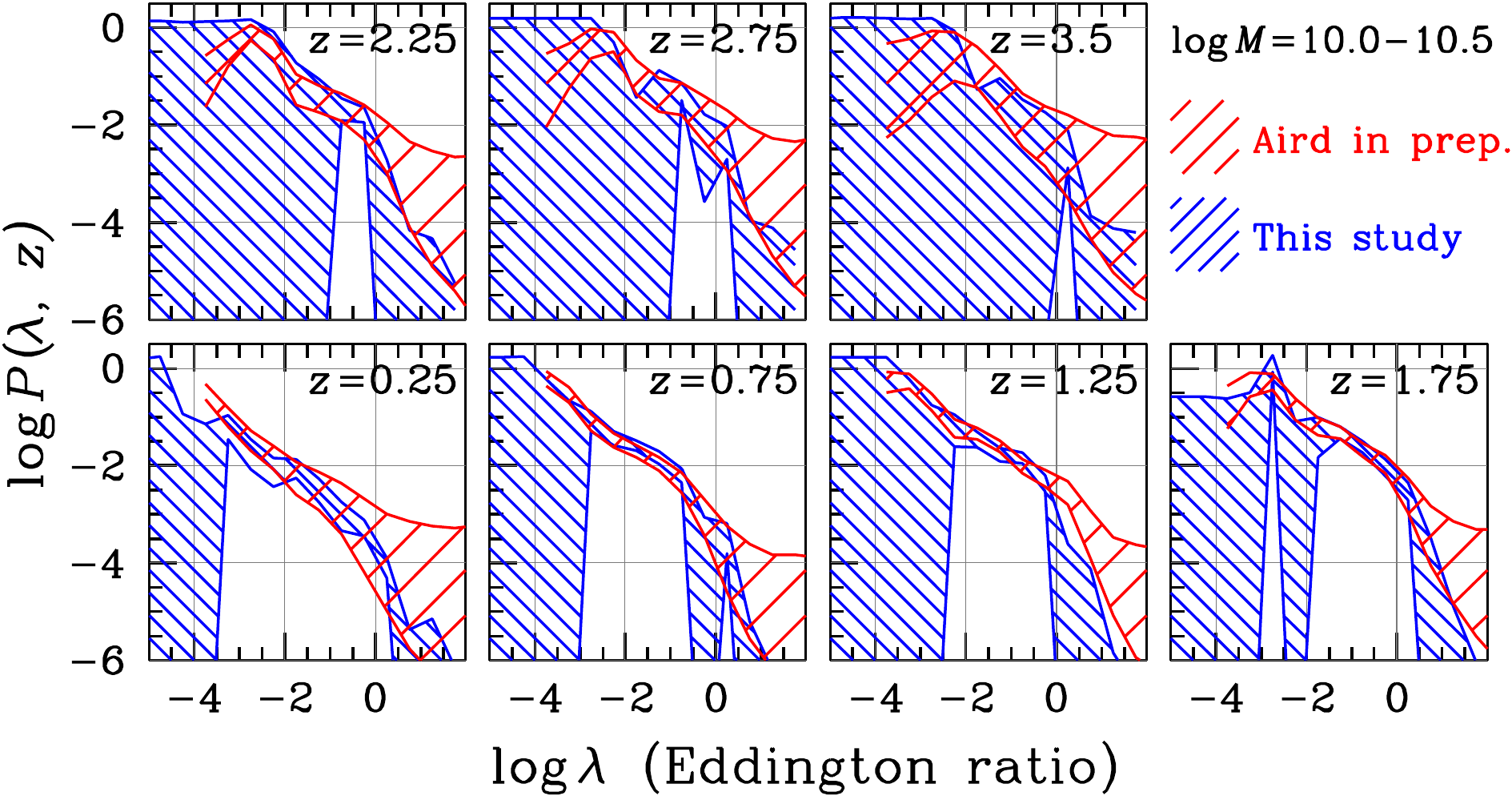}
\includegraphics[height=0.85\columnwidth]{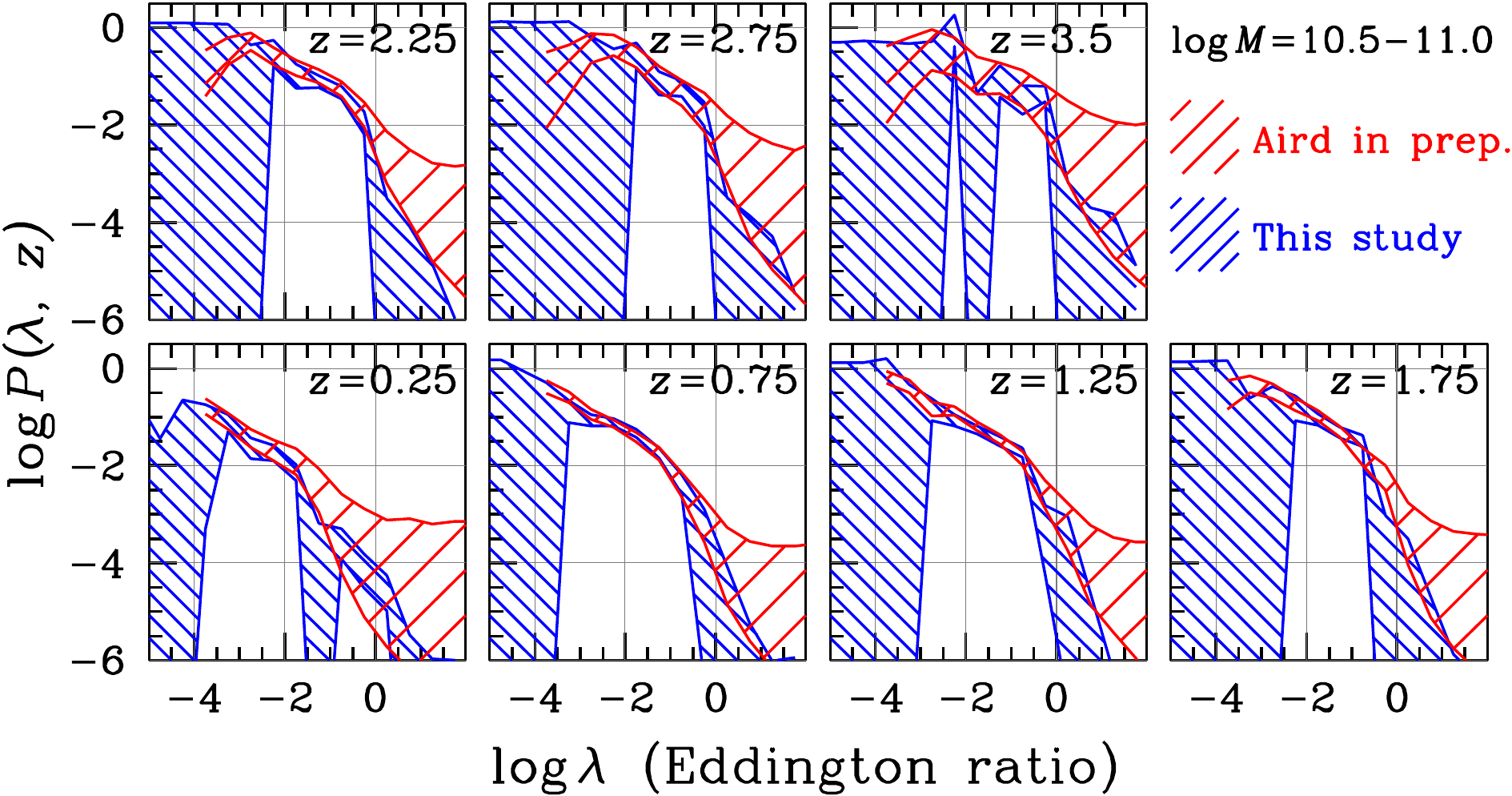}
\end{center}
\caption{Specific accretion-rate distribution,  $P(\lambda, z, M)$, of
X-ray selected  AGN in different  stellar mass and redshift  bins. The
top  set of  panels corresponds  to  the stellar  mass interval  $\log
M/M_{\odot}=10.0-10.5$ and the  bottom set of panels is  for the range
$\log M/M_{\odot}=10.5-11.0$.  In both panels the blue hatched regions
are our  constraints to  the specific accretion-rate  distribution and
the red  hatched regions are  those of Aird  et al.  (in  prep.).  The
extent  of  the  hatched  regions  correspond  to  the  5th  and  95th
percentiles  of  the   $P(\lambda,  z,  M)$  probability  distribution
function.  }\label{fig:plz_aird}
\end{figure*}

In a study that developed parallel to our work, Aird et al. (in prep.)
investigated  the   specific  accretion-rate  distribution   of  X-ray
selected  AGN,  using  as  starting  point  galaxy  samples  from  the
CANDELS-3DHST    (GOODS-S,   GOODS-N,    AEGIS,   COSMOS)    and   the
COSMOS/UltraVISTA field.  In the common fields between the Aird et al.
(in prep.) and our studies  the same X-ray observations are used. Aird
et  al.  (in  prep.)  used  the  Bayesian mixture  model described  in
\cite{Aird2017} to determine  the specific accretion-rate distribution
of  the  galaxy population  in  different  stellar  mass and  redshift
bins. The comparison  between the specific accretion-rate distribution
derived in our work and in Aird et al.  (in prep.)  is shown in Figure
\ref{fig:plz_aird}.   Overall the agreement  is good,  considering the
different  multiwavelength  data  used  in  the two  studies  and  the
independent  methodologies for  inferring the  specific accretion-rate
distribution of  AGN. At the  high-$\lambda$ end of  the distributions
the Aird  et al.   (in prep.) results  are systematically  higher that
ours, although we emphasise that  the tension is not significant given
the  large uncertainties.  It  is likely  that this  difference arises
because of the different  methodologies, the smoothness priors adopted
by Aird  et al.  (in prep.),  and possibly the  propagation of stellar
mass and redshift errors in our work.

\section{The X-ray Luminosity Function of AGN}\label{sec:agnLF}

A by-product of our  analysis deriving the $P(\lambda,z)$ distribution
are new observational constraints  on the X-ray luminosity function of
AGN by  combining {\it Chandra}  and {\it XMM} surveys  with different
characteristics   in   terms   of    area   and   depth   (see   Table
\ref{table_data}). This section  presents these constraints using both
a parametric  and a non-parametric  approach. It also  compares direct
estimates of the X-ray  luminosity function with that reconstructed by
convolving  the   $P(\lambda,z)$  distributions  derived   in  Section
\ref{sec_results}  with the  stellar mass  function of  galaxies using
Equation \ref{eq:lambda2phi}.

The parametric X-ray luminosity  function estimation is using the LADE
model as  described in Section  \ref{sec_method}. We also  account for
the contribution  of non-AGN among  X-ray sources using  the Schechter
parameterisation of  the X-ray luminosity function  of normal galaxies
presented  in   Section  \ref{sec_method}.   For   the  non-parametric
estimate,  a dimensional  grid  in X-ray  luminosity  and redshift  is
defined and  the AGN  space density is  assumed to be  constant within
each grid  pixel (see Section  \ref{sec_method}).  We account  for the
contribution   of   normal    galaxies   as   described   in   Section
\ref{sec_method},  i.e.,  by  adopting   a  Schechter  model  for  the
luminosity  function of  these sources  and fixing  the  parameters to
those listed in Table \ref{tab:xlf_results}.

\begin{figure*}
\begin{center}
\includegraphics[height=1.0\columnwidth]{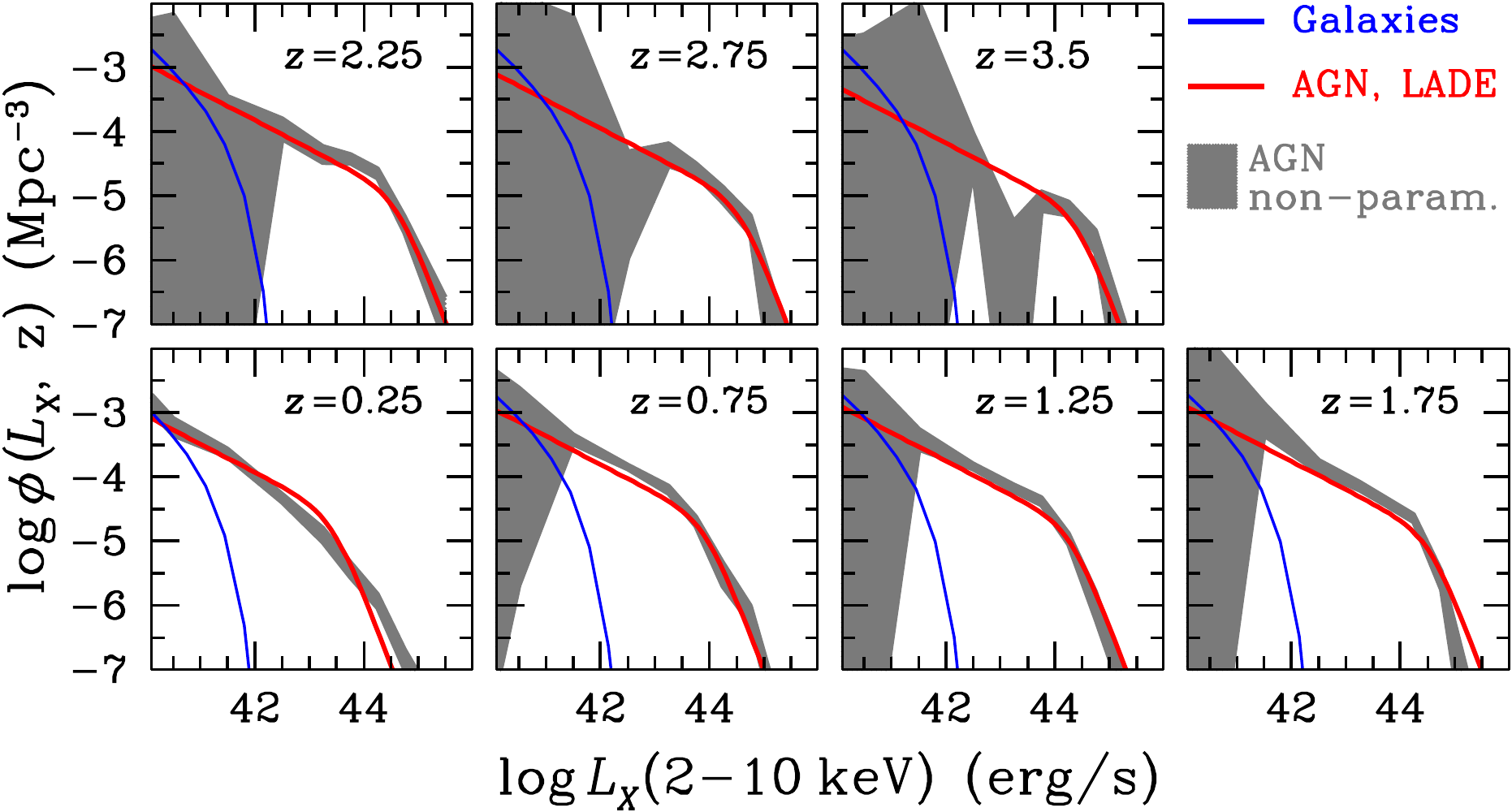}
\end{center}
\caption{X-ray  luminosity  function.   Each  panel corresponds  to  a
different   redshift  interval,  $z=0.0-0.5$,   $0.5-1.0$,  $1.0-1.5$,
$1.5-2.0$,  $2.0-3.0$, $3.0-4.0$.  The  mean redshift  of each  bin is
indicated in each panel.  The grey shaded region is the non-parametric
estimate  of the space  density of  extragalactic X-ray  selected AGN,
i.e., accounting for normal  galaxies within X-ray sources. The extent
of the grey shaded regions corresponds to the 5th and 95th percentiles
around the  median of the space-density probability  distribution at a
given  luminosity bin. The  red curves  represent the  parametric LADE
(Luminosity      and      Density      Evolution;     see      section
\protect\ref{sec_method})  model  for   the  AGN  luminosity  function
derived from the soft-band selected sample of \protect\cite{Aird2015}.
The blue  hatched regions correspond the X-ray  luminosity function of
normal  galaxies  determined  by  \protect\cite{Aird2015}  assuming  a
Schechter       function       parametrisation      (see       section
\protect\ref{sec_method}).}\label{fig:xlf_comp1}
\end{figure*}

Figure \ref{fig:xlf_comp1} compares  the parametric and non-parametric
constraints to the X-ray luminosity function of both AGN and galaxies.
Also  plotted in  this  Figure for  comparison  is the  LADE model  of
\cite{Aird2015}  for  their soft-band  selected  sample.  This  figure
shows  the level  of agreement  between parametric  and non-parametric
luminosity  function  estimates,   the  contribution  of  the  non-AGN
population to  the estimated space  densities and the  comparison with
previous constraints from the literature.

\begin{figure*}
\begin{center}
\includegraphics[height=1.0\columnwidth]{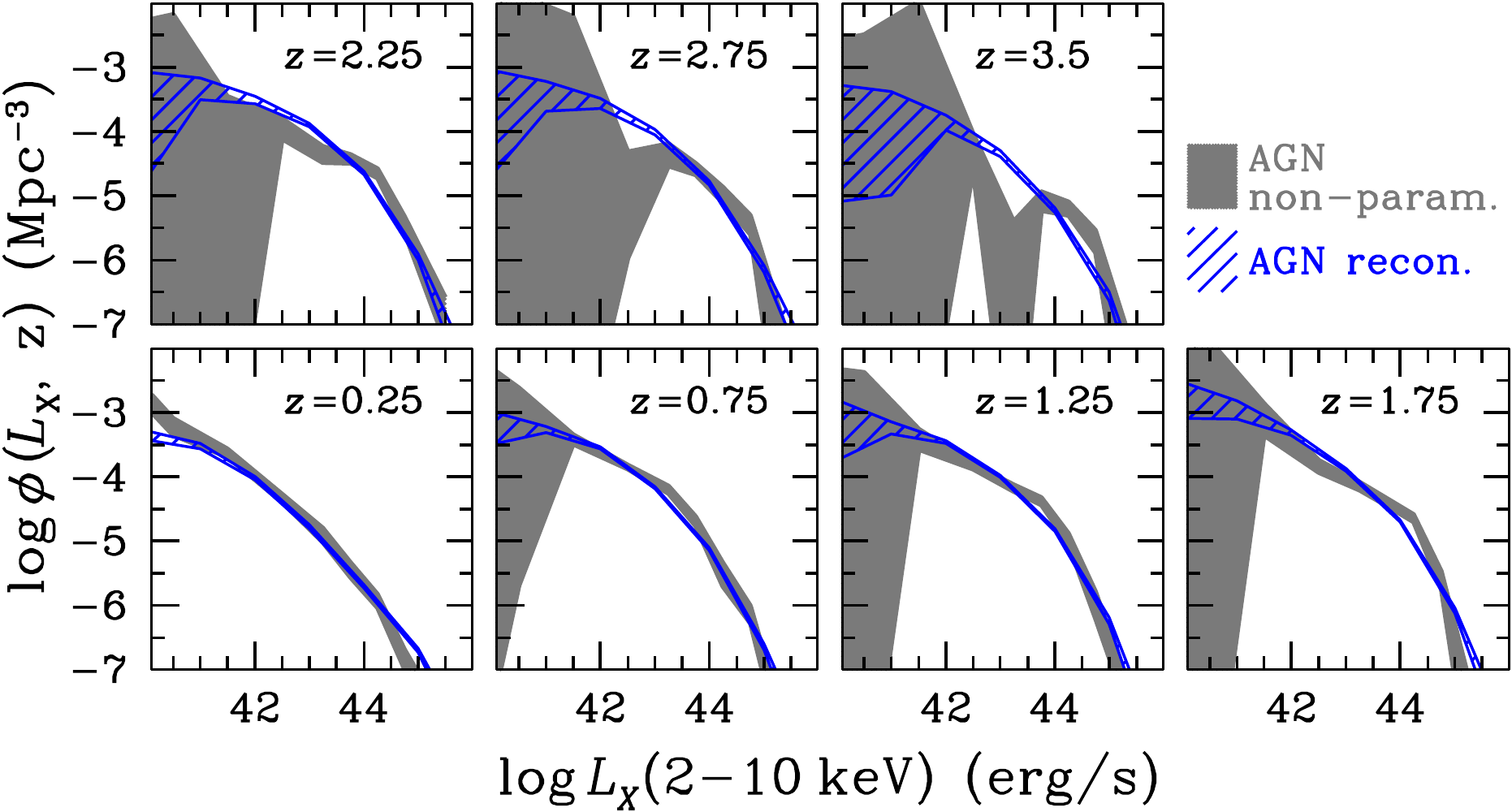}
\end{center}
\caption{AGN X-ray  luminosity function.  Each panel  corresponds to a
different   redshift  interval,  $z=0.0-0.5$,   $0.5-1.0$,  $1.0-1.5$,
$1.5-2.0$,  $2.0-3.0$, $3.0-4.0$.  The  mean redshift  of each  bin is
labelled in each  panel.  The grey shaded region  is the non-parametric
estimate  of  the  AGN  space  density, i.e.,  accounting  for  normal
galaxies  within  X-ray sources.  The  extent  of  the shaded  regions
corresponds to the  5th and 95th percentiles around  the median of the
space-density probability distribution at  a given luminosity bin. The
blue hatched regions correspond the reconstructed AGN X-ray luminosity
function  derived  by   convolving  the  $P(\lambda,z)$  distributions
produced  in  Section  \ref{sec_results}  with the  mass  function  of
galaxies. The  extend of blue  hatched regions corresponds to  the 5th
and 95th percentiles around the median.  }\label{fig:xlf_comp2}
\end{figure*}

Figure \ref{fig:xlf_comp2}  compares the reconstructed  AGN luminosity
function based on the  $P(\lambda,z)$ distributions derived in Section
\ref{sec_results}  with the  non-parametric X-ray  luminosity function
derived after accounting for  the contribution of normal galaxies. The
agreement between the two estimates is reasonable.

\bibliographystyle{mnras}
\bibliography{/home/age/soft9/BIBTEX/mybib} 

\bsp	
\label{lastpage}
\end{document}